\documentclass[usenatbib]{mnras}

\usepackage{newtxtext,newtxmath}
\usepackage[T1]{fontenc}

\usepackage{graphicx}
\usepackage{amsmath}
\usepackage{natbib}

\newcommand{\kev}{\,\rm keV}

\newcommand{\cm}{\,{\rm cm}}

\newcommand{\pc}{\,{\rm pc}}    
\newcommand{\kpc}{\,{\rm kpc}}
\newcommand{\Mpc}{\,{\rm Mpc}}

\newcommand{\s}{\,{\rm s}}

\newcommand{\yr}{\,{\rm yr}}
\newcommand{\Myr}{\,{\rm Myr}}
\newcommand{\Gyr}{\,{\rm Gyr}}
\newcommand{\erg}{\,\rm erg}
\newcommand{\ergs}{\,\rm erg\ s^{-1}}

\newcommand{\g}{\,{\rm g}}
\newcommand{\K}{\,{\rm K}}

\newcommand{\msun}{\,{\rm M_{\odot}}}
\newcommand{\zsun}{\,{\rm Z_{\odot}}}

\newcommand{\tadv}{t_{\rm flow}}
\newcommand{\cs}{c_{\rm s}}
\newcommand{\vc}{v_{\rm c}}

\newcommand{\rhotag}{\frac{{\rm d} \ln \rho}{{\rm d} \ln r}}

\newcommand{\Mdotcrit}{\Mdot_{\rm crit}}

\newcommand{\mach}{\mathcal{M}}

\newcommand{\Mstar}{M_*}
\newcommand{\Mdot}{{\dot M}}
\newcommand{\nH}{n_{\rm H}}

\newcommand{\kms}{\,\rm km\ s^{-1}}

\newcommand{\vr}{v_{r}}
\newcommand{\vphi}{v_{\phi}}

\newcommand{\Mvir}{M_{\rm vir}}

\newcommand{\vvir}{v_{\rm vir}}
\newcommand{\Rvir}{r_{\rm vir}}
\newcommand{\Tvir}{T_{\rm vir}}

\newcommand{\Rsonic}{r_{\rm sonic}}

\newcommand{\tcool}{t_{\rm cool}}
\newcommand{\Rcool}{r_{\rm cool}}

\newcommand{\tff}{t_{\rm ff}}

\newcommand{\tflow}{\tadv}

\newcommand{\Mhalo}{M_{\rm halo}}

\newcommand{\Rcirc}{R_{\rm c}}
\newcommand{\rvir}{r_{\rm vir}}

\newcommand{\tfive}{t(10^5\K)}
\newcommand{\disprho}{\langle\delta\rho/\bar\rho\rangle}

\title[hot and rotating CGM inflows]{Accretion onto disk galaxies via hot and rotating CGM inflows}

\author[Stern et al.]{
\parbox{\textwidth}{
Jonathan Stern$^{1}$, 
Drummond Fielding$^{2}$,
Zachary Hafen$^{3}$,
Kung-Yi Su$^{2}$,
Nadav Naor$^{1}$, 
Claude-Andr\'e Faucher-Gigu\`ere$^{4}$,
Eliot Quataert$^{5}$
and James Bullock$^3$
}
\vspace{0.4cm}\\
\parbox{\textwidth}{
$^1${School of Physics \& Astronomy, Tel Aviv University, Tel Aviv 69978, Israel}\\
$^2${Center for Computational Astrophysics, Flatiron Institute, 162 Fifth Avenue, New York, NY 10010, USA}\\
$^3${Department of Physics \& Astronomy, 4129 Reines Hall, University of California, Irvine, CA 92697, USA}\\
$^4${Department of Physics \& Astronomy and CIERA, Northwestern University, 1800 Sherman Ave, Evanston, IL 60201, USA}\\
$^5${Department of Astrophysical Sciences, Princeton University, Princeton, NJ 08544, USA}
}}
\vspace{-40pt}

\date{\vspace{-20pt}Accepted XXX. Received YYY; in original form ZZZ}

\pubyear{2020}

\begin{document}
\label{firstpage}
\pagerange{\pageref{firstpage}--\pageref{lastpage}}
\maketitle

\begin{abstract}
Observed accretion rates onto the Milky-Way and other local spirals fall short of that required to sustain star formation for cosmological timescales. A potential avenue for this unseen accretion is a rotating inflow in the volume-filling hot phase ($\sim10^6\,{\rm K}$) of the circumgalactic medium (CGM), as suggested by some cosmological simulations. 
Using hydrodynamic simulations and a new analytic solution valid in the slow-rotation limit, we show that a hot inflow spins up as it approaches the galaxy, while remaining hot, subsonic and quasi-spherical. Within the radius of angular momentum support ($\sim15\kpc$ for the Milky-Way) the hot flow flattens into a disk geometry and then cools from $\sim10^6\K$ to $\sim10^4\K$ at the disk-halo interface. Cooling affects all hot gas, rather than just a subset of individual gas clouds, implying that accretion via hot inflows does not rely on local thermal instability in contrast with `precipitation' models for galaxy accretion. 
Prior to cooling and accretion the inflow completes $\approx\tcool/\tff$ radians of rotation, where $\tcool/\tff$ is the cooling time to free-fall time ratio in hot gas immediately outside the galaxy. The ratio $\tcool/\tff$ may thus govern the development of turbulence and enhancement of magnetic fields in gas accreting onto low-redshift spirals.
We show that if rotating hot inflows are common in Milky-Way size disk galaxies, as predicted, then signatures of the expected hot gas rotation profile should be observable with X-ray telescopes and FRB surveys. 

\end{abstract} 


\section{Introduction}\label{s:intro}

Observation of neutral gas surrounding the Milky Way and nearby spirals suggest accretion rates of $0.05-0.2\msun\yr^{-1}$, falling short of the $1-2\msun\yr^{-1}$ required to sustain observed star formation rates (SFRs) for cosmological timescales \citep{Sancisi08,Putman12,Kamphuis22}. This `missing accretion' is often attributed to predominantly ionized gas clumps with temperature $\sim10^4\K$  \citep[e.g.,][]{Voit17}, observable mainly in UV absorption. It is however unclear if this phase can provide the necessary fuel for star formation, due to both uncertainties in converting UV absorption features to net accretion rates \citep[e.g.][]{Fox19}, and 
since hydrodynamic instabilities may disrupt and evaporate cool gas clumps before they reach the galaxy \citep{Heitsch09,Armillotta17,Tan23,Afruni23}. 
An alternative, less explored possibility is that accretion proceeds via a subsonic inflow in the volume-filling hot phase ($\sim10^6\K$) of the circumgalactic medium (CGM),  similar to classic `cooling flow' solutions
 discussed in the context of the intracluster medium \citep[ICM,][]{Mathews78,Fabian84}. Such hot CGM inflows are evident in modern cosmological simulations such as FIRE \citep{Stern21a, Hafen22} and TNG (\citealt{ZuHone23}; see also figure~9 in \citealt{Nelson19}).

Since the hot CGM is expected to have a net rotation 
\citep[e.g.,][]{Roskar10,Stevens17,Oppenheimer18,DeFelippis20, Truong21,Huscher21, Nica21},
an inflow will cause it to spin up. 
\cite{Stern20} used an idealized one-dimensional model to show that in Milky-Way mass halos, 
such a rotating hot inflow will remain hot down to the radius where the rotation velocity approaches the circular velocity $\vc=\sqrt{GM(<r)/r}$, at which point the gas cools to $\sim10^4\K$ and joins the ISM disk.  \cite{Hafen22} demonstrated that this picture applies, and is the dominant accretion mode onto $z\sim0$ Milky-Way mass galaxies in the FIRE-2 cosmological zoom simulations \citep{Hopkins18}. They further showed that 
the flow forms a coherently spinning disk prior to accretion onto the galaxy, and that this coherence may be a necessary condition for the formation of thin disk galaxies, consistent with conclusions from related FIRE-2 analyses \citep{Stern21a,Stern21b,Yu21, Yu22,Gurvich23}. 
It thus appears that a deep understanding of the physics of hot and rotating inflows could be crucial for understanding the evolution of local star forming disks. 
 
In the present paper, we complement the cosmological simulation-based analysis of hot and rotating CGM inflows in \cite{Hafen22}, by deriving an idealized, two-dimensional axisymmetric solution for inflowing and rotating hot CGM. 
Deriving an idealized solution allows identifying its dependence on system parameters and boundary conditions, and provides a basis for assessing the effects of additional physics. Our derivation is built on previous 1D
hot inflow solutions\footnote{Also called `cooling flows', though note this term has been used also for flows in which the gas does not remain hot \citep[e.g.,][]{McQuinn18}.} 
 which accounted for rotation in a highly approximate manner \citep{Cowie80, Birnboim03, NarayanFabian11, Stern20}. These 1D studies assumed the centrifugal force is directed outward in the spherical radius direction, so the solution remained spherically symmetric. 
Here we assume the centrifugal force is directed outward in the cylindrical radius direction, and derive a 2D axisymmetric solution which captures the transition from a quasi-spherical flow at large scales where rotational support is weak to a disk geometry at small scales where rotational support dominates. The idealized nature of our approach implies that insights may be applicable also to other astrophysical disks fed by spherical inflows, such as AGN disks in galaxy centers \citep[e.g.,][]{Quataert00} or protoplanetary disks in the center of star-forming clouds \citep[e.g.,][]{Fielding15}.

The inflowing hot CGM solution derived herein  focuses on the limit where feedback heating is subdominant to radiatively cooling, thus differing qualitatively from radially-static hot CGM models \citep[also known as `thermal balance' models, e.g.,][]{McCourt12,Sharma12,Voit17,Faerman17,Faerman20,Pezzulli17,Sormani18}, and from hot outflow models \citep{Thompson16, Schneider20}. Thermal balance models assume that radiative cooling is equal to feedback heating, thus inhibiting the hot inflow, while outflow models require that feedback heating dominates. 
Observational evidence for thermal balance is strong in the ICM, since the star formation rate (SFR) at the cluster center is small relative to the inflow rate $\Mdot$ implied by the X-ray emission $L_X$. This is the well-known `cooling flow problem', where $\Mdot\approx L_X/\vc^2\gg {\rm SFR}$ (see \citealt{McDonald18} for a recent revisit). There is, however, no similar cooling flow problem in disc galaxies. Upper limits on $L_X$ from the hot CGM of Milky-Way mass galaxies are a few $\times10^{40}\ergs$ \citep{Li13b,Li14,Anderson15,Comparat22}, and recent results based on eROSITA data indicate the actual emission may be comparable to this value \citep{Chadayammuri22}. For $\vc\approx200\kms$ this $L_X$ implies $\Mdot\approx1 \msun\yr^{-1}\sim{\rm SFR}$, in contrast with $\Mdot\gg{\rm SFR}$ deduced for the ICM. Similarly, estimates of $\Mdot$ in the Milky-Way halo suggest $\Mdot\approx0.1-1\msun\yr^{-1}$ \citep{Miller15, LiBregman17, Stern19}, again inconsistent with the $\Mdot\gg{\rm SFR}$ derived for the ICM. More massive spirals in which the hot CGM is detected in individual objects with $L_X\gtrsim10^{41}\ergs$ have SFR $\approx10\msun\yr^{-1}$ and hence also satisfy $\Mdot\sim{\rm SFR}$ \citep{Anderson16, Bogdan17, Das19}. A cooling flow problem akin to that in the ICM does not exist in the CGM of disk galaxies, allowing for the possibility that the hot CGM is inflowing.

The paper is organized as follows. In section~\ref{s:analytic} we discuss the structure and properties of hot and rotating CGM using analytic considerations, while in section~\ref{s:numeric} we derive a numerical solution. In section~\ref{s:additional physics} we consider the effect of additional physical mechanisms which were not included in the basic analysis, and in section~\ref{s:obs} we derive several observables of hot rotating CGM. Implications of our results are discussed in section~\ref{s:discussion} and section~\ref{s:summary} provides a summary. 

\section{The structure of hot and rotating CGM -- analytic considerations}\label{s:analytic}

The flow equations for radiating, ideal gas with adiabatic index $\gamma=5/3$ subject to an external gravitational potential $\Phi$ are 
\begin{eqnarray}
\label{e:mass}
\nabla \cdot (\rho \vec{v}) &=& -\frac{\partial \rho}{\partial t}\\
\label{e:momentum}
\left(\frac{\partial }{\partial t} + \vec{v} \cdot \nabla\right) {\vec{v}} &=& -\frac{1}{\rho}\nabla P - \nabla \Phi\\
\label{e:K}
\left(\frac{\partial }{\partial t} + \vec{v}\cdot\nabla\right) \ln K &=& -\frac{1}{\tcool} 
\end{eqnarray}
where $\rho, P$ and $\vec{v}$ are respectively the gas density, pressure, and velocity. We use $K\equiv P\rho^{-5/3}$ for the `entropy' (up to an exponent and a constant) and $\tcool$ for the cooling time, defined as
\begin{equation}\label{e:tcool}
 \tcool = \frac{3}{2}\frac{P}{\nH^2\Lambda} ~,
\end{equation}
where $\nH$ is the hydrogen density, $(3/2)P$ is the energy per unit volume, and $\Lambda$ is the cooling function defined such that $\nH^2\Lambda$ is the energy lost to radiation per unit volume per unit time.  Equations (\ref{e:mass})--(\ref{e:K}) neglect conduction, viscosity and magnetic fields, the potential effect of which will be assessed below. We also do not include a heating term in equation~(\ref{e:K}) since we search for a solution in the limit that heating is subdominant to cooling (see introduction).

\subsection{Hot CGM without angular momentum}\label{s:no AM}

\newcommand{\vcn}{v_{\rm c, 200}}
\newcommand{\Mdotn}{\Mdot_1}
\newcommand{\Lambdan}{\Lambda_{-22}}
\newcommand{\rn}{r_{10}}

We start with a brief review of steady-state ($\partial/\partial t=0$) hot inflow solutions without angular momentum, which were studied extensively mainly in the context of the inner ICM \cite[classic `cooling flows', e.g.,][]{Mathews78} and adapted to galaxy-scale halos by \cite{Stern19}. 
When angular momentum is neglected spherical symmetry can be assumed, and hence eqns.\ (\ref{e:mass})--(\ref{e:K}) reduce to 
\begin{eqnarray}
\label{e:mass rad}
4\pi r^2 \rho v_r &=& \Mdot \\
\label{e:momentum rad}
\frac{1}{2}v_r^2 &=& -\frac{1}{\rho}\frac{d \left(P_{\rm th}+P_{\rm turb}\right)}{dr} - \frac{\vc^2}{r}\\
\label{e:K rad}
v_r\frac{d\ln K}{dr} &=& -\frac{1}{\tcool}  
\end{eqnarray}
where $r$ is the spherical radius, 
$\Mdot$ is the mass flow rate (constant with radius in steady-state, down to the radius where stars form), $P_{\rm turb}=\rho \sigma_{\rm turb}^2$ is the turbulent pressure, and $\sigma_{\rm turb}$ is the turbulent velocity. Multiplying both sides of eqn.~(\ref{e:K rad}) by the free-fall time $\tff=\sqrt{2}r /\vc$ (defined as in   \citealt{McCourt12}) we get
\begin{equation}
\frac{\sqrt{2} v_r}{\vc}\frac{d\ln K}{d\ln r} = -\frac{\tff}{\tcool} ~.
\end{equation}
This equation indicates that if $\tcool\gg\tff$, then either the flow is isentropic with $d\ln K/dr\approx0$ as in the \cite{Bondi52} solution, or that the inflow velocity is small, i.e.,
\begin{equation}\label{e:CF condition}
\frac{v_r}{\vc} \sim \left(\frac{\tcool}{\tff}\right)^{-1} \ll 1 ~.
\end{equation}
The solutions discussed here correspond to the latter type of solutions where $v_r\ll \vc$, known as cooling flow solutions in the ICM literature. Hydrodynamic simulations show that initially static gas with $\tcool\gg\tff$ converges onto a cooling flow solution within a timescale $\tcool$, rather than onto an isentropic flow \citep[e.g.,][]{Stern19}. 

To derive an analytic approximation, one can neglect in eqn.~(\ref{e:momentum rad}) the small inertial term $v_r^2$ and the turbulent term which is also expected to be small $P_{\rm turb}\sim (v_r/\vc)^2 P_{\rm th}$ (see section \ref{s:turbulence} below). Further approximating the gravitational potential as isothermal with some constant $\vc$ then gives \citep[see section 2 in][]{Stern19}: 
\begin{eqnarray}
\label{e:rad solution}
  \cs^2 &=&  \frac{10}{9}\vc^2  \nonumber\\
  T &=&  2.0\cdot 10^6 \,\vcn^2 \K   \nonumber\\
  \nH &=& 0.8 \cdot 10^{-3} \, \rn^{-1.5} \vcn \Mdotn^{0.5} \Lambdan^{-0.5}\cm^{-3} \nonumber\\
  \tcool &=& 370 \, \rn^{1.5}\vcn \Mdotn^{-0.5} \Lambdan^{-0.5} \Myr \nonumber\\
  -v_r  =  \frac{r}{\tcool} &=& 27 \,\rn^{-0.5}\vcn^{-1} \Mdotn^{0.5} \Lambdan^{0.5} \kms \nonumber\\
  -\mach_r =  \sqrt{\frac{9}{20}}\frac{\tff}{\tcool} &=& 0.13\, \rn^{-0.5}\vcn^{-2} \Mdotn^{0.5} \Lambdan^{0.5} 
\end{eqnarray}
where $\cs$ is the sound speed, $T$ is the temperature, $\mach_r \equiv v_r/\cs$ is the radial Mach number of the flow, and we normalized by the following numerical values: $\rn=r/10\kpc$, $\vcn=\vc/200\kms$, $\Mdotn=\Mdot/1\msun\yr^{-1}$,  and $\Lambdan=\Lambda/10^{-22}\erg \cm^3\s^{-1}$. Eqn.~(\ref{e:rad solution}) treats $\Mdot$ as the free parameter, though one can also treat the CGM mass or density profile normalization as a free parameter, and then $\Mdot$  follows from the density relation in eqn.~(\ref{e:rad solution}).

The numerical values used in eqn.~(\ref{e:rad solution}) are appropriate for the Milky Way CGM: $\Mdot$ is taken to be roughly half the star formation rate (SFR) of $\approx1.5-2\msun\yr^{-1}$ \citep{BlandHawthorn16}, as expected in steady-state where the ISM mass is constant with time, and $\approx40\%$ of the stellar mass formed is ejected back into the ISM via winds and supernovae \citep[e.g.][]{Lilly13}. This $\Mdot$ is also consistent with X-ray absorption and emission constraints on the hot CGM of the Milky-Way (see introduction). The value of $\Lambda$ is appropriate for $T=2\cdot10^6\K$ gas with  metallicity $Z_\odot/3$, as measured for the Milky-Way CGM \citep{Miller15}.

Eqn.~(\ref{e:rad solution}) reveals several properties of the non-rotating solution. The inflow velocity $v_r$ is equal to $r/\tcool$, so the accretion time equals the cooling time, as expected in a cooling flow. This also implies that the entropy drops linearly with decreasing radius (see eqn.~\ref{e:K rad}). Additionally, the inflow has a temperature which is independent of radius and roughly equal to the halo virial temperature, despite radiative losses. This is a result of compressive heating during the inflow balancing radiative cooling. 

The solution in eqn.~(\ref{e:rad solution}) also highlights that the parameter $\tcool/\tff$ sets the Mach number of the flow, and thus also the sonic radius of the flow where $|\mach_r|=1$:
\begin{equation}\label{e:Rsonic}
\Rsonic \approx r(\tcool=0.7\tff) = 0.17\, \vcn^{-4} \Mdotn \Lambdan \kpc ~,
\end{equation}
where we approximated $\sqrt{9/20}\approx0.7$. Note that near and within the sonic radius the assumption of a quasi-hydrostatic flow is invalid, so the estimate of $\Rsonic$ is approximate.
Equation~(\ref{e:Rsonic}) indicates that $\Rsonic$ is well within the galaxy for Milky-Way parameters, though it can be on CGM scales in lower mass galaxies where $\vc$ is lower, or at higher redshift where $\Mdot$ is higher. In this paper we focus on systems with $\Rsonic$ smaller than the galaxy scale, so the quasi-hydrostatic approximation is valid throughout the CGM. 

Another important scale of cooling flows is the cooling radius $\Rcool$ at which $\tcool$ equals the system age or the time since the last heating event. 
This scale is not part of the steady-state solution, since a cooling flow develops on a timescale $\tcool$ and thus steady-state is achieved only at $r<\Rcool$. A time-dependent solution for how the outer boundary of the inflow at $r\sim\Rcool$ expands as $\Rcool$ grows was derived by \cite{Bertschinger89}. For the above parameters $\tcool=10\Gyr$ occurs at $r=110\kpc$, so when necessary we assume $\Rcool=100\kpc$. The present study however focuses on smaller radii of $r\lesssim40\kpc$ where the dynamical effects of angular momentum are most pronounced,  so $\tcool$ is short relative to cosmological timescales and thus steady-state is more likely to be achieved. 
This inner CGM region is also less susceptible to cosmological effects not included in our analysis, such as non-spherical accretion and satellite galaxies \citep{Fielding20b}.

\subsection{Rotating hot CGM inflows -- the circularization radius}\label{s:yes AM}

\newcommand{\Rcyl}{R}

Given some net angular momentum in the hot gas, for example due to torques applied by neighboring halos, the rotation velocity will increase as the gas inflows. We can hence define a circularization radius $\Rcirc$ as the radius where the rotational velocity equals the circular velocity and the flow becomes rotationally supported:
\begin{equation}
\Rcirc\equiv R(v_\phi=\vc) ~,
\end{equation}
where here and henceforth we use $R$ for the cylindrical radius (and $r$ for the spherical radius). One can also express $\Rcirc$ using the specific angular momentum of the hot gas $j$: 
\begin{equation}\label{e:Rcirc0}
j = \vc(\Rcirc)\Rcirc ~.
\end{equation}
Cosmological considerations can be used to estimate a typical $\Rcirc$. In a $\Lambda$CDM universe, a given dark matter halo is expected to have a spin parameter $\lambda$ which on average equals:
\begin{equation}
 \lambda \equiv \frac{J}{\sqrt{2}\Mvir \vvir \Rvir} \sim 0.035
\end{equation}
\citep[e.g.,][]{Bullock01,rodriguezpuebla16}, 
where $\Mvir$, $\vvir$, $\Rvir$, and $J$ correspond to the halo mass, virial velocity, virial radius, and angular momentum, correspondingly. Assuming the hot CGM has roughly the same spin as the dark matter halo as suggested by cosmological simulations \citep[e.g.,][]{Stewart13,Stewart17,DeFelippis20,Hafen22}, and assuming that near the disk $\vc=f_{\vc}\vvir$ with $f_{\vc}\gtrsim1$ we get
\begin{equation}\label{e:Rcirc}
 \Rcirc \approx \sqrt{2}\lambda f_{\vc}^{-1}\Rvir \sim 15f_{\vc}^{-1}\left(\frac{\Rvir}{300\kpc}\right)\kpc ~.
\end{equation}
Comparison of eqn.~(\ref{e:Rcirc}) with eqn.~(\ref{e:Rsonic}) implies that $\Rcirc\gg\Rsonic$ in Milky-Way halos. Thus, a hot CGM inflow in a Milky-Way halo is expected to become rotation-supported well before it transitions into a supersonic flow. This conclusion is also apparent if we estimate the radial Mach number near $\Rcirc$. Using eqn.~(\ref{e:rad solution}) we have 
\begin{equation}\label{e:MW t_ratio}
\frac{\tcool}{\tff}(\Rcirc) = 6.3\,\left(\frac{\Rcirc}{15\,{\rm kpc}}\right)^{0.5}\vcn^{2} \Mdotn^{-0.5} \Lambdan^{-0.5} ~,
\end{equation}
and hence 
\begin{equation}
\mach_r(\Rcirc)\approx 0.7(\tcool/\tff)^{-1} \approx 0.1 ~.
\end{equation}

The difference between CGM with $\Rsonic<\Rcirc$ and CGM with $\Rsonic>\Rcirc$ was discussed by \cite{Stern20}, and is related to the classic distinction between `hot mode' and `cold mode' accretion \citep{White78,Birnboim03,Fielding17}. In this paper we focus on systems with $\Rsonic<\Rcirc$ so the hot accretion mode dominates. 

\subsection{Rotating hot CGM inflows -- fluid equations and outer boundary condition}\label{s:yes AM2}

Accounting for angular momentum, and assuming steady-state and axisymmetry, the $r$ and $\theta$ components of the momentum equation~(\ref{e:momentum}) reduce to 
\begin{eqnarray}
\label{e:HSE r}
\frac{\partial{P}}{\partial r}  &=&  -\rho\frac{\vc^2}{r} + \rho\Omega^2 r \sin^2\theta  \\
\label{e:HSE theta}
\frac{\partial{P}}{\partial \theta} &=& \rho\Omega^2 r^2 \sin\theta\cos\theta 
\end{eqnarray}
where $\theta$ is the angle relative to the rotation axis and $\Omega=\vphi/(r\sin\theta)$ is the angular frequency ($v_r,v_\theta,v_\phi$ are the velocity vector components). 
We neglect the inertial $v_r^2$ term since its magnitude relative to the other terms is of order $\mach_r^2\approx(\tcool/\tff)^{-2}$. We similarly neglect the $v_\theta^2$ and $v_r v_\theta$ terms, since motion in the $\theta$ direction is a result of the combination of radial and rotational motions, and hence $v_\theta$ is of the same order as $v_r$ or smaller. 
The momentum equation in the $\phi$ direction is 
\begin{equation}\label{e:AM}
v_r\frac{\partial}{\partial r} (\Omega r^2 \sin^2\theta)  = -v_\theta\frac{\partial}{r\partial \theta}(\Omega r^2 \sin^2\theta) ~,
\end{equation}
which indicates that the specific angular momentum $j=\Omega r^2 \sin^2\theta$ is conserved along flowlines, as expected under our assumption of axisymmetry.  
The mass and entropy equations (\ref{e:mass}) and (\ref{e:K}) reduce to 
\begin{eqnarray}
 \label{e:mass 2d}
\frac{1}{r^2}\frac{\partial}{\partial r} (\rho v_r r^2)\ + & \frac{1}{r\sin\theta}\frac{\partial}{\partial \theta}(\rho v_\theta \sin\theta) & =\ 0\\
\label{e:K with AM}
v_r\frac{\partial\ln K}{\partial r}\ + & v_\theta\frac{\partial\ln K}{r\partial \theta} & =\ -\frac{1}{\tcool} ~.
\end{eqnarray}
 
 \newcommand{\Rcircdisk}{R_{\rm c,max}}
 \newcommand{\Rcircbase}{R_{\rm c}}
 \renewcommand{\Rcirc}{\Rcircdisk}
 
At large radii where the centrifugal terms in eqns.~(\ref{e:HSE r})--(\ref{e:HSE theta}) are small, the solution will approach the spherical no-angular momentum solution discussed in section~\ref{s:no AM}. 
In this limit we also expect $v_\theta\rightarrow 0$, so eqn.~(\ref{e:AM}) implies that angular momentum is independent of radius, i.e., it preserves the relation between $j$ and $\theta$ that exists in the outer boundary of the flow at $r\sim\Rcool$. We denote this boundary condition as
\begin{equation}\label{e:outer BC}
j_1(r,\theta) = \vc(\Rcircdisk)\Rcircdisk \sin^2\theta F(\theta) ~,
\end{equation}
where $\Rcircdisk\equiv\Rcircbase(\theta=\pi/2)$ and thus $\vc(\Rcircdisk)\Rcircdisk$ is the specific angular momentum at $\theta=\pi/2$ (see eqn.~\ref{e:Rcirc0}), while $F$ is some function that satisfies $F(\pi/2)=1$.
The subscript `1' denotes that this relation is valid at large radii where $v_\theta\rightarrow0$. Equivalently, the angular frequency at large radii is
\begin{equation}\label{e:outer BC2}
\Omega_1(r,\theta) = \frac{\vc(\Rcircdisk)\Rcircdisk}{r^{2}}F(\theta) ~.
\end{equation}
Eqn.~(\ref{e:outer BC}) implies that the circularization radius of a flowline which originates at polar angle $\theta$ is
$\approx\Rcircdisk \sin^2\theta F(\theta)$, with this expression being exact for constant $\vc$. We show below that flowlines accrete onto the disk at a cylindrical radius equal to their circularization radius. To avoid flowline intersection,  $\sin^2\theta F(\theta)$ is assumed to monotonically increase at $0\leq\theta\leq\pi/2$, and thus the midplane flowline has the largest circularization radius. 

The function $F$ can be estimated using the results of non-radiative cosmological simulations, which provide the initial conditions in the hot gas before a radiatively driven inflow develops. \cite{Sharma12b}  
 found that $\Omega$ is weakly dependent on $\theta$ in a sample of $10^{11}-10^{13}\msun$ halos in such simulations,   
with differences of $\approx15\%$ between the midplane and the rotation axis. This suggests $F(\theta)\approx1$, so we use $F(\theta)=1$ as our fiducial value.

We note that the \cite{Sharma12b}  simulations also suggest $j \propto r^{0.5} -r^{0.7}$ at $0.1<r/\rvir<1$ (see their fig.~2), i.e., the initial specific angular momentum profile of the hot gas increases outwards. We thus expect this increasing $j$ profile to be replaced by a flat $j\propto r^0$ profile (eqn.~\ref{e:outer BC}) once the inflow develops. Assuming the outer boundary of the inflow expands with time as $\Rcool$ grows (see section~\ref{s:no AM}), then we expect the normalization of the $j$ profile and hence $\Rcirc$ to increase with time, as the inflow originates from larger radii where $j$ is larger. This expected evolution however occurs on a cosmological timescale, and thus does not invalidate our steady-state assumption in the inner CGM which is achieved on the shorter cooling timescale at small CGM radii. Thus, for the purpose of deriving this steady-state solution in the inner CGM we treat $\Rcirc$ as a constant.

\subsection{Analytic solution in the slow rotation limit}\label{s: analytic sol}

\newcommand{\C}{\mathcal{C}}
\newcommand{\rad}[0]{\,{\rm rad}}
\newcommand{\trot}{t_{\rm rot}}

In this section we derive a solution to equations~(\ref{e:HSE r})--(\ref{e:K with AM}) which is accurate to lowest order in the effects of rotation. 
A similar approach was employed to study meridional flows in the Sun \citep{Sweet50, Tassoul07} and in Bondi flows \citep{Cassen76}.

The dynamical effects of rotation on hot CGM inflows increase with decreasing $r$, and become dominant at $r\lesssim\Rcirc$. To find the solution in the slow rotation limit we thus keep only terms which depend on $\Rcirc/r$ to the lowest order. It is straightforward to show that there are no terms of order $(r/\Rcirc)^{-1}$, since the lowest order of $\Omega$ is proportional to $(r/\Rcirc)^{-2}$ (eqn.~\ref{e:outer BC2}) and rotation enters the other flow equations only via the term $\Omega^2 r^2$ (eqns.~\ref{e:HSE r}--\ref{e:HSE theta}). We thus define a perturbation parameter
\begin{equation}\label{e:epsilon}
    \epsilon  = \left(\frac{r}{\Rcirc}\right)^{-2} ~,
\end{equation}
and search for a solution of the form
\begin{eqnarray}\label{e:forms}
 P_1(r,\theta) &=& P_0(r) \left[1 + \epsilon(r)f_P(\theta)\right] \nonumber\\
 \rho_1(r,\theta) &=& \rho_0(r) \left[1+ \epsilon(r) f_\rho(\theta)\right] \nonumber\\
 v_{r,1}(r,\theta) &=& v_{r,0}(r) \left[1+ \epsilon(r) f_{v_r}(\theta)\right] \nonumber\\
 v_{\theta,1}(r,\theta) &=& v_{r,0}(r) \epsilon(r) f_{v_\theta}(\theta) \nonumber\\
 \Omega_1(r,\theta) &=& \frac{\vc}{\Rcirc}\epsilon(r) ~.
\end{eqnarray}
Here, a subscript `$0$' denotes the non-rotating solution (eqn.~\ref{e:rad solution}), a subscript `$1$' denotes the approximate solution which we wish to find, and $f_P, f_\rho, f_{v_r}, f_{v_\theta}$ are some functions of $\theta$. The motivation for the form of $v_{\theta,1}$ will become apparent below. 
The solution for $\Omega_1$ is equivalent to eqn.~(\ref{e:outer BC2}) assuming $F(\theta)=1$ as suggested by non-radiative cosmological simulations. The implications of other forms of $F(\theta)$ on the solution are discussed below. 

We emphasize that the assumption of mild rotation is on top of the assumption of quasi-hydrostatic conditions, which allowed neglecting the quadratic velocity terms. Together, these assumptions imply that we assume the following conditions on timescales in the system:
\begin{equation}\label{e:timescales}
t_{\rm sc}\approx \tff,\ \ \tcool\gg\tff,\ \ \trot^2 \gg \tff^2 ~,
\end{equation}
where $t_{\rm sc}$ is the sound crossing time which is approximately equal to $\tff$ since the flow is quasi-hydrostatic, and $\trot=\Omega^{-1}$ is the rotation time. The squares on the relation between $\trot$ and $\tff$ follow from the relative size of the neglected centrifugal terms, which are of order $(\Omega r/\vc)^2 \sim (\trot/\tff)^{-2}$ (eqns.~\ref{e:HSE r}--\ref{e:HSE theta}).  

\newcommand{\dv}{\mathrm{d}}

Using eqn.~(\ref{e:forms}) in eqn.~(\ref{e:HSE theta}), and keeping only first-order terms in $\epsilon$ we get
\begin{equation}
 P_0\frac{\dv f_P}{\dv \theta} = \rho_0\vc^2\sin\theta\cos\theta   ~,
\end{equation}
where $P_0$ can be derived  from $P=(3/5)\rho\cs^2$ and eqn.~(\ref{e:rad solution}):
\begin{equation}\label{e:P0}
P_0=\frac{2}{3}\rho_0\vc^2~.
\end{equation}
We thus get
\begin{equation}\label{e:P1}
 f_P = \frac{3}{4}\sin^2\theta + \C ~,
\end{equation}
where $\C$ is a constant of integration to be determined below. 
Next, the first order terms in eqn.~(\ref{e:HSE r}) give
\begin{equation}
  f_P P_0\frac{\dv{(\ln\epsilon})}{\dv r} + f_P\frac{\dv{P_0}}{\dv r}  = -  f_\rho\rho_0\frac{\vc^2}{r}  + \rho_0\frac{\vc^2}{r}\sin^2\theta  ~. \\
\end{equation}
Using eqns.~(\ref{e:rad solution}), (\ref{e:epsilon}), and (\ref{e:P0}) then gives
\begin{equation}\label{e:rho1}
f_\rho = \frac{11}{4}\sin^2\theta + \frac{7}{3}\C ~.
\end{equation}
We can also define $T_1=T_0(1 + \epsilon f_T)$. Since $T\propto P/\rho$ we get to first order in $\epsilon$ that
\begin{equation}\label{e:T1}
f_T = f_P - f_\rho = -2\sin^2\theta-\frac{4}{3}\C ~,
\end{equation}
and similarly defining  $K_1=K_0(1+\epsilon f_K)$ and using $K\propto P/\rho^{5/3}$ we get
\begin{equation}\label{e:K1}
f_K = f_P - \frac{5}{3}f_\rho = -\frac{23}{6}\sin^2\theta-\frac{26}{9}\C ~.
\end{equation}
Eqns.~(\ref{e:P1}), (\ref{e:rho1}) and (\ref{e:T1}) indicate that the pressure and density increase when traversing from the rotating axis to the midplane at a fixed $r$, while the temperature decreases. The increase in pressure in the midplane is due to the higher density, which overcomes the lower effective gravity in the midplane which tends to decrease the pressure. 

In the entropy equation~(\ref{e:K with AM}), the second term on the left hand side is of order $\epsilon^2$ (see eqns.~\ref{e:forms}) and can be neglected. The first order terms of this equation are hence
\begin{equation}
 v_{r,0}f_K\frac{\partial \ln \epsilon}{\partial r}
+  v_{r,0}f_{v_{r}}\frac{\partial\ln K_0}{\partial r} 
= -\frac{1}{t_{\rm cool,0}}\left[f_\rho- (1+l)f_T\right] 
\end{equation}
where we use $\tcool\propto T/\rho\Lambda$, and approximate the temperature dependence of the cooling function as a power-law $\Lambda=\Lambda(T_0, Z)(T/T_0)^{-l}$. Further using $K_0\propto T_0/\rho_0^{2/3} \propto r$ and $t_{\rm cool,0}=-r/v_{r,0}$ based on eqn.~(\ref{e:rad solution}), and the above relations for $f_K, f_\rho, f_T$, we get
\begin{equation}\label{e:vr1}
f_{v_{r}} = \left(-\frac{35}{12}+2l\right)\sin^2\theta - \left(-\frac{19}{9}+\frac{4}{3}l\right)\C\approx -\frac{23}{12}\sin^2\theta -\frac{13}{9}\C~.
\end{equation}
In the approximation on the right we use $l=0.5$, appropriate for gas with $T\sim10^6\K$ and a characteristic CGM metallicity of $Z\approx0.3 Z_\odot$ \citep{Miller15}. 
Last, we use the continuity equation~(\ref{e:mass 2d}) to derive $v_\theta$, which we cast in the form $v_{\theta,1}=v_{r,0}\epsilon f_{v_{\theta}}$ (see eqn.~\ref{e:forms}). Keeping only first order terms we get
\begin{equation}
\frac{\rho_0\epsilon v_{r,0}}{r\sin\theta}\frac{\partial}{\partial \theta}\left(f_{v_{\theta}}\sin\theta\right) = 
 -\frac{f_\rho+f_{v_{r}}}{r^2}\frac{\partial}{\partial r}\left(\epsilon\rho_0 v_{r,0} r^2\right) ~.
\end{equation}
Using the definition of $\epsilon$ and that $\rho_0 v_{r,0} r^2$ is independent of $r$, we then get
\begin{equation}
 \frac{\partial}{\partial \theta}\left(\sin\theta f_{v_{\theta}}\right) = 
  2\sin\theta (f_\rho+f_{v_{r}})
\end{equation}
so
\begin{eqnarray}\label{e:vtheta1 with const}
 f_{v_{\theta}} &=& \frac{1}{\sin\theta} \int \left(\frac{5}{3}\sin^3\theta +\frac{16}{9}\C\sin\theta\right) d\theta \nonumber\\
 &=& \frac{1}{9\sin\theta} \left[\cos\theta\left(5\cos^2\theta - 15- 16\C \right) + \mathcal{D}\right] ~,
\end{eqnarray}
where $\mathcal{D}$ is another constant of integration. 
We further require $v_\theta(\pi/2)=v_\theta(0)=0$, in order to avoid a discontinuity at the rotation axis and to enforce symmetry with respect to the midplane. This gives $\mathcal{D}=0$ and $\C=-5/8$, and hence
\begin{equation}\label{e:vtheta1}
 f_{v_{\theta}} =  -\frac{5}{18}\sin 2\theta ~.
\end{equation}
Note that since $v_{r,0}$ is negative, then $v_{\theta,1}=v_{r,0}f_{v_\theta}\epsilon$ is positive for $\theta<\pi/2$ and negative for $\theta>\pi/2$, indicating that rotation diverts the flow towards the disc plane, as expected.

To summarize our solution we use the derived $\mathcal{C}=-5/8$ in eqns.~(\ref{e:P1}), (\ref{e:rho1}), (\ref{e:T1}) and (\ref{e:vr1}) and get
\begin{eqnarray}\label{e:solution}
 P_1        &=& P_0(r)\left(1 + \frac{\Rcirc^2}{r^2}\left(\frac{3}{4}\sin^2\theta-\frac{5}{8}\right)\right) \nonumber\\
 \rho_1     &=& \rho_0(r) \left(1+ \frac{\Rcirc^2}{r^2}\left(\frac{11}{4}\sin^2\theta-\frac{35}{24}\right)  \right)\nonumber\\
 T_1        &=& T_0 \left(1 -\frac{\Rcirc^2}{r^2} \left(2 \sin^2\theta-\frac{5}{6}\right)\right)\nonumber\\
 v_{r,1}      &=& v_{r,0}(r) \left(1 - \frac{\Rcirc^2}{r^2}\left(\frac{23}{12} \sin^2\theta-\frac{65}{72}\right)\right)\nonumber\\
 v_{\theta,1} &=& -v_{r,0}(r)\cdot\frac{5}{18}\frac{\Rcirc^2}{r^2}\sin(2\theta)\nonumber\\
 \Omega_1   &=& \frac{\vc \Rcirc}{r^2} 
\end{eqnarray}
where the zero-order terms are given by eqn.~(\ref{e:rad solution}). 
For a given $\vc$, the solution in eqn.~(\ref{e:solution}) depends on three parameters: $\Mdot$ and $\Lambda(T_0, Z)$ (or equivalently CGM mass and metallicity) which set the non-rotating solution, and $\Rcirc$ which sets the corrections due to rotation.

The solution in eqn.~(\ref{e:solution}) is for an outer boundary condition in which $\Omega_1$ is independent of $\theta$ (i.e., $F(\theta)=1$). In  Appendix~\ref{a:F} we give several solutions for $\Omega_1(\theta)\propto \sin^n(\theta)$ with integer $n$,  i.e., the rotation frequency at the outer boundary increases with angle from the rotation axis. In these solutions, the $\theta$-dependent term in the solution for $P_1$ is multiplied by a factor of $\sin^{2n}(\theta)/(n+1)$ relative to that in eqn.~(\ref{e:solution}), while the corresponding term for $T_1$ is multiplied by a factor of $\sin^{2n}(\theta)\cdot(n+2)/(n+1)$. Thus, the result above that $P$ and $\rho$ increase towards the midplane, while $T$ decreases, holds also  when $\Omega_1$ increases with $\theta$. These deviations from spherical symmetry however tend to become weaker, and more concentrated near the midplane, with increasing $n$.

\subsection{Number of revolutions in CGM inflows}\label{s:Nrot}

The number of revolutions around the rotation axis completed by a flowline can be derived from the ratio 
$v_\phi/v_r$ implied by eqn.~(\ref{e:solution}):
\begin{equation}
\frac{v_\phi}{v_r} = \frac{\Omega r \sin\theta}{v_r} = \frac{\vc\Rcirc\sin\theta}{r v_{r,0}}\left(1+\mathcal{O}(\epsilon)\right) ~.
\end{equation}
Using $v_{r,0}=r/\tcool(1+ \mathcal{O}(\epsilon))$ and $\tff=\sqrt{2}r/\vc$ we thus get 
\begin{equation}\label{e:vphi to vr}
\frac{v_\phi}{v_r} =  \sqrt{2}\frac{\tcool}{\tff}\frac{\Rcirc}{r}\sin\theta\left(1 +\mathcal{O}\left(\epsilon\right)\right)~.
\end{equation}
It is thus evident that in solutions with larger $\tcool/\tff$ the flowlines are more tightly wound, i.e.\ the flow rotates more prior to accreting onto the galaxy. The total number of radians a flowline rotates can be approximated with the following integral
\begin{equation}\label{e:Nrot}
    \int \Omega dt = \int \frac{v_\phi}{r\sin\theta} \frac{dr}{v_r} \approx  \sqrt{2}\int \frac{\tcool}{\tff}\frac{\Rcirc}{r^2}d r \approx 2\frac{\tcool}{\tff}(\Rcirc)~,
\end{equation}
where in the first approximation we used eqn.~(\ref{e:vphi to vr}) 
and in the second approximation we used $\tcool/\tff\propto r^{1/2}$ (eqn.~\ref{e:rad solution}) and integrated over the range $r/\Rcirc=1-7$. The upper bound of this range corresponds to the cooling radius (see section~\ref{s:no AM}), though since the integrand scales as $r^{-3/2}$ most of the rotation happens near $\Rcirc$, and the choice of upper limit does not significantly affect the result. Also, the lower bound of this range implies that neglecting the $O(\epsilon)$ term is not formally justified. Below we validate this approximation with a more accurate numerical calculation.

Equation~(\ref{e:Nrot}) suggests that the number of rotations is set by $\tcool/\tff$ near $\Rcirc$, where $\tcool/\tff$ is estimated by the non-rotating solution (eqn.~\ref{e:rad solution}). This can be understood intuitively since the cooling time tracks the inflow time ($\tcool=r/\left|v_r\right|$) while the free fall time tracks the rotation time near $\Rcirc$ ($\tff\approx r/\vc=r/v_\phi(\Rcirc)$).  For the MW halo in which $\tcool/\tff(\Rcirc)\approx6$ (eqn.~\ref{e:MW t_ratio}), we get that an accreting element rotates $\approx12$ radians prior to accretion. Furthermore, since $\tcool/\tff$ increases with $\vc$ (see eqn.~\ref{e:rad solution}) we generally expect the amount of rotation to increase with galaxy mass.

It is informative to extend the result in eqn.~(\ref{e:Nrot}) also to galaxies with halo masses $\ll10^{12}\msun$, where $\tcool<\tff$ and the volume-filling phase is cool and in free-fall with $v_r\approx -\vc$ (e.g., \citealt{Stern20}). Using $-v_r=\min(r/\tcool,\vc)$ in eqn.~(\ref{e:Nrot}) we thus get
\begin{equation}\label{e:Nrot complete}
    \int \Omega dt \approx \max (2\frac{\tcool}{\tff}(\Rcirc),1)
\end{equation}
Equation~(\ref{e:Nrot complete}) demonstrates that only in the hot accretion mode where pressure-support slows down  accretion relative to free-fall, the CGM has time to rotate significantly before accreting. In contrast, free-falling cold flows rotate merely by $\approx1$ radian prior to accretion. 

The ratio $\tcool/\tff$ at $r=\Rcirc$ fully determines the hot rotating inflow solution, up to a scaling of the physical dimensions, as we show in Appendix~\ref{a:Mdotcrit} using non-dimensional analysis of the flow equations. 
We similarly show that the solution is fully determined up to a scaling by the ratio $\Mdot/\Mdotcrit$, where $\Mdotcrit$ is the critical accretion rate in which $\tcool/\tff=1$ at $r=\Rcirc$ \citep{Stern20}.

\begin{figure*}
    \centering
    \includegraphics{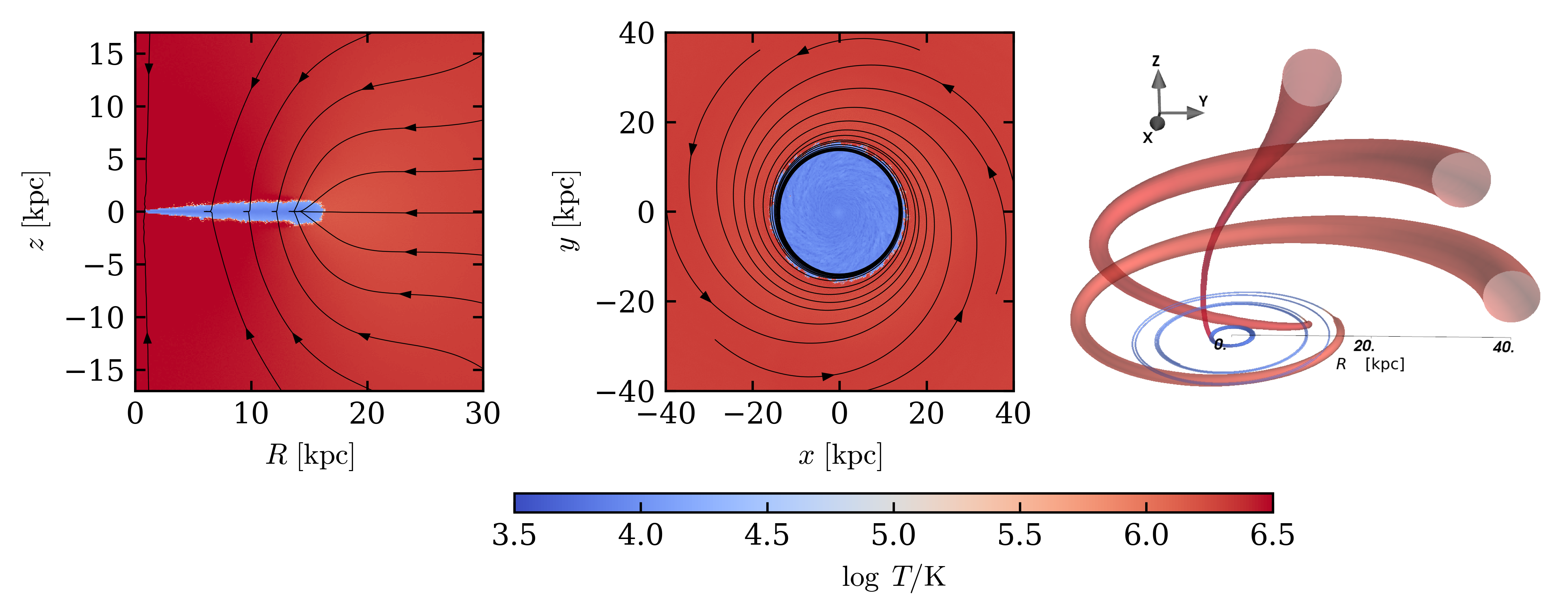}
\vspace{-0.5cm}

    \caption{Temperature map (color) and flowlines (black lines) in a hot rotating CGM inflow. Left and middle panels show the solution in the cylindrical $\Rcyl-z$ plane and the $x-y$ plane. The right panel depicts three specific flowlines as 3D `tubes', where the cross-section along each tube scales as $(\rho v)^{-1}$ and hence illustrates the compression of the flow.  Note that the hot $\sim10^6\K$ phase inflows along helical paths, and cools to $\sim10^4\K$ just prior to joining the ISM disk. 
    }
    \label{f:map}
\end{figure*}

\newcommand{\jn}{j/j_{\rm max}}
\newcommand{\Racc}{R_{\rm disk}}
\newcommand{\Mdotd}{\Mdot_{\rm disk}}

\subsection{Disk Accretion Radius}\label{s:Racc}

The distribution of disk radii at which gas accretes is an important input parameter for chemical evolution models and dynamical models of galaxy disks \citep[e.g.,][]{Schonrich09,Krumholz18, Wang22}. In this section we derive this distribution for hot and rotating CGM inflows.

Since the flow is spherical at large radii it holds that
\begin{equation}\label{e:AMD spherical}
\frac{d\Mdot}{d\theta_0}= \frac{1}{2}\Mdot\sin\theta_0 ~,
\end{equation}
where we use $\theta_0$ to denote the polar angle of a flowline at large radii. In hot rotating CGM inflows flowlines both conserve angular momentum (eqn.~\ref{e:AM}), and accrete onto the disk at their circularization radius $\Rcircbase(\theta_0)$ (see numerical calculation below). The value of $\Rcircbase$ can be estimated from eqn.~(\ref{e:outer BC}), which for constant $\vc$ and $F(\theta_0)=1$ implies 
\begin{equation}\label{e:Racc}
\Rcircbase(\theta_0) = \Rcirc\sin^2\theta_0 ~.
\end{equation}
We thus get
\begin{eqnarray}\label{e:Mdotd}
\Mdotd(<\Rcyl) &=& 2\int_0^{\theta_0(\Rcyl)} \frac{d\Mdot}{d\theta_0'}d\theta_0' = \Mdot(1-\cos\theta_0) \nonumber\\
&=& 
\begin{Bmatrix} 
\Mdot\left(1-\sqrt{1-\Rcyl/\Rcirc}\right) & & \Rcyl\leq\Rcirc \\
\Mdot & & \Rcyl\geq\Rcirc
\end{Bmatrix} ~,\nonumber\\
\end{eqnarray}
where the factor of $2$ before the integral accounts for the two sides of the disk. 
Eqn.~(\ref{e:Mdotd}) indicates that the median accretion radius is $(3/4)\Rcirc$, or $11.25\kpc$ for $\Rcirc=15\kpc$,  suggesting accretion weighted towards large disk radii. 
A similar conclusion of accretion mainly from large disk radii was deduced in FIRE cosmological simulations of Milky-Way like galaxies, in which the accretion is also dominated by hot rotating inflows \citep{Trapp22}. 

\begin{figure}
    \includegraphics{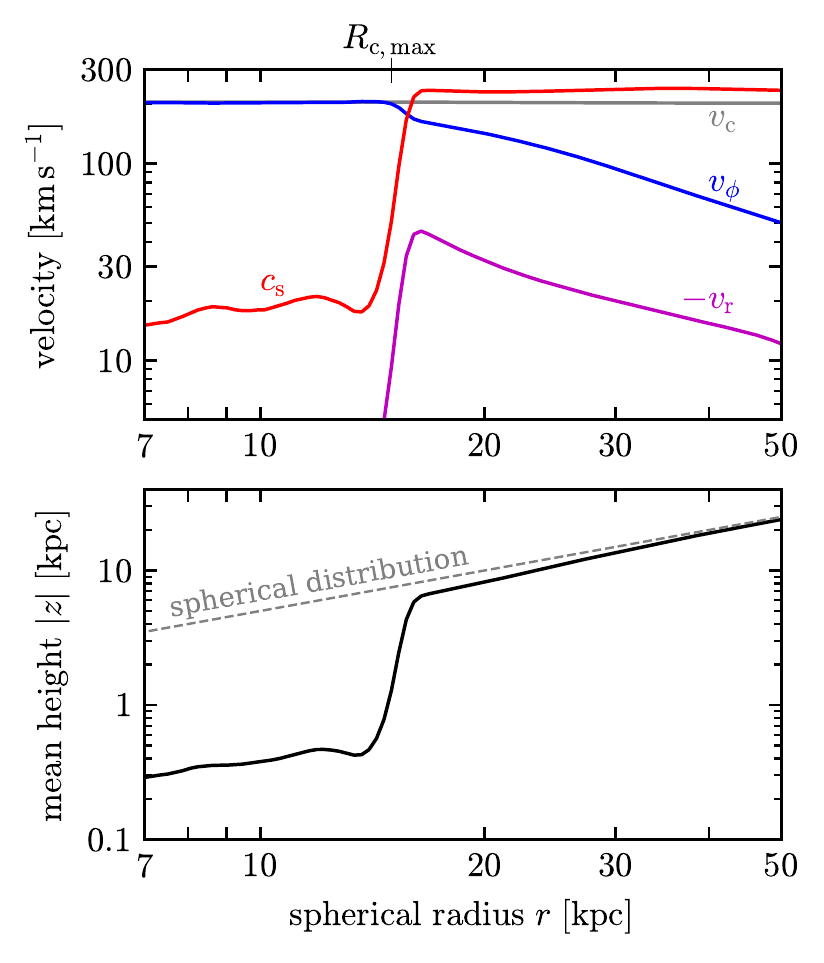}
    \caption{Radially-averaged kinematics and geometry of a hot rotating CGM inflow.  \textit{Top:} Lines show sound speed (red), rotation velocity (blue), and inflow velocity (magenta). \textit{Bottom:} Black line shows the average absolute height above the midplane. Rotation velocity increases inward due to conservation of angular momentum. At radii $r<\Rcircdisk$} where the inflow becomes fully rotation supported ($v_\phi=\vc$) the hot inflow cools out, the inflow halts ($|v_r|\rightarrow0$), and the geometry  transitions from quasi-spherical to a disk.
\label{f:eulerian view}
\end{figure}

\section{The structure of hot and rotating CGM -- numerical solution}\label{s:numeric}

In this section we derive numerical solutions for hot and rotating CGM inflows. To this end we run 3D hydrodynamic simulations until they converge onto a steady-state, where mass continuously flows through the hot CGM, cools and accretes onto the disk, and forms stars. Using this method to find the numerical solution has the advantage that it demonstrates that the solution is an attractor. The properties of the numerical solution are then compared to the approximate analytic solution derived in the previous section. 

\subsection{Setup}\label{s:setup}

We use the meshless finite-mass (`MFM') mode of GIZMO \citep{hopkins15}, a Lagrangian method with no inter-element mass flux, which enables us to track the history of each resolution element. 
The code accounts for self-gravity of the gas and stars, to which we add an acceleration term $-(\vc^2/r)\hat{r}$ with $\vc=200\kms$. This term approximates the gravitational field in the inner halo due to unmodelled dark matter and stars. 
Optically thin radiative cooling is calculated using the $z=0$ tables from \cite{Wiersma09} down to $T=10^4\K$, while optically thick radiative cooling to lower temperatures is disabled. 
All gas resolution elements  with $n_{\rm SF}>10 \cm^{-3}$ are converted into stellar particles. All stellar feedback processes are disabled. 

The density, temperature and radial velocity of gas is initialized with a spherical, non-rotating hot inflow solution from \cite{Stern19}, to which we add rotation corresponding to some $\Rcirc$. This solution is found by integrating the 1D spherically-symmetric and steady-state flow equations, starting at $\Rsonic=0.1\kpc$ and proceeding outward. The integration uses the same $\vc=200\kms$ and cooling function with $Z=0.3\zsun$ as in the simulation. The 1D solution has $\Mdot=1\msun\yr^{-1}$ and at radii $r\gg\Rsonic$ is well approximated by eqn.~(\ref{e:rad solution}), with $T=2\cdot10^6\K$ and $\Lambda_{-22}=0.3$.
We then randomly select initial positions in $(r,\phi,\theta)$ for the initial location of gas resolution elements, such that the radial mass profile reproduces that in the spherically symmetric solution. 
To add a net rotation to the gas, all resolution elements at $r>\Rcirc$ are initialized with $v_\phi=200\sin\theta (r/\Rcirc)^{-1}\kms$, with $\Rcirc=15\kpc$. This addition of rotation implies that the initial conditions are not in steady-state, since the initial pressure structure does not account for rotation support. We show below that  the simulation adjusts to a new steady-state within a cooling time of $<1\Gyr$, with a somewhat larger $\Mdot$ of $1.7\msun\yr^{-1}$. 

The mass of individual resolution elements is set to $m_b=1000\msun$ for elements at $r<100\kpc$. 
This mass resolution implies a characteristic size of $\approx(3m_b/4\pi\nH)^{1/3}\approx 0.2n_{-3}^{-1/3}\kpc$ for typical hot gas densities of $\approx10^{-3}n_{-3}\cm^{-3}$ near the disk scale, and smaller sizes for the denser cool gas.  For comparison, the 
height of the $\approx10^4\K$ gaseous disk which forms from the cooling of the hot gas is $\approx1\kpc$\footnote{This follows since the disk height to radius ratio is $\approx\cs/\vc$, and hence for a gaseous disk temperature $10^4\K$ we get $(\cs/\vc)\Rcirc\approx1\kpc$.}. 

Beyond $100\kpc$ the gas does not participate in the inflow since cooling times are too long, but needs to be included in the simulation in order to confine gas at smaller radii from expanding outward (in a realistic halo this confinement is achieved either by such hot gas with a long cooling time or by the ram pressure of infalling gas outside the accretion shock). To avoid  investing too much computing time in this confining outer gas, we sample the spherically-symmetric solution beyond $100\kpc$ with resolution elements which masses  increase by a factor of three every factor of $\sqrt{2}$ in radius, out to $3.2\Mpc$ where the sound-crossing time equals $10\Gyr$. In total the CGM is simulated with $3.1\times10^7$ resolution elements.

\newcommand{\Rd}[0]{R_{\rm d}}

We also add a galaxy to the initial conditions, using the MakeDisk code \citep{Springel05_2} with the following parameters. The stellar disk is initialized with mass $\Mstar=10^8\msun$, cylindrical radial scale length of $\Rd=\Rcirc/4=3.75\kpc$ spanning $0.03-4\Rd$, and a vertical scale length of $0.1\Rd$. The gaseous disk has a mass $M_{\rm disk~gas}=0.2\Mstar$, and the same exponential distribution as stars, and the bulge has a mass $M_{\rm bulge}=2\cdot10^7\msun$ and scale length $0.1\kpc$. 
We include in the MakeDisk calculation the same isothermal gravitational field used in the hydro simulation, and stellar and gas particles in the disk have the same $m_b=1000\msun$ resolution as in the CGM.  The choice of galaxy parameters is inconsequential as long as the initial mass is small compared to the accreted mass, which at $t=1\Gyr$ is $\Mdot t\sim10^9\msun$.  
The simulation is run for $3.5\Gyr$, with snapshots saved every $5\Myr$. At all times and radii the gravitational field is dominated by the included isothermal gravitational field with $\vc=200\kms$, rather than by the simulated gas and stars. 

Our setup is loosely based on the setup of \cite{Su19,Su20}, which simulated the behavior of gas in group and cluster-sized halos. A similar setup to ours for Milky-Way mass halos was employed by \cite{Kaufmann06} using an SPH code. This code was later found to over-predict artificial clumping of the cool gas \citep{Agertz07,Kaufmann09}. Our use of the MFM code addresses this numerical issue (see \citealt{hopkins15} for code tests). Additionally, since in our simulation \textit{all} the hot gas cools once it inflows past $\Rcirc$ (see below), the numerical details may affect the distribution of clumps and their typical sizes, but not the total mass which cools.

\newcommand{\thetalarge}{\theta_0}
\newcommand{\tacc}{t_{\rm acc}}

\newcommand{\tffeff}[0]{t_{\rm ff;\ eff}}

\subsection{Overview of results}

Figure~\ref{f:map} shows temperature maps in the simulation at $t>1\Gyr$, after the hot CGM phase converged onto an axisymmetric steady-state solution  within $r\approx40\kpc$. 
Steady-state and axisymmetry are evident 
from the small dispersion in hot CGM properties with time and $\phi$, as shown below.  
The left and middle panels respectively show the $\Rcyl-z$ plane and the $x-y$ plane (mass-weighted over $-10<z<10\kpc$). 
The figure shows that the hot gas fills the volume except in the disk region at $\Rcyl\lesssim \Rcirc=15\kpc$ and $\left| z \right| \lesssim 1\kpc$. 
Black lines depict flowlines in the two planes, derived as described below.  Three of these flowlines are also depicted in the right panel as 3D `tubes', rendered using the Mayavi software \citep{ramachandran2011mayavi} on snapshot data. The tube cross-section scales as $(\rho v)^{-1}$ and thus illustrates the compression of the flow. 

Figure~\ref{f:map} shows that the flowlines in the hot gas are helical, with the hot gas spiraling onto the galaxy. While inflowing, the gas initially remains hot with $T\sim\Tvir$, and then cools to $\sim10^4\K$ just prior to joining the ISM disk. This cooling is accompanied with strong compression of the flow, as evident from the sharp decrease in the width of the flow tubes in the right panel (tube thickness should drop to $\lesssim 0.01$ pixels upon cooling according to the $(\rho v)^{-1}$ scaling, though is plotted with one pixel for visibility).

Figure~\ref{f:eulerian view} plots radial shell-averaged velocities in the simulation after steady-state is achieved. 
The top panel shows that at radii $r>\Rcirc$ the sound speed $\cs$ (red) approximately equals $\vc$ (gray), indicating the hot gas is to first-order supported against gravity by thermal pressure, as indicated also by the slow inflow velocities of $-v_r\ll\vc$ (magenta line). At radii $r>\Rcirc$ the rotation velocity increases inward roughly as $r^{-1}$ due to conservation of angular momentum, reaching $v_\phi=\vc$ at $\Rcirc$. Within $\Rcirc$ the gas is fully rotationally supported and cool with $c_s\ll\vc$, and the radial velocity drops to zero. The associated change in geometry is evident in the bottom panel, which plots the average absolute height above the midplane $|z|$ in different radial shells. The gas distribution is close to spherically-symmetric at $r>\Rcirc$, in contrast with a thin disk distribution at $r<\Rcirc$.

\subsection{Accretion of the hot CGM onto the cool ISM}

\begin{figure}
    \includegraphics{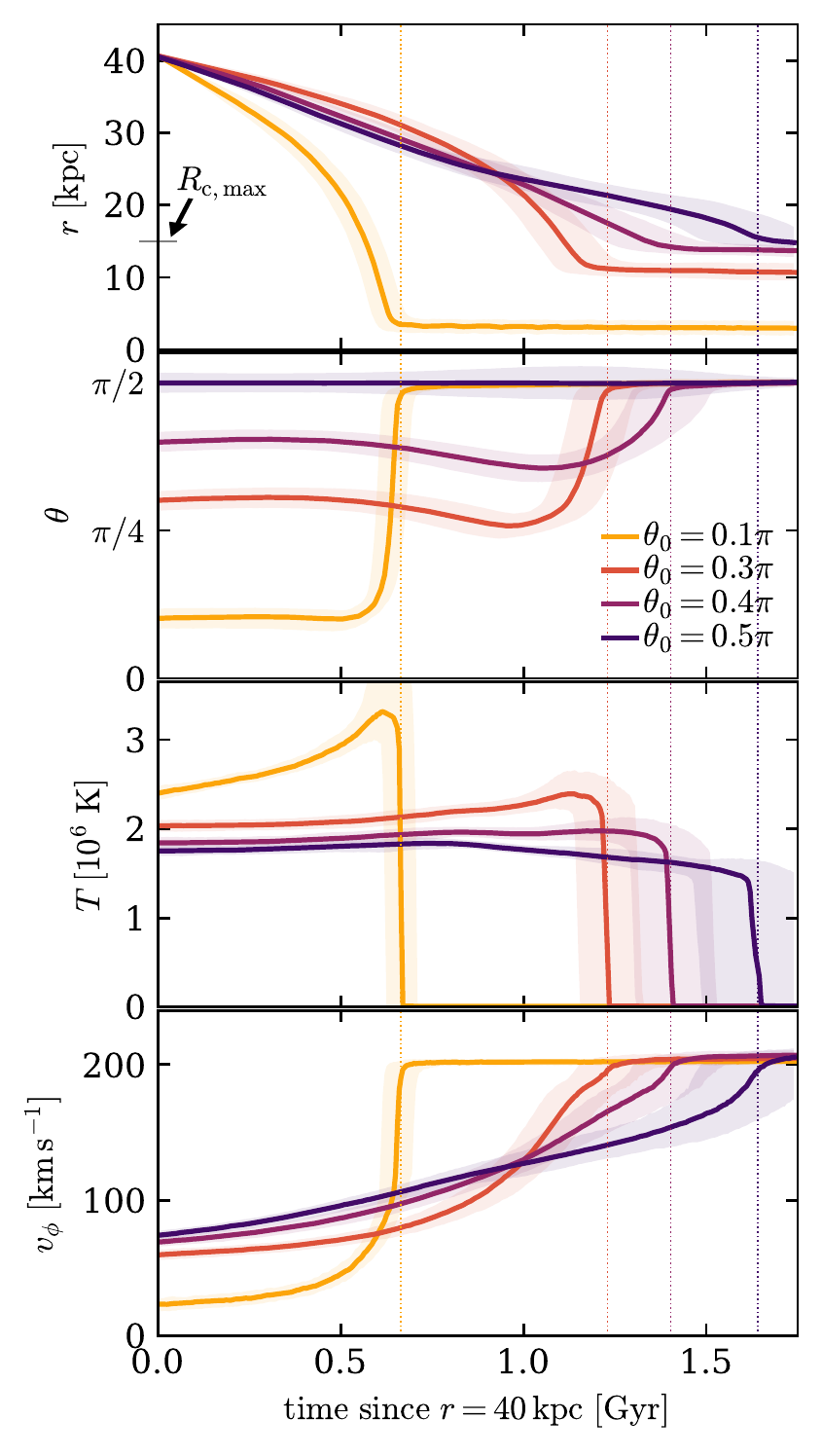}
    \vspace{-0.5cm}

    \caption{Gas properties along flowlines in hot rotating CGM inflows, versus time since a fluid element is at $r=40\kpc$. Panels show spherical radius, polar angle, temperature, and rotation velocity. Different lines and bands correspond to medians and $16^{\rm th}-84^{\rm th}$ percentiles of flowlines with different polar angles at large radii $\theta_0$. Vertical dotted lines indicate times where $T$ drops to $10^5\K$. Initially the flowlines have roughly constant  $\theta\approx\theta_0$. 
   About $200\Myr$ prior to cooling the flow geometry flattens ($\theta\rightarrow\pi/2$), and the temperature increases mainly in flowlines with small $\thetalarge$. 
 Cooling occurs when $\vphi$ reaches $\vc\approx200\kms$, i.e., at the circularization radius of the flowline $R_{\rm c}(\theta_0)$, indicating a transition from quasi-thermal pressure support against gravity to rotational support. 
Dispersion in hot gas properties prior to cooling is small, demonstrating the hot inflow is steady and axisymmetric.
}
\label{f:track properties}
\end{figure}

\begin{figure*}
    \centering
    \includegraphics{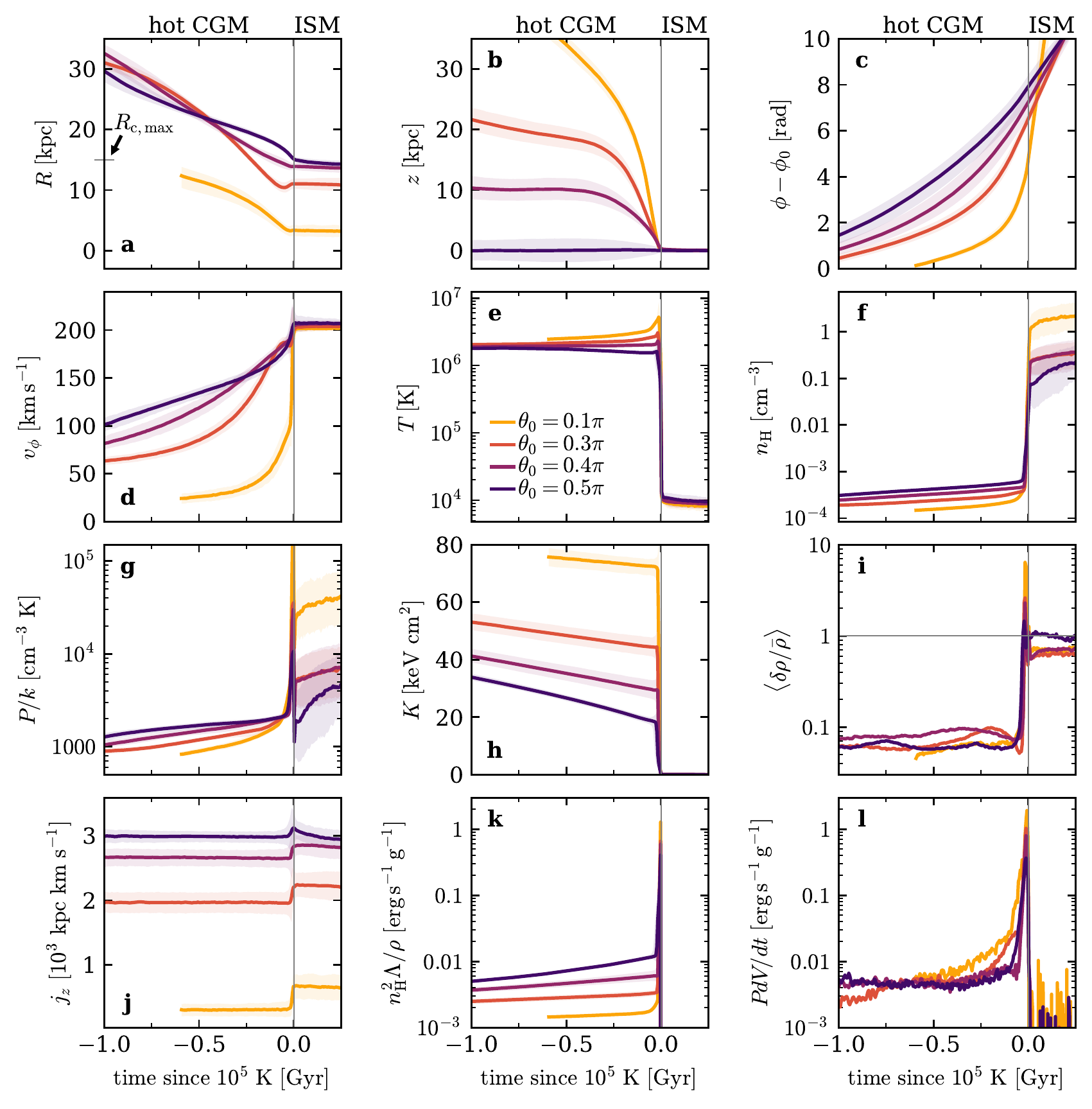}
    \caption{
Gas properties along flowlines in hot rotating CGM inflows, versus time since gas in the flowline cools. The time of cooling is equal to the time of accretion onto the ISM. 
From left to right and top to bottom the panels plot cylindrical radius, height above the midplane, total rotation since $r=40\kpc$, rotational velocity, density, temperature, pressure, entropy, density dispersion, specific angular momentum, radiative losses per unit mass, and compressive heating rate per unit mass. 
Colored lines and bands correspond to medians and $16^{\rm th}-84^{\rm th}$ percentile ranges of flowlines with different polar angles at large radii $\theta_0$, except in panel {\textbf l} where lines denote mean values.
At early times $t-\tfive<-200\Myr$ the temperature remains roughly constant at $\approx2\cdot10^6\K$ (panel {\textbf e}) since compressive heating and radiative cooling roughly balance (panels {\textbf k}--{\textbf l}). Also, density fluctuations are small (panel {\textbf i}), and the specific angular momentum is conserved (panel {\textbf j}). About $200\Myr$ prior to cooling the geometry of the flow starts to flatten into a disk  (panel \textbf{b}). At $t\approx\tfive$ rotation velocity reaches the circular velocity (panel \textbf{d}), densities increase by $\times300$ or more (panel \textbf{f}), and density fluctuations become significant (panel \textbf{i}). 
 }
    \label{f:track properties t5}
\end{figure*}

Figure~\ref{f:track properties} provides a Lagrangian view of hot inflowing CGM, by plotting median properties of resolution elements versus time since the element was at $r=40\kpc$. The properties depend on the initial polar angle of the flowline $\thetalarge$, so for each $\thetalarge$ in $(0.1\pi,0.3\pi,0.4\pi,0.5\pi)$ we group all resolution elements that reside at $40<r<41\kpc$ and $|\theta-\thetalarge|<0.025\pi$ at times $1<t<1.5\Gyr$. Then, for each $\thetalarge$ group we plot the median and $16-84$ percentile ranges of $r$, $\theta$, $T$, and $\vphi$. The $16-84$ percentile range thus accounts for the dispersion both with $\phi$ and with $t$, and specifically a small $16-84$ percentile range indicates that the solution is axisymmetric and in steady-state. 

Fig.~\ref{f:track properties} shows that at $r\gtrsim25\kpc$ (early times in this plot), the gas is hot and inflowing, with a somewhat larger inflow velocity in the $\theta_0=0.1\pi$ flowline near the rotation axis. Rotation is sub-Keplerian ($\vphi<\vc\approx200\kms$) but growing with time, as also indicated by Fig.~\ref{f:eulerian view}. The value of $\theta$ remains roughly constant and equal to $\theta_0$ for a given flowline. 
Then, when the flowlines reach radii of $r\lesssim20\kpc$,
the gas initially heats up and is diverted to the midplane ($\theta=\pi/2$), and then abruptly cools. The initial heating is more pronounced near the rotation axis, and is a result of the compression due to the change in geometry from spherical to disk-like (see below).  Cooling occurs simultaneously with $\theta$ reaching $\pi/2$, $\vphi$ reaching $\vc\approx200\kms$, and $r$ becoming constant, i.e.~when the hot gas has achieved full rotational support and transitioned to a flat, disk geometry. 
We thus identify the time of cooling also as the time of accretion from the hot CGM onto the ISM. A similar relation between cooling and accretion was identified by \cite{Hafen22} in cosmological zoom simulations of $z\sim0$ Milky Way-mass galaxies from the FIRE project. 

Note also that prior to cooling the $16-84$ percentile ranges in individual flowlines are small, indicating the hot CGM is axisymmetric and in steady-state. Furthermore, cooling and accretion happens to all hot inflow resolution elements, indicating that cooling is a global transition in the inflow, rather than occuring only in individual gas clumps as in precipitation models \citep[e.g.][]{Maller04, Voit17}. 

Figure~\ref{f:track properties t5} plots gas properties along flowlines versus $t-\tfive$, where $\tfive$ is defined as the time at which the temperature in the flowline equals $10^5\K$. This time is also marked with vertical lines in Fig.~\ref{f:track properties}, and as mentioned above is equivalent to the time of accretion onto the ISM. 
The twelve panels show cylindrical radius $\Rcyl$, $z$, total rotation $\phi-\phi_0$ where $\phi_0\equiv \phi(r= 40\kpc)$, rotation velocity, density, temperature, pressure, entropy, density dispersion, specific angular momentum, radiative cooling rate per unit mass $\nH^2\Lambda/\rho$, and compressive heating rate per unit mass $PdV/dt$ where $V=\rho^{-1}$ is the specific volume. 
The figure shows that while the flow is hot (at $t<\tfive$), the density and pressure of the hot inflow mildly increase with time, 
while the temperature is roughly constant 
and the entropy decreases. 
The increase in density is due to the contraction of the inflow, which also causes a compressive heating ate of order $PdV/dt\approx0.005\erg\s^{-1}\g^{-1}$. 
This compressive heating offsets comparable radiative losses (compare panels {\textbf k} and {\textbf l}), 
thus keeping the temperature roughly constant while the entropy decreases. 
Also, panel {\textbf i} shows that density fluctuations are small ($\disprho\ll1$) 
as in a non-rotating cooling flow \citep{Balbus89,Stern19}, and panel {\textbf j} shows that the specific angular momentum is conserved since the system is axisymmetric. 
Panel {\textbf c} 
shows that
flowlines rotate $4-8$ radians prior to cooling, or roughly one full revolution.

Fig.~\ref{f:track properties t5}{\textbf f} shows that at $t\approx\tfive$ when the flow abruptly cools, the density increases by a factor of $\approx$$300$ for the flowline in the midplane ($\thetalarge=0.5\pi$), and by a larger factor in flowlines with smaller $\thetalarge$. 
At the same time $j_z$ slightly increases (panel~{\textbf j}), likely as a result of torques by stars and preexisting disk gas.
Also apparent is that density fluctuations become strong just before the gas cools (at $t - \tfive \approx -25\Myr$), and remain of order unity after cooling, in contrast with the weak density fluctuations when the flow is hot (panel {\textbf i}). The transition to a disk geometry starts somewhat earlier,  at $t-\tfive\approx -250\,\Myr$ (panel {\textbf b}, and also $\theta$ panel in Fig.~\ref{f:track properties}).

The eventual drop in temperature from $T\approx2\cdot10^6\K$ to $T\approx10^4\K$ at $\tfive$ is an inevitable result of the inflow halting due to rotation support, which stops compressive heating. Absent any heating sources, the gas cools on a cooling timescale, which is $\sim10\Myr$ at $\tfive$. This short cooling timescale is a result of the flattening to a disk geometry  which increases the density to $\nH\gtrsim0.01\cm^{-3}$ (see zoom-in on $t\approx\tfive$ in Figure~\ref{f:track properties t5 zoom}). This layer where the hot inflow cools  corresponds to the disc-halo interface \citep[e.g.,][]{Fraternali08,Marasco12, Fraternali17}, also known as `extraplanar gas', which is further addressed in the discussion.

The result that density fluctuations remain small when the flow is hot (Fig.~\ref{f:track properties t5}{\textbf i}) is potentially due to
the accretion process occurring on a timescale comparable or shorter than $\tcool$ on which thermal perturbations develop. For example, at $0.5\Gyr$ prior to accretion (i.e., at $t-\tfive=-0.5\Gyr$) we find $\tcool=1.7, 2.9, 5.3$ and $12\Gyr$ in the four flowlines shown in Fig.~\ref{f:track properties t5}, 
with longer $\tcool$ closer to the rotation axis due to the higher temperature and lower densities. The $0.5\Gyr$ remaining until accretion is thus insufficient for the thermal instability to grow, despite that the hot gas is formally unstable. 
A similar argument explains why significant density perturbations do not develop spontaneously in non-rotating cooling flows 
\citep{Balbus89}. This conclusion is also consistent with the simulations in \cite{Sormani19}, which found that condensations develop only when a heating term is added to part of the hot gas, thus increasing the accretion time relative to a cooling flow. 

\subsection{Deviations from spherical symmetry in hot inflowing CGM}\label{s:asymmetry}

\begin{figure}
    \centering
    \includegraphics{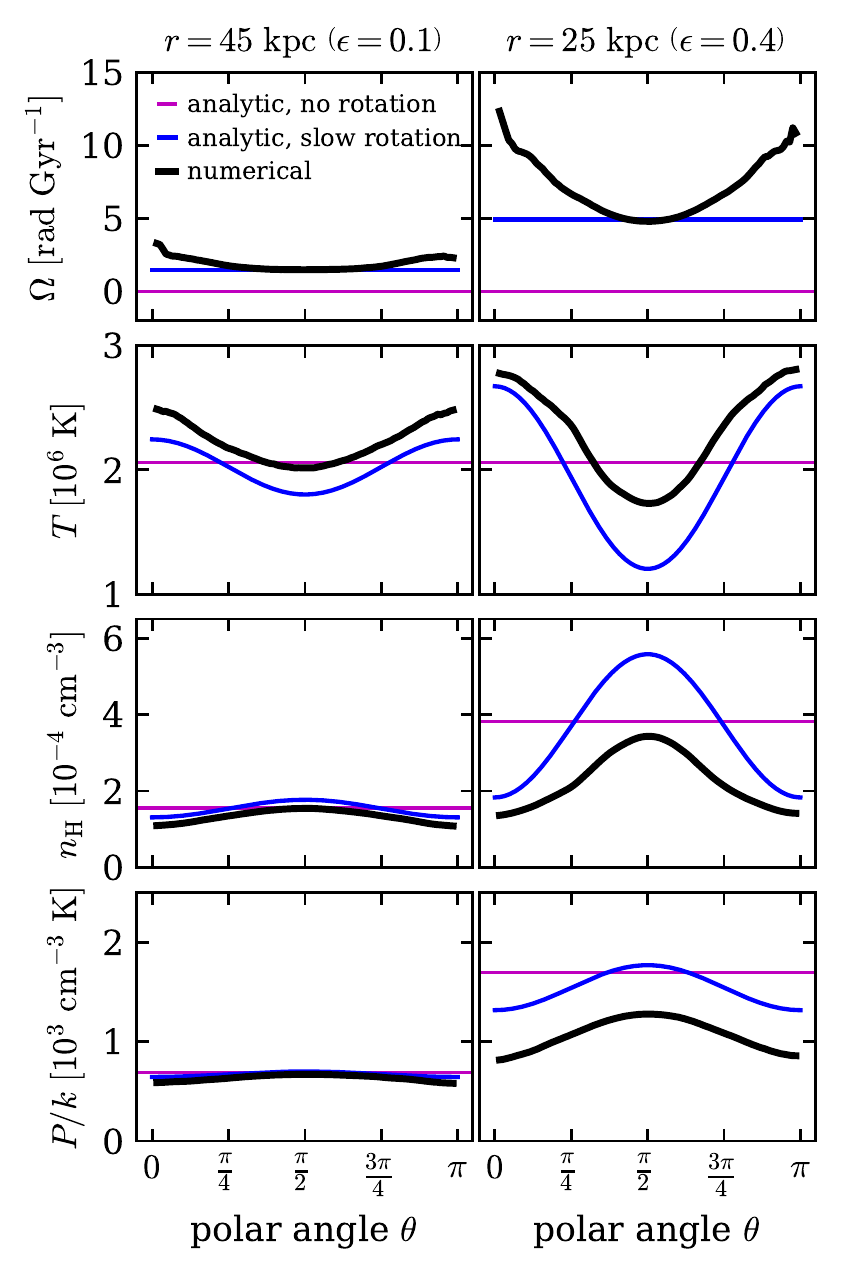}
    \caption{Deviations from spherical symmetry in hot rotating CGM inflows. Panels show from top to bottom the hot gas angular frequency, temperature, hydrogen density, and pressure, versus angle from the rotation axis $\theta$, at $r=45\kpc$ (\textit{left}) and $r=25\kpc$ (\textit{right}). Black lines are based on the simulation after steady-state is achieved (the same simulation as in Figs.~\ref{f:map}--\ref{f:track properties t5}). Magenta lines plot the analytic non-rotating solution (eqn.~\ref{e:rad solution}), while blue lines plot the analytic slow-rotating solution  (eqn.~\ref{e:solution}) which accounts only for lowest-order terms in $\epsilon=(r/\Rcirc)^{-2}$ (noted on top). 
In the rotating solutions density and pressure increase towards the midplane, while temperature decreases. }
    \label{f:comparison theta}
\end{figure}

Rotational support in the hot CGM is expected to induce lower gas densities and higher gas temperatures along the rotation axis relative to the midplane \citep{Barnabe06, Pezzulli17, Sormani19}. In this section we quantify these deviations from spherical symmetry in our simulation of a hot rotating inflow, and compare it to the analytic approximation deduced above (eqn.~\ref{e:solution}).

Figure~\ref{f:comparison theta} plots the dependence of hot CGM properties on polar angle $\theta$ at radii of $45\kpc$ and $25\kpc$. From top to bottom the different rows plot angular frequency, temperature, hydrogen density and thermal pressure. 
Magenta lines denote the non-rotating analytic solution (eqn.~\ref{e:rad solution}), blue lines denote the slow rotating analytic solution (eqn.~\ref{e:solution}), and black lines the solution in the simulation after steady-state is achieved (the same solution used in Figs.~\ref{f:map}--\ref{f:track properties t5}). 
The perturbation parameter $\epsilon=(r/\Rcirc)^{-2}$ defined in eqn.~(\ref{e:epsilon}) is noted at the top. The slow-rotating analytic solution accounts only for the lowest-order terms in this quantity. 

Fig.~\ref{f:comparison theta} demonstrates how the properties of the hot gas deviate from spherical symmetry due to the rotation, and more so at radii approaching $\Rcirc$ where rotation support is more significant. At $r=25\kpc$ in the numerical solution, the temperature at the rotating axis is almost a factor of two lower than in the midplane, while the density is a factor of two higher. Note also that the slow-rotating analytic solution rather accurately reproduces the simulation at $r=45\kpc$ where $\epsilon=0.1$. At $r=25\kpc$ where $\epsilon=0.4$ the analytic solution is qualitatively consistent with the trends of $T, \nH$, and $P$ versus $\theta$, though there are quantitative differences potentially since high-order terms in $\epsilon$ are neglected in the analytic solution.

\subsection{Revolutions in inflow versus $\tcool/\tff$}

\begin{table}
    \centering
\caption{Parameters of simulations used in Figure~\ref{f:time ratio}.}
    \begin{tabular}{c|c|c|c|c}
         $\vc$ & $\Rcirc$ & $\dot{M}$ & $Z_{\rm CGM}$ &  $\tcool/\tff$ at $r=\Rcirc$ $^{({\rm a})}$ \\
          $[{\rm km}\,{\rm s}^{-1}]$ & [{\rm kpc}] & $[\msun\yr^{-1}]$ & $[\zsun]$ & \\
         \hline
         $200^{({\rm b})}$ & $15$ & $1.7$   & $0.3$ & $9.3$   \\ 
         $200$ & $15$ & $3.2$   & $0.3$ & $6.5$   \\ 
         $200$ & $10$ & $2.7$   & $0.1$ & $8.5$   \\ 
         $200$ & $10$ & $8.3$   & $0.1$ & $4.9$   \\
         $200$ & $1$   & $1.8$   & $0.1$ & $3.3$   \\ 
         $200$ & $1$   & $4.2$   & $3.0$ & $0.5$   \\ 
         $200$ & $1$   & $5.4$   & $20$  & $0.2$   \\ 
         $230$ & $18$ & $2.9$   & $0.3$ & $12.2$ \\ 
         $210$ & $18$ & $3.5$   & $0.3$ & $8.1$   \\ 
         $150$ & $10$ & $1.3$   & $0.3$ & $3.6$   \\  
         $150$ & $10$ & $2.7$   & $0.3$ & $2.5$   \\  
         $150$ & $10$ & $8.8$   & $0.3$ & $1.4$   \\  
         $100$ & $5$   & $0.2$   & $0.3$ & $2.4$   \\  
         $100$ & $5$   & $0.8$   & $0.3$ & $1.3$   \\  
    \end{tabular}
    \begin{flushleft}
$^{({\rm a})}$ Derived from the four other parameters using eqn.~(\ref{e:rad solution}). \\
$^{({\rm b})}$ Fiducial simulation used also in other figures.
\end{flushleft}
\label{t:sims}
\end{table}

\begin{figure}
    \centering
  \includegraphics{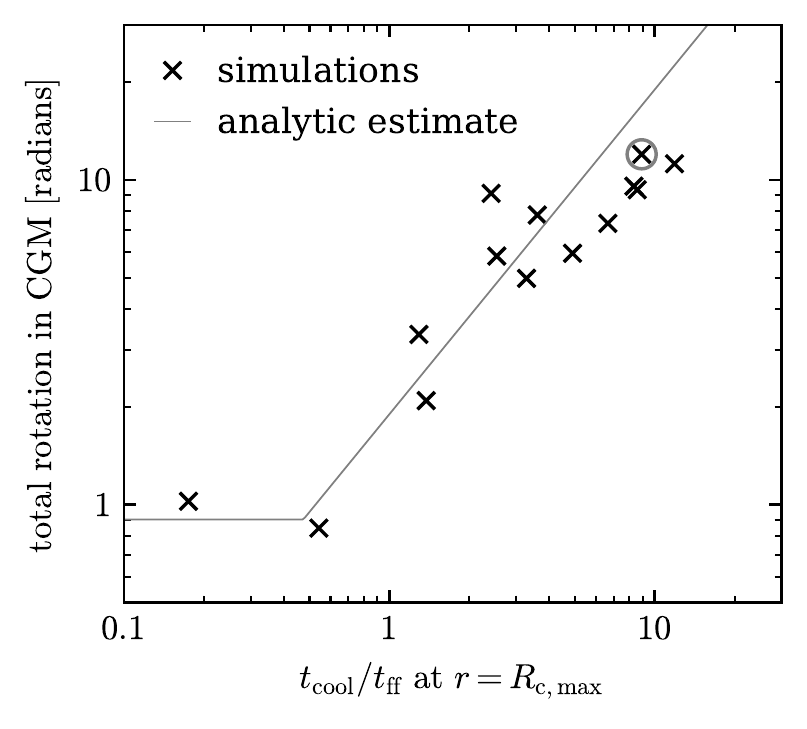}
    \caption{Total rotation completed by a fluid element in the CGM prior to accreting onto the ISM. Markers denote mean values in simulations with different $\tcool/\tff$ (measured at $r=\Rcirc$, see table~\ref{t:sims}), while the thin line denotes the analytic estimate from eqn.~(\ref{e:Nrot complete}). The fiducial simulation shown in Figs.~\ref{f:map}--\ref{f:comparison theta} is marked with a circle. Hot CGM with longer $\tcool$ have slower inflow velocities, and hence fluid elements rotate more prior to accretion. }
    \label{f:time ratio}
\end{figure}

To test the relation between  total rotation in the inflow and $\tcool/\tff$ measured at $\Rcirc$ (section~\ref{s:Nrot}), we run several simulations with different combinations of $\Mdot$, $Z_{\rm CGM}$, and $\vc$ which yield different $\tcool/\tff$ via eqn.~(\ref{e:rad solution}). The parameters of the simulations are listed in Table~\ref{t:sims}. For $\Mdot$ we use the value measured through a shell at $2\Rcirc$ in snapshots after the simulations achieve steady-state, which is typically $10-75\%$ larger than $\Mdot$ in the initial conditions (see \S\ref{s:setup}). On these snapshots we also measure the average rotation $\Delta \phi = \int \Omega dt$ a fluid element completes as it inflows from $10\Rcirc$ to $\Rcirc$, and plot them in Figure~\ref{f:time ratio}. The figure shows that the simulations roughy follow  the analytic estimate from eqn.~(\ref{e:Nrot complete}), confirming that the number of rotations in hot rotating CGM scales with $\tcool/\tff$.

\subsection{Disk accretion radius}\label{s:AMD}

\begin{figure}
    \centering
    \includegraphics{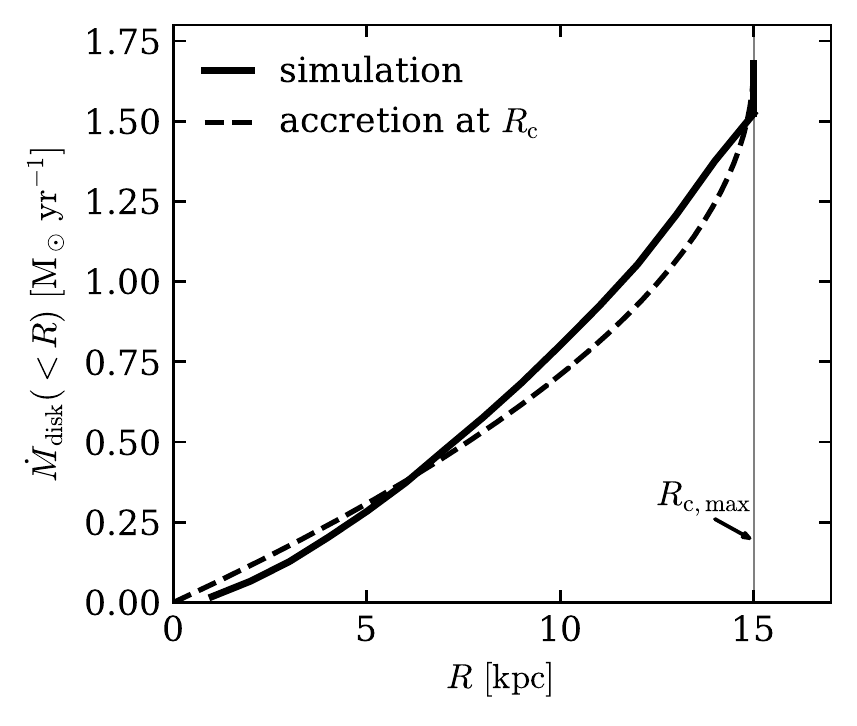}
    \vspace{-0.4cm}
    \caption{Accretion rate onto the disk versus disk radius in hot rotating inflows, for $\Rcirc=15\kpc$ and total accretion rate $\Mdot=1.7\msun\yr^{-1}$. The solid line is the accretion rate within a cylindrical disk radius $\Rcyl$ in the fiducial simulation. The step at $\Rcyl=\Rcirc$ corresponds to horizontal accretion from the disk edge. 
   The dashed line is the analytic estimate, which assumes each flowline accretes at its circularization radius (eqn.~\ref{e:Mdotd}), and provides a good match to the simulation result. This predicted $\Mdotd$ profile in hot rotating inflows can be useful for chemical evolution models and dynamical models of galaxy disks \citep[e.g.,][]{Schonrich09,Krumholz18}. }
    \label{f:Racc}
\end{figure}

Figure~\ref{f:Racc} plots the accretion rate onto the disk within cylindrical disk radius $\Rcyl$. Solid line is based on the fiducial simulation, which we measure by calculating the mass flux through two disk-shaped surfaces with radius $\Rcyl$ located at $z=\pm1\kpc$, in a snapshot after the simulation achieved steady-state. The step at $\Rcyl=\Rcirc=15\kpc$ corresponds to planar accretion from the disk edge, derived from the mass flux through a closing vertical surface spanning $-1<z<1\kpc$ at $\Rcyl=\Rcirc$. 
The simulation result is close to the analytic estimate in eqn.~(\ref{e:Mdotd}), which is based on the assumption that each flowline in the hot inflow accretes at its circularization radius.

\section{Additional considerations}\label{s:additional physics}

In this section we consider the effect of additional physical mechanisms and effects which were not included in the simulation above. 

\subsection{Viscosity}

Viscous forces in the flow may in principle cause angular momentum transport along directions where there is shear in the flow. 
In our solution $\Omega\propto r^{-2}$ up to corrections of order $\epsilon$ (eqn.~\ref{e:solution}) and thus there is shear in the radial direction. Here we show that for standard kinematic viscosity the expected angular momentum transport due to this shear is small. 

The specific angular momentum of the flow is $j=\Omega r^2\sin^2\theta=\vc\Rcirc\sin^2\theta$ (eqn.~\ref{e:solution}). Viscous forces in the radial direction will cause an angular momentum loss per unit time of
\begin{equation}
    \frac{dj}{dt} = \nu r^2 \sin^2\theta\frac{ d^2\Omega}{d r^2} = 6\nu \Omega \sin^2\theta ~,
\end{equation}
where $\nu$ is the kinematic viscosity \citep[e.g.,][]{Fabian05}
\begin{equation}
    \nu = 0.56\,T_6^{5/2}\left(\frac{\nH}{10^{-3}\cm^{-3}}\right)^{-1} \xi_\nu\kpc \kms
\end{equation}
and $\xi_\nu$ is the reduction of the viscosity relative to the Spitzer value. We can estimate the fractional angular momentum loss due to viscous forces by multiplying $dj/dt$ by the flow time $\approx\tcool$ and dividing by $j$:
\begin{eqnarray}
 \frac{\Delta j_{\rm visc}}{j}\approx\frac{dj}{dt}\cdot\frac{\tflow}{j}= \frac{6\Omega\nu\tcool}{\vc\Rcirc}
 \approx 0.092\,
 \xi_\nu\rn\vcn^5 \Mdotn^{-1}
\end{eqnarray}
where we used the analytic solution in eqn.~(\ref{e:solution}) and neglected corrections of order $(\Rcirc/r)^2$. For typical estimates of $\xi_\nu\sim0.1$ \citep{Narayan01} this value is substantially smaller than unity, 
indicating that viscous forces can generally be neglected.

\subsection{Turbulence}\label{s:turbulence}
\newcommand{\sturb}{\sigma_{\rm turb}}
\newcommand{\tdisp}{t_{\rm diss}}

Assuming that turbulence is seeded at large CGM radii, for example by cosmological accretion or due to stirring by subhalos, what would be the fate of these turbulent motions in the inner CGM inflow explored here?  
In a non-rotating inflow, we expect a balance between dissipation of turbulence on a timescale $\tdisp=r/\sturb$ and `adiabatic heating' of turbulence due to the contraction of the inflow on a timescale $\tflow=r/v_r$ \citep{Robertson12}. This balance suggests that contracting turbulent fluids converge to $\tflow\sim\tdisp$ and hence $\sturb\sim v_r$, since more rapid turbulent motions will dissipate while slower turbulence will heat up \citep{Murray15, Murray17}. In a steady-state cooling flow where $\tflow\approx\tcool$, 
we thus expect also $\tdisp\sim\tcool$. 
Using $\tff\sim r/\vc$ and $\vc\approx\cs$, it thus follows that
\begin{equation}\label{e:turb}
\frac{\sturb}{\cs}\sim \frac{\tff}{\tcool} ~.
\end{equation}
Since $\tff <\tcool$, eqn.~(\ref{e:turb}) suggests that turbulence is subsonic, i.e., turbulent support is subdominant to thermal support, as assumed in section~\ref{s:no AM}. Furthermore, this relation suggests that relative importance of turbulent motions decreases with increasing $\tcool/\tff$.  

Note that eqn.~(\ref{e:turb}) is based on the assumption that the dominant turbulence driving mechanism at inner CGM radii is adiabatic `heating' of pre-existing turbulence in the inflow. 
This assumption is similar to the underlying assumption of our solution that the dominant thermal heating mechanism 
is compression of the CGM inflow, rather than other heating sources such as feedback. 

\begin{figure}
\includegraphics{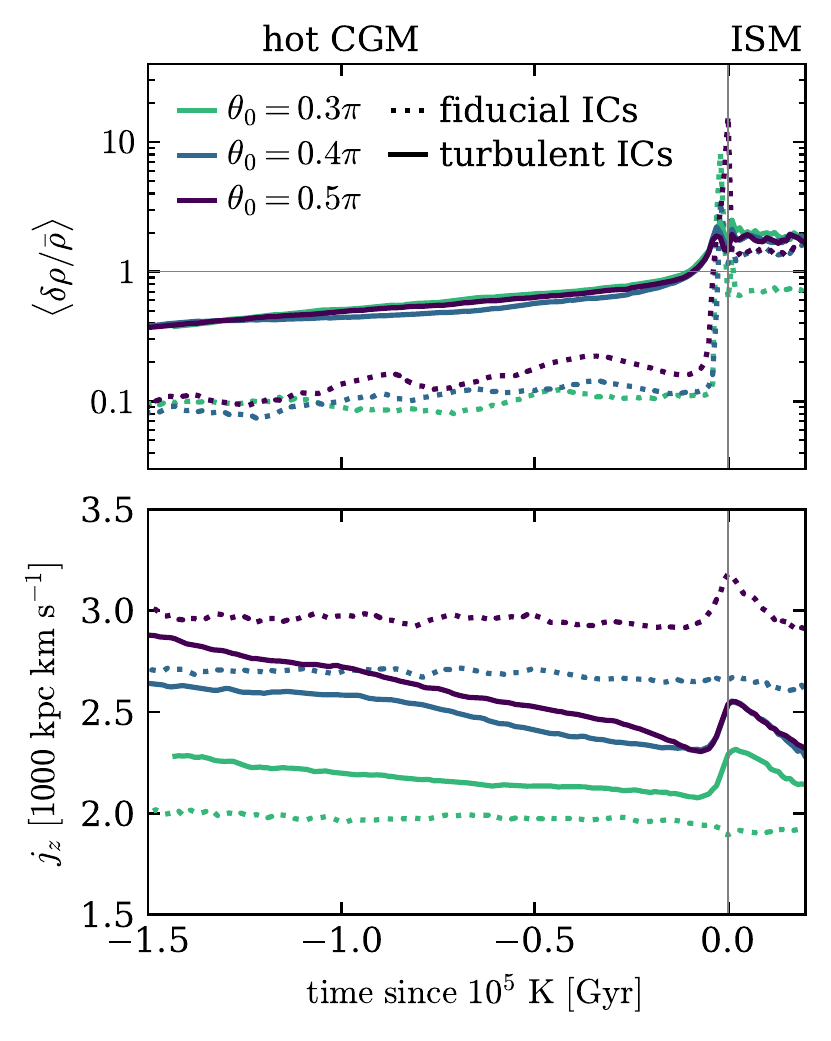}
\caption{The effect of turbulence on hot rotating CGM inflows. The panels show density fluctuations (\textit{top}) and specific angular momentum (\textit{bottom}) along flowlines, versus time since the hot CGM cools and accretes onto the ISM. Dashed lines denote the simulation with fiducial initial conditions (ICs) shown also in Figs.~\ref{f:map}--\ref{f:comparison theta}, while solid lines denote a simulation with turbulent ICs. 
Line color denotes the initial polar angle of the flowline as in Fig.~\ref{f:track properties t5}. Note the mild angular momentum loss along flowlines in the turbulent simulation, in contrast with the constant angular momentum in the fiducial simulation. 
}
\label{f:turb}
\end{figure}

In a rotating inflow, turbulence may also be induced by the shear between adjacent shells. Radial displacements due to such turbulent motions are not subject to restoring Coriolis forces, since the non-perturbed solution is angular momentum conserving (Fig.~\ref{f:track properties t5}) and hence the epicyclic frequency $\kappa$ is zero. \cite{Balbus96} showed that for a rotationally-dominated disk with $\kappa=0$ such turbulence develops with an e-folding time roughly equal to an orbit time. Given that hot CGM inflows onto Milky-Way like galaxies complete $\lesssim2$ orbits before accretion (Fig.~\ref{f:time ratio} with $\tcool/\tff\approx10$) it is unclear if shear-induced turbulence has sufficient time to develop. 

As a preliminary test of the effect of turbulence on our hot rotating CGM solution, we run another simulation with similar initial conditions as in our fiducial simulation, to which we add a turbulent velocity field with amplitude $\sturb(t = 0) = 30\kms$ and a lognormal power spectrum peaking at a wavelength of $50\kpc$ and a logarithmic width of $1$.
Figure~\ref{f:turb} compares the results of this simulation after steady-state is achieved with the results of the fiducial simulation shown in Figs. \ref{f:map}--\ref{f:comparison theta}.  
The top panel shows that adding turbulence increases the density fluctuations in the hot inflow, though they are still below unity at $t<\tfive$. The bottom panel demonstrates the effect of turbulence on the specific angular momentum in the flow. The turbulent simulation shows a smaller difference in $j_z$ between different flowlines than in the fiducial simulation, likely as a result of angular momentum transfer between flowlines.
Also evident is the mild ($\lesssim 20\%$) decrease in $j_z$ in the turbulent simulation over the last $1.5\Gyr$ prior to cooling, in contrast with the constant $j_z$ in the fiducial simulation.
This is potentially a result of transport of angular momentum outward during the inflow due to the turbulent viscosity between adjacent shells. 
Fig.~\ref{f:turb} thus implies that turbulence, at the strength explored, has only a mild effect on the angular momentum content of the inflow, and specifically $\gtrsim80\%$ of the angular momentum in the inflowing hot CGM accretes onto the ISM\footnote{The turbulent simulation has $\kappa\approx0.2$ at $r\gtrsim\Rcirc$, in contrast with $\kappa=0$ in the fiducial simulation. This finite $\kappa$ is potentially sufficient to prevent any further development of hydrodynamic turbulence due to restoring Coriolis forces \citep{Balbus96}.}.

The development of turbulence in hot and rotating CGM inflows also affects the angular momentum distribution of accreting gas, since a large $\sturb$ necessary implies a broad angular momentum distribution. \cite{Hafen22} argued that a narrow angular momentum distribution in accreting gas may be necessary for the formation of thin disk galaxies seen in the low-redshift Universe. This points to the importance of understanding turbulence in hot rotating CGM. We defer a detailed analysis to future work.

\subsection{Magnetic fields}\label{s:B}
The contraction and rotation in hot rotating CGM inflows is expected to enhance magnetic fields present at the outer radius of the inflow. In appendix~\ref{a:B fields} we estimate this enhancement using the rotating inflow solution derived above, assuming ideal MHD and ignoring potential dynamical effects of the magnetic field on the flow. Defining $r_0$ as the outer radius of the inflow, and assuming an isotropic seed field $B_r(r_0)=B_\theta(r_0)=B_\phi(r_0)\equiv B_0$, we find that
\begin{eqnarray}\label{e:B}
\frac{B_r}{B_0} &=& \left(\frac{r}{r_0}\right)^{-2} \nonumber \\
\frac{B_\theta}{B_0} &=& \left(\frac{v_r r}{v_0 r_0}\right)^{-1}\nonumber \\
\frac{B_\phi}{B_0} &=& \sqrt{2}\frac{\tcool}{\tff}\frac{r_0^2\Rcirc}{r^3}\sin\theta
\end{eqnarray}
where we also defined $v_0 \equiv v_r(r_0)$. The values of $B_r/B_0$, $B_\theta/B_0$, and $B_\phi/B_0$ are accurate up to corrections of order $(\Rcirc/r)^2$. 

The enhancement of the magnetic field in eqn.~(\ref{e:B}) can be compared to the enhancement in a non-rotating spherical inflow, which was derived by \cite{Shapiro73}. They found $B_r\propto r^{-2}$ and $B_\theta \propto B_\phi \propto (v_r r)^{-1}$, which can be understood intuitively as a result of conservation of magnetic flux through patches moving with the flow. Rotation thus mainly affects the enhancement of $B_\phi$, due to the winding of the field. Eqn.~(\ref{e:B}) shows that the enhancement of $B_\phi$ is a product of ($\tcool/\tff \cdot \Rcirc/r)$  and $(r_0/r)^{2}$, where the former tracks the number of radians rotated by the inflow (eqn.~\ref{e:Nrot}) and the latter tracks the contraction of an inflowing shell. 

For $r_0/\Rcirc\sim6$ and $\tcool/\tff\sim6$ as expected in the hot Milky-Way CGM (section~\ref{s:analytic}), eqn.~(\ref{e:B}) suggests an increase in $B_\phi$ of order  $\sim 200$ by the time the hot gas reaches $\Rcirc$, just prior to accreting onto the galaxy. For comparison, the thermal pressure increases over the same range of radii as $P(\Rcirc)/P(r_0)\approx (\Rcirc/r_0)^{-3/2}\sim15$ (eqn.~\ref{e:rad solution}), and thus the ratio of thermal to magnetic pressure $\beta\propto P/B^2$ is expected to decrease by a large factor of $\sim 3000$. Current upper limits on the magnetic field in the inner CGM of $\sim L^\star$ galaxies at $z\lesssim0.3$ suggest magnetic pressure is subdominant to the thermal pressure ($\beta>1$), at least along the major axis where most of the accretion is expected \citep{Prochaska19, Lan20,Heesen23}. It thus follows that if the hot CGM is accreting as suggested in this work, seed magnetic fields are sufficiently small that they do not dominate even after the large enhancement induced by contraction and rotation. 

The eventual cooling of the inner hot CGM onto the ISM will further enhance $B$. Fig.~\ref{f:track properties t5}{\textbf f} suggests that the gas density increases by a factor of $\sim1000$ as it cools, which would increase $B$ by a further factor $\sim1000^{2/3}=100$ in the limit of ideal MHD.  
Another potentially interesting  implication of the hot CGM solution concerns the development of turbulence due to the magnetic-rotational instability (MRI). 
The MRI amplitude growth rate is $\sim\Omega$  \citep[e.g.][]{Balbus98,Masada08}, so the result that $\int\Omega dt\approx\tcool/\tff$ (Fig.~\ref{f:time ratio}, eqn.~\ref{e:Nrot}) implies that prior to accretion MRI can grow by $e^{\tcool/\tff}$, i.e.~a factor of $10^4$ for $\tcool/\tff\approx10$. 
The solution may thus change considerably as $\tcool/\tff$ exceeds some critical value where MRI becomes fully developed. We defer analysis of accretion via magnetic hot rotating CGM inflows to future work.

\section{Observational implications}\label{s:obs}

\newcommand{\keV}{\,{\rm keV}}

\newcommand{\tobs}{\theta_{\rm obs}}

In this section we discuss several observational signatures of hot and rotating CGM inflows. 
While it would be challenging to detect the predicted slow radial velocities of a few tens of $\kms$ (eqn.~\ref{e:rad solution}), the predicted rotation velocities are faster, reaching $\vc\sim200\kms$, and indeed evidence for such fast hot gas rotation has been detected in the Milky-Way CGM \citep{HodgesKluck16}.  Since an inflow imparts a specific rotation profile in the hot gas which is flatter ($\vphi\sim r^{-1}$, eqn.~\ref{e:outer BC})
than the rotation profile expected  prior to the development of an inflow 
($\vphi\sim r^{-0.5}$, \citealt{Sharma12,Pezzulli17,Sormani18}), 
measuring the rotation profile could be used to support or rule out the existence of inflows in the hot gas.

The hot gas rotation pattern can potentially be detected directly
using X-ray emission line centroiding (\ref{s:centroid}), or 
indirectly, by identifying the lower densities and higher temperatures along the rotation axis relative to the midplane (\ref{s:Xray}--\ref{s:FRB}). The latter indirect signature would however have to be distinguished from qualitatively similar trends induced by feedback \citep{Truong21, Truong23, Nica21, Yang23}.

Given the idealized nature of the solution, signal strength estimates below are at the order of magnitude level. More realistic calculations based on cosmological simulations would be a useful next step.

\begin{figure}
\includegraphics{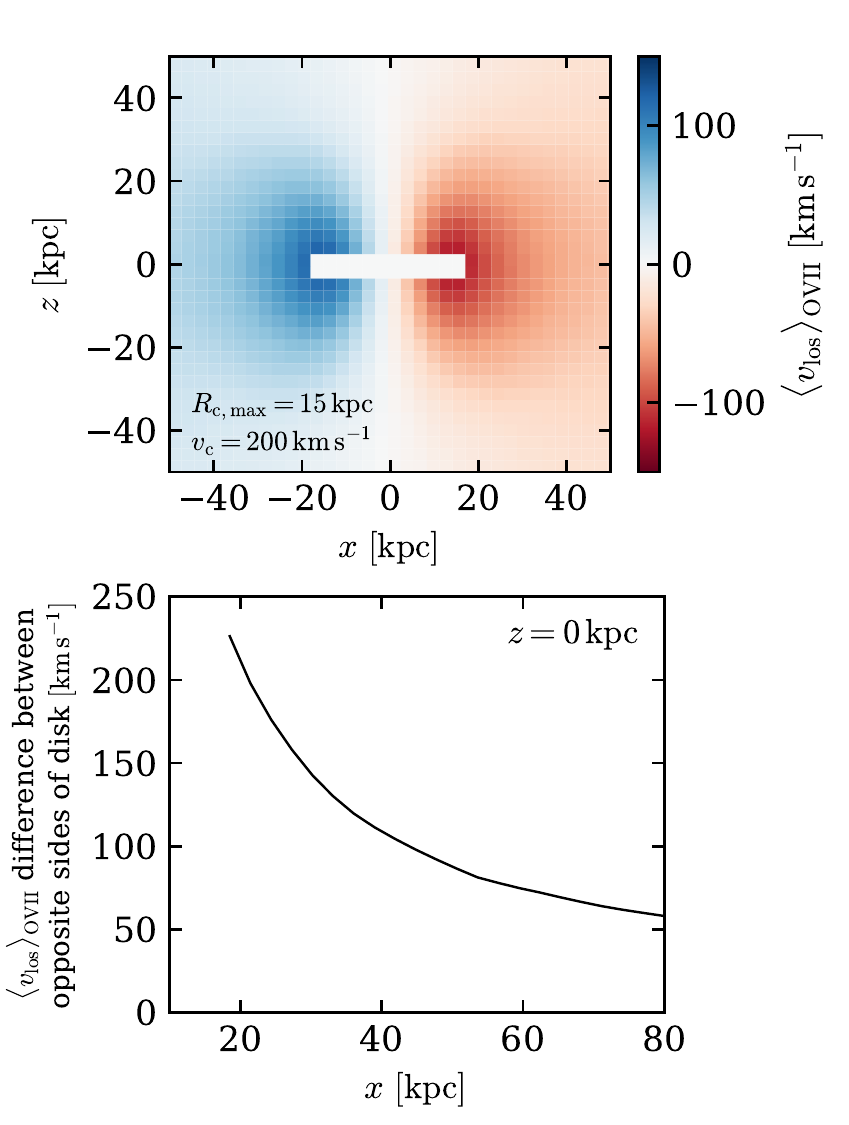}
\caption{
\textit{Top:} 
Predicted centroid shift of the OVII~$0.56\keV$ emission line in a hot rotating CGM inflow feeding a disk galaxy. CGM rotation assumed to be viewed edge-on. Pixel size is $3\kpc\times3\kpc$, corresponding to $15''$ resolution at a distance of $40\Mpc$. The calculation is based on the fiducial simulation, with the central disk masked. 
\textit{Bottom:} difference between line-of-sight velocity
 at opposite sides of the disk, in the midplane.
Detecting the predicted $\vphi\sim r^{-1}$ rotation profile in the hot CGM would support the hypothesis that it is inflowing. }
\label{f:obs velocity}
\end{figure}

\subsection{Measuring hot gas rotation using line centroiding}\label{s:centroid}

\cite{HodgesKluck16} analyzed the centroids of 37 O{\sc vii} absorption lines in the Milky-Way CGM, in the context of a phenomenological hot CGM model with $\nH\propto r^{-1.5}$ and uniform inflow and rotational velocities. They deduced $v_r=-15\pm20\kms$ and   $\vphi=180\pm41\kms$, or $v_r/\vc=-0.06\pm0.08$ and $\vphi=0.75\pm0.17$ for a circular velocity of $\vc=240\kms$ measured near the Sun. These values are comparable to those expected in hot rotating inflows near the disk, since $\vphi/\vc\approx \Rcirc/r$ and $\vr/\vc\approx \tff/\tcool$ (Fig.~\ref{f:eulerian view}, eqns.~\ref{e:solution} and \ref{e:rad solution}). Specifically, we calculate the O{\sc vii} absorption-averaged $\vphi$ and $v_r$ in our fiducial simulation, along sightlines starting at $(R,z)=(8\kpc,0\kpc)$ with the same distribution of Galactic latitudes and longtitudes as in the \citeauthor{HodgesKluck16} sample. We find a median $\vphi/\vc=0.62$ and $v_r/\vc=-0.1$, consistent with their results. The \citeauthor{HodgesKluck16} observations of hot CGM kinematics in the Milky-Way halo are thus consistent with a rotating inflow. 

A more accurate test of the predicted hot CGM rotational velocity profile can be performed with an X-ray microcalorimeter with high spectral resolution. Top panel of Fig.~\ref{f:obs velocity} plots a projection of the line of sight velocity, weighted by the emissivity of the OVII~He$\alpha$ line, assuming CGM rotation is viewed edge on. For this calculation we use a simulation with $\Mdot=3\msun\yr^{-1}$, $\vc=200\kms$, and $\Rcirc=15\kpc$. These parameters are chosen to simulate the signal around NGC~891 which has a similar circular velocity and size as the Milky-Way, but a higher SFR of $\approx4\msun\yr^{-1}$  \citep{Popescu04}. The projection is done on a snapshot after the simulation converged onto a steady-state. Line emissivity is calculated based on the gas density and temperature in the simulation using the pyXSIM package \citep{ZuHone16}\footnote{version 4.2.0}. Pixel size in this panel is $3\kpc$, corresponding to the planned $15''$ resolution of  the proposed Line Emission Mapper probe (LEM, \citealt{LEM}) for a target at a distance of $40\Mpc$\footnote{At smaller redshifts the target line emission would be drowned by line emission from the Milky-Way, see \cite{LEM}.}. 
The bottom panel shows the line centroid 
difference between the approaching and receding sides of the CGM, in the midplane. This difference reaches $200\kms$ near the disk, higher than the planned centroiding accuracy of $\lesssim70\kms$ planned for LEM. 
X-ray telescopes with high spectral resolution may thus be able to measure the rotation velocity profile in the hot CGM, and test whether it is consistent with the inflow solution. 

\subsection{Angle dependence of X-ray emission and temperature}\label{s:Xray}

\begin{figure}
\vspace{-0.3cm}
\includegraphics{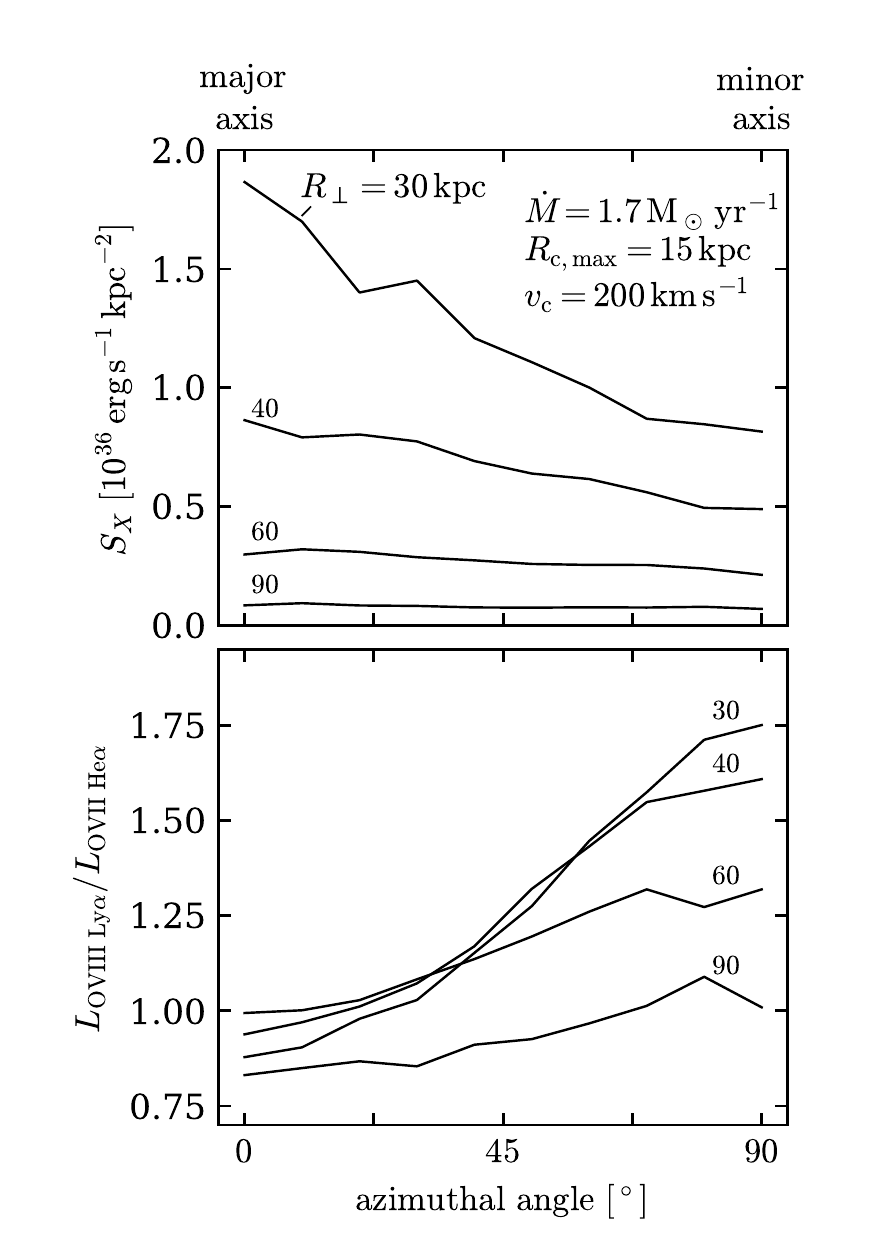}
\vspace{-0.4cm}
\caption{X-ray emission from hot rotating inflows versus azimuthal angle (defined as the sightline orientation relative to galaxy major axis), for an edge-on galaxy. Different lines denote different impact parameters. Calculations are based on the fiducial simulation. \textit{Top:} predicted soft X-ray surface brightness. Surface brightness decreases with angle due to the lower densities near the rotation axis induced by angular momentum support. \textit{Bottom:} O{\sc viii}~Ly$\alpha$/O{\sc vii}~He$\alpha$  emission line ratio. The ratio increases with angle due to higher temperatures near the rotation axis (see Fig.~\ref{f:comparison theta}). 
}
\label{f:emission}
\end{figure}

Angular momentum support induces deviations from spherical symmetry in the hot CGM density and temperature
(section~\ref{s:asymmetry}, Fig.~\ref{f:comparison theta}). These are potentially detectable by measuring the dependence of CGM X-ray emission on the `azimuthal angle', defined as the orientation of the sightline with respect to the galaxy major axis \citep[e.g.,][]{Kacprzak15}. 
The top panel in Figure~\ref{f:emission} shows the predicted soft X-ray surface brightness versus azimuthal angle and impact parameter $R_\perp$, assuming CGM rotation is oriented edge-on in the plane of the sky. Surface brightness is calculated using pyXSIM on the NGC~891-like simulation used also in Fig.~\ref{f:obs velocity}. 
The figure shows that the soft X-ray brightness ($\propto \nH^2$) increases towards the major axis at small $R_\perp$, since rotation induces a higher CGM density near the midplane (eqn.~\ref{e:solution}, Fig.~\ref{f:comparison theta}). 

The bottom panel in Fig.~\ref{f:emission} shows the luminosity ratio of the O{\sc viii}~Ly$\alpha$ 0.65\kev\ and O{\sc vii}~He$\alpha$ 0.56\kev\ emission lines. The ratio increases towards the minor axis, by up to $75\%$ at $R_\perp=30\kpc$. This follows since this emission line ratio depends mainly on CGM temperature which is higher along the rotation axis (see Fig.~\ref{f:comparison theta}).

\subsection{Dispersion Measure}\label{s:FRB}

\begin{figure}
\includegraphics{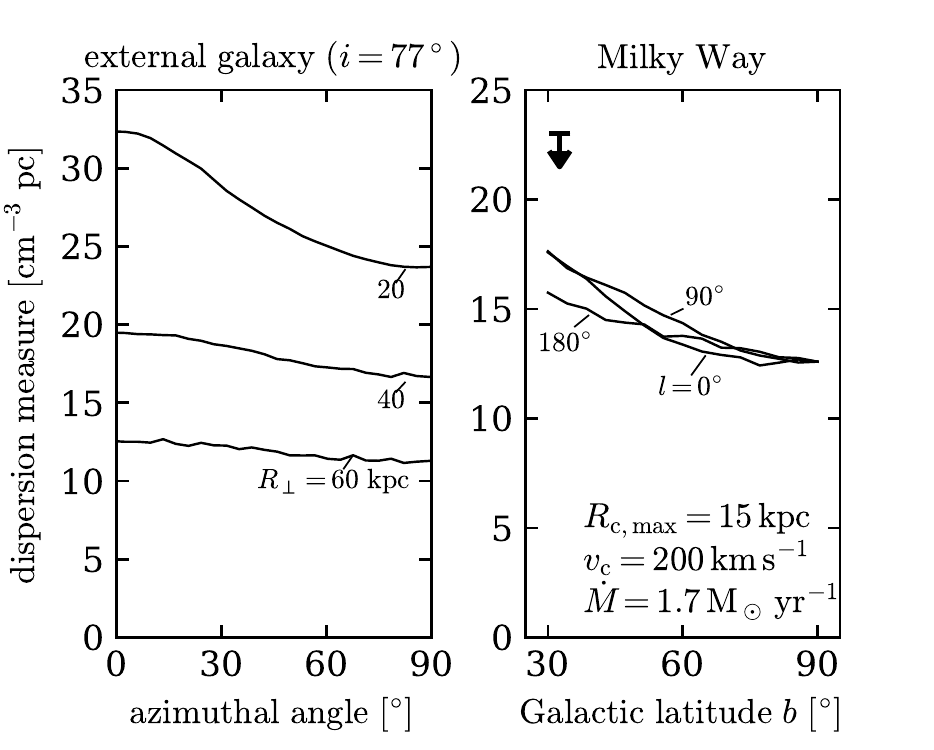}
\caption{Predicted dispersion measures from a CGM forming a hot rotating inflow, for sightlines through an external galaxy CGM (\textit{left}, assuming M31-like inclination) and through the Milky-Way CGM (\textit{right}), based on the fiducial simulation. External galaxy sightlines are given as a function of impact parameter and azimuthal angle, while Milky-Way sightlines as a function of Galactic coordinates. Integration is limited to $r<100\kpc$ where a hot inflow is possible. The upper limit from Anderson \& Bregman (2010), derived from sightlines to pulsars in the LMC, is noted in the right panel. The decrease in dispersion measure with increasing azimuthal angle and $b$ is due to the lower  densities near the rotation axis.  }
\label{f:FRB}
\end{figure}

Observations of the dispersion measures to pulsars in the Magellanic clouds and towards extragalactic  Fast Radio Bursts (FRBs) constrain electron column densities in the CGM \citep[e.g.][]{Anderson10, XYZ19, Williams23, Ravi23}. 
Figure~\ref{f:FRB} plots predicted dispersion measures for an external galaxy and for the Milky-Way CGM, based on our fiducial simulation. For external galaxy sightlines we assume an inclination $i=77^\circ$ similar to M31, and integrate the electron density along sightlines with different impact parameters $R_\perp$ and different azimuthal angles. We start and end the integration at a spherical radius $r=100\kpc$, roughly equal to $\Rcool$ (section~\ref{s:no AM}) beyond which the hot inflow solution does not apply. Fig.~\ref{f:FRB} shows that 
the highest dispersion measures of $\approx30\cm^{-3}\pc$ are found at small impact parameters and small azimuthal angles where densities are highest (see eqn.~\ref{e:solution}). 
For Milky-Way sightlines the dispersion measures were calculated by integrating from $(R,z)=(8\kpc,1\kpc)$ out to $r=100\kpc$, for different Galactic latitudes $b$ and Galactic longtitudes $l$. We start at $z=1\kpc$ to avoid the contribution of the cool disk, while the exact choice of outer limit does not significantly affect the result since most of the contribution comes from small radii. 
The predicted dispersion measures are $12-18\cm^{-3}\pc$ and increase towards lower $b$, again due to higher densities near the disk plane. 

Fig.~\ref{f:FRB} also shows the upper limit of $23\,{\rm cm}^{-3}\,{\rm pc}$ for the CGM dispersion measure from  \cite{Anderson10}, deduced based on sightlines to pulsars in the LMC  (though note caveat in \citealt{Ravi23}, which may imply this upper limit is too restrictive). This upper limit is consistent with our prediction of $14\cm^{-3}\pc$ in sightlines with $|b|=30^\circ$ and $l=270^\circ$ when integrating out to an LMC distance of $r=50\kpc$.

\section{Discussion}\label{s:discussion} 

\subsection{Comparison to previous hot CGM models}\label{s:prev models}

In the solution described in this work the hot CGM is inflowing, since ongoing feedback heating is assumed to be small relative to radiative losses. This is in contrast with the typical assumption employed by models of the low-redshift CGM, where the hot CGM phase is static due to feedback heating balancing radiative losses \citep[e.g.][]{Sharma12, Voit17, Faerman17,Faerman20}.
Several models have accounted also for rotation in the hot CGM \citep{Pezzulli17, Sormani18, Afruni22}, though also in these latter models the gas is assumed not to flow in the radial direction. 
 A third possibility is that feedback heating exceeds radiative losses, and the hot CGM forms an outflow \citep[e.g.,][though note these studies neglected a pre-existing CGM]{Thompson16, Schneider20}. As the mechanics of CGM heating by stellar and AGN feedback are currently not well understood, and existing X-ray constraints do not rule out an inflow as in the ICM (see introduction), it is currently unclear which of these three paradigms is more accurate. 

Despite the qualitative distinction between hot inflows explored here and static hot CGM models explored by previous studies, both types of models satisfy similar hydrostatic equilibrium constraints. This follows since the inflow solution is highly subsonic, and hence deviations from hydrostatic equilibrium in the radial direction are small, of order
$(\tcool/\tff)^{-2}$. Indeed, these deviations are neglected in the analytic solution derived in section~\ref{s: analytic sol}.  
The assumption of an inflow however
enforces conservation of mass, energy and angular momentum between adjacent shells (eqns.~\ref{e:AM}--\ref{e:K with AM}), while there are no similar constraints on static models. The allowed space of inflowing hot CGM solutions is thus significantly smaller than that of static solutions. 
For example, deducing the entropy profile in static models requires employing another assumption, such that the entropy is independent of radius \citep{Faerman20}, or that $\tcool/\tff$ is independent of radius \citep{Voit17}. In the inflow solution derived here the entropy profile is fully determined by the flow equations. 

Inflow solutions are also more specific than static models in the predicted rotation profile of the hot CGM. \cite{Sormani18} used several constraints to derive $\vphi(r,\theta)$ in their static hot CGM model, mainly the typical CGM angular momentum distribution in cosmological simulations and O{\sc vii} absorption-based estimates of Milky-Way CGM rotation from \cite{HodgesKluck16}. These constraints are however rather sparse and leave considerable freedom for different choices of $\vphi(r,\theta)$, as discussed in \citeauthor{Sormani18}
In contrast, in the inflow solution derived here angular momentum is conserved along flowlines (eqn.~\ref{e:AM}), and thus to fully determine $\vphi(r,\theta)$ one only has to specify the angular momentum of each flowline. In our fiducial model, the midplane flowline has an angular momentum corresponding to the typical spin of dark matter halos (section~\ref{s:yes AM}), while the angular momentum of other flowlines is set by requiring  a constant angular frequency at the inflow outer boundary at $r\sim\Rcool\sim100\kpc$, as suggested by non-radiative cosmological simulations (section~\ref{s:yes AM2}). The predicted $\vphi(r,\theta)$ is consistent with the Milky-Way O{\sc vii} absorption measurements (section~\ref{s:centroid}). Also, the inflow solution predicts a flat angular momentum profile at $\Rcirc < r\lesssim\Rcool$ (eqn.~\ref{e:outer BC}), as often seen in FIRE cosmological simulations of Milky-Way mass galaxies at $z\sim0$ \citep[see figure 13 there]{Stern21a}.

\subsection{Accretion via hot inflows versus `precipitation'}
Accretion onto the Milky Way and nearby spirals 
is likely dominated either by `precipitation', where local thermal instability in the hot phase creates $\sim10^4\K$ gas clumps which lose buoyancy and accrete \citep[e.g.,][]{Fall85, Maller04, Voit15, Armillotta16}, or by the hot $\sim10^6\K$ inflows discussed in this work (see introduction). Both accretion modes would be considered `hot accretion' in the context of the classic distinction between the hot and cold accretion modes, since they both originate in the hot phase of the CGM \citep[e.g.][]{Nelson13}. However, in the scenario studied here the CGM inflow remains at $\sim10^6\K$ down to a cylindrical radius $\Rcirc\sim15\kpc$ and height of $\sim\kpc$ above the midplane, at which point all the hot gas cools and joins the ISM, rather than just a subset of localized clouds. Hot inflows are thus a type of `quiet accretion' \citep{Putman12} -- accretion which becomes accessible to cool gas observations only when it cannot be kinematically distinguished from  pre-existing disk gas. 

We note  that at outer CGM radii of $r\gtrsim\Rcool\sim100\kpc$ a hot inflow is not expected since cooling times are long (section~\ref{s:no AM}), so any infall would likely be dominated by cool $\sim10^4\K$ gas. This cool inflow at large radii can potentially join a hot inflow at small CGM radii if the cool clouds are disrupted by hydrodynamic instabilities \citep[e.g.,][]{Tan23, Afruni23}.

\subsection{Hot inflows require that CGM of typical $\sim L^*$ spirals have previously expanded due to feedback}

In steady state we expect a CGM inflow rate of $\Mdot\approx0.6\,{\rm SFR}$, which is $\approx1\msun\yr^{-1}$ for the Milky-Way (section~\ref{s:no AM}). We show here that if this $\Mdot$ originates in a hot inflow, the CGM density must be lower than expected in a baryon-complete halo. 

Integrating the density profile in eqn.~(\ref{e:rad solution}) out to the virial radius and solving for $\Mdot$ we get:
\begin{equation}\label{e:Mdot to MCGM}
\Mdot = 6.6 \frac{M_{\rm CGM}}{10^{11}\msun}\left(\frac{R_{\rm CGM}}{300\kpc}\right)^{-3}\vcn^{-2} \Lambdan\msun\yr^{-1}~,
\end{equation}
where we normalized $M_{\rm CGM}$ by the the baryon-complete CGM mass of $0.16\Mhalo-M_{\rm galaxy}=10^{11}\msun$ for $M_{\rm halo}=10^{12}\msun$ and a galaxy mass $M_{\rm galaxy}=6\cdot10^{10}\msun$. The CGM size $R_{\rm CGM}$ is normalized by $\Rvir\approx300\kpc$. This predicted $\Mdot$ is higher than the $\approx1\msun\yr^{-1}$ required to sustain star formation in local disks.
It thus follows that for local discs to be fed by hot CGM inflows, gas originally associated with the halo must have expanded beyond $\Rvir$. Such an expanded CGM is supported by recent thermal Sunyaev-Zeldovich (tSZ) maps of nearby spirals, which indicate that the baryon budget of the halo is spread over a size of $R_{\rm CGM}\gtrsim 500\kpc$ \citep{Bregman22}. Using this larger estimate of $R_{\rm CGM}$ in eqn.~(\ref{e:Mdot to MCGM}) would suggest  $\Mdot\lesssim1.5\msun$, consistent with observed SFRs. The large $R_{\rm CGM}$ deduced by \cite{Bregman22} thus supports the scenario that local disk galaxies are fed by hot CGM inflows. 

There is however an inherent challenge in a scenario where discs are fed by an inflow from an expanded hot CGM. An expanding CGM is apparently contradictory to an inflowing CGM, and requires feedback heating to dominate radiative cooling, in contrast with the above assumption that feedback heating is subdominant. This apparent contradiction can be circumvented if the inflow and expansion are separated either in time or in space. 
For example, it is plausible that feedback was strong high redshift causing the CGM to expand, and has since subsided so the hot CGM develops an inflow at low redshift. Such an evolution in feedback strength is predicted by FIRE simulations of Milky-Way mass galaxies \citep{Muratov15,FaucherGiguere18,Stern21a,Pandya21}, and is also consistent with observed stellar winds which are strong in $z\sim2$ galaxies but  weak in typical $\sim$$L^\star$ galaxies at $z\sim0$  \citep{Heckman17}. Alternatively, black hole feedback or stellar feedback which occur in `bursts' separated by more than $\sim1\Gyr$ would allow a hot inflow to develop between bursts, since this timescale is the cooling time in the inner CGM on which a hot inflow develops. A third possibility is that feedback mainly heats the outer CGM, allowing the inner CGM to form an inflow. While this may seem counter-intuitive, it is potentially possible if feedback is focused on the rotation axis at small CGM radii and isotropizes only at large CGM radii, thus allowing the hot inner CGM to inflow from the midplane. Another option is that feedback energy is propagated by weak shocks or sound waves which dissipate and dump heat only in the outer CGM. The distribution of feedback energy in space and time has been explored extensively in the context of the ICM \citep[e.g.,][]{Yang16, Martizzi19,DonahueVoit22}, but less so in the context of the CGM. A thorough exploration would allowing understanding under which conditions the hot CGM inflow scenario is viable. 

\subsection{Disk-halo interface}

The cooling layer of the hot inflow solution found above, in which the gas temperature drops from $\approx2 \cdot 10^6\K$ to $\lesssim 10^4\K$, occurs at $|z|\lesssim0.5\,{\rm kpc}$ in our fiducial simulation (see Figs.~\ref{f:map} and \ref{f:track properties t5 zoom}). This simulated cool layer is thinner than the ‘extraplanar’ layers of neutral and ionized $\sim10^4\,{\rm K}$ gas detected around nearby spirals, which extend to $|z|\approx$ several kpc \citep[e.g.,][]{Sancisi08, Gaensler08}. Also, the mass flow rate through this layer in our simulation is $\Mdot\approx1\msun\yr^{-1}\sim{\rm SFR}$, in contrast with $\sim10\times{\rm SFR}$ in observed extraplanar gas layers \citep[e.g.,][]{Marasco19}. 

Extraplanar gas is often assumed to be dominated by fountain flows \citep{Shapiro76, Bregman80}, in which cool clouds driven upward by feedback penetrate into the hot CGM and then fall back onto the disk \citep{Fraternali06,Fraternali08,Marasco12, Fraternali17}. Fountain flows thus both increase the vertical extent of the cool gas and the total mass circulating through the extraplanar gas layer. These models usually assume the hot CGM is not inflowing, which as mentioned above requires a delicate balance between feedback heating and cooling, a balance especially hard to achieve in the inner CGM where cooling times are short \citep[see also discussion of this issue in ][]{Sormani19}.  
It would thus be beneficial to study how fountain flows interact with the inflowing hot CGM solution derived herein. Specifically, one could use the angular momentum structure of a hot inflow derived above to understand how fountain orbits are affected by angular momentum exchange with the hot CGM, an interaction that has observable implications \citep[e.g.,][]{Fraternali17}. Also it would be useful to understand how this fountain -- hot CGM interaction affects the accretion rate profile onto the disk (compare Figure~\ref{f:Racc} above with figure~9 in \citealt{Marasco12}). 

We note that some of the above studies argued that fountain flows are required for accretion from the hot CGM, since they trigger cool cloud condensation in the hot gas. Our simulations show that the hot CGM cools and can fuel star formation even without fountain flows. Cooling in our solution is however a steady process which affects the entire hot inflow as it reaches the disk-halo interface, rather than a result of condensation of localized cool clouds.

\subsection{Truncation of thin disks at $\Rcyl\approx\Rcirc$}\label{s:truncation}

For disks fed by hot rotating inflows, $\Rcirc$ corresponds to the maximum cylindrical radius at which gas cools and accretes onto the ISM (Fig.~\ref{f:Racc}). This is also evident as a sharp edge of the cool gas disk at $R=\Rcirc$ in the temperature maps shown in Fig.~\ref{f:map}, and is a result of hot CGM inflows cooling and accreting only when $\vphi=\vc$ (Fig.~\ref{f:track properties t5}{\textbf d}). 
Similar behavior has been identified in FIRE cosmological simulations of low-redshift Milky-Way mass galaxies which are also fed by hot rotating inflows \citep{Hafen22, Trapp22}. 

\renewcommand{\Rd}{R_{\rm d}}
\newcommand{\Rt}{R_{\rm trunc}}

The predicted maximum accretion radius at $\Rcirc$ is similar to the observed  truncation radius $\Rt$ of nearby disk galaxies, beyond which the stellar and H{\sc i} surface densities drop \citep[e.g.,][]{VanderKruit07}. Such a truncation is observed in $60-80\%$ of thin discs at $\Rt\approx 3.5-4 \Rd$ where $\Rd$ is the disk scale length \citep{Kregel02, Comeron12, MartinNavarro12}\footnote{The derived fraction of thin disks which exhibit a truncation is based on discs observed edge-on, since at low inclination stellar halos make the truncation harder to observe \citep{MartinNavarro14}.}. 
When a H{\sc i} warp is present it also often starts at $R\approx\Rt$ \citep{VanderKruit07}.  
Using $\Rt=3.5 R_{\rm d}\approx0.04\Rvir$ based on the $R_{\rm d}-\Rvir$ relation from \cite{Kravtsov13} and $\Rcirc\approx0.05\Rvir$ 
(eqn.~\ref{e:Rcirc}), we get $\Rcirc\approx\Rt$. 
It is thus plausible that observed disc truncations are ultimately a result of the abrupt cut in gas accretion beyond this radius, as suggested by \cite{Trapp22}. In the context of hot rotating CGM inflows this maximum radius of accretion is set by $\Rcirc$. 

\newcommand{\kb}{k_{\rm B}}
\newcommand{\ptom}{\langle p_*/m_*\rangle}

\section{Summary}\label{s:summary}

In this work we derive an axisymmetric, steady-state solution for hot and rotating CGM inflows, focusing on Milky-Way mass galaxies at $z\sim0$. 
We demonstrate that such accretion flows transition from a spherical, hot ($\sim10^6\K$) radial inflow 
to a cool ($\lesssim10^4\K$) disk supported by rotation. This cooling occurs at the disc-halo interface, within a cylindrical radius equal to the maximum circularization radius of the flow $\Rcirc\sim15\kpc$ and at heights $|z|\lesssim\kpc$ above the disk. 
Such hot inflows are expected in the CGM if radiative cooling has dominated over feedback heating for a cooling timescale. This condition is easier to satisfy at inner CGM radii where cooling times can be substantially shorter than the Hubble time.

We find both a new analytic solution for hot inflows in the slow rotation limit, which is valid in the limit $(r/\Rcirc)^{2}\gg1$ (section~\ref{s: analytic sol}),
and a numerical solution applicable also at $r\lesssim\Rcirc$ (section~\ref{s:numeric}). These solutions provide an idealized version of the hot CGM inflows identified by \cite{Hafen22} in FIRE simulations of Milky-Way mass galaxies at $z\sim0$.  
The main properties of these solutions can be summarized as follows:

\begin{enumerate}
\item 
Due to a balance between radiative cooling and compressive heating, hot inflows remain hot down to where they  
become rotationally supported, at which point the inflow and compressive heating stop and the entire flow cools (Figs.~\ref{f:map}--\ref{f:track properties t5}). Hot inflows thus differ qualitatively from `precipitation' in which accretion proceeds via a subset of clouds formed due to  thermal instability. 
\item Angular momentum of accreting gas is conserved during the inflow due to the axisymmetry of the solution (Fig.~\ref{f:track properties t5}{\textbf j}), yielding an accretion flow which feeds the disk mainly from its outskirts (Fig.~\ref{f:Racc}, section~\ref{s:Racc}). Conservation of angular momentum also suggests $\vphi\propto r^{-1}$ in the hot CGM at radii $\Rcirc<r<\Rcool$ (Fig.~\ref{f:eulerian view}), flatter than the $\vphi\propto r^{0.5}-r^{0.7}$ suggested by non-radiative cosmological simulations \citep{Sharma12b}. This predicted rotation profile could potentially be detected 
with X-ray microcalorimeters which can centroid emission lines to $\lesssim100\kms$ (Fig.~\ref{f:obs velocity}). 
\item Gas accreting via hot CGM inflows revolves $\approx\tcool/\tff$ radians prior to accretion (Fig.~\ref{f:time ratio}), where $\tcool/\tff$ is estimated in gas just outside the galaxy ($\sim6$ in the Milky-Way). This in contrast with only $\approx1$ radian of revolution in gas accreting via cold flows. Enhancement of magnetic fields and development of turbulence in the hot CGM thus likely depend on the value of $\tcool/\tff$ (sections~\ref{s:turbulence}--\ref{s:B}). 
\item Rotational support induces deviations from spherical symmetry in the density and temperature structure of hot CGM inflows (Fig.~\ref{f:comparison theta}), qualitatively similar to the rotating but radially-static models of \cite{Sormani18}. These deviations may be detectable with X-ray telescopes (Fig.~\ref{f:emission}), or with dispersion measures derived from FRB surveys (Fig.~\ref{f:FRB}). 

\item Observed SFRs in local spirals constrain typical hot inflow accretion rates to $\Mdot\lesssim1-2\msun\yr^{-1}$. In the absence of ongoing feedback heating as assumed here, such accretion rates require that the halo baryon budget is spread over $\gtrsim2\Rvir$ in order to reduce CGM densities and cooling rates. \cite{Bregman22} recently reported evidence for such expanded CGM using tSZ maps around nearby spirals. 
\item By analyzing the centroids of O{\sc vii} absorption lines in the Milky-Way CGM, \cite{HodgesKluck16} deduced a hot CGM rotation velocity of $\vphi\approx180\pm40\kms$ and a marginal inflow velocity of $\vr\approx-15\pm20\kms$. Both values are consistent with those expected if the hot Milky-Way CGM forms a rotating inflow (section~\ref{s:centroid}).
\end{enumerate}

The solution derived herein provides an analytic basis for understanding the structure of the inner hot CGM, in the presence of angular momentum and radiative cooling. It would be beneficial for future studies to further develop this framework with additional relevant physical processes and complications applicable to the real Universe, including polar-angle dependent feedback heating, fountain flows, turbulence (briefly discussed in section~\ref{s:turbulence}), magnetic fields (\ref{s:B}), and misalignments between the rotation axes of the hot CGM, disk, and dark matter halo.

\section*{Acknowledgements}
We thank the anonymous referee for a highly detailed and insightful report that significantly improved the paper. 
JS thanks S.\ Peng Oh and M.~Voit for useful discussions. 
JS was supported by the Israel Science Foundation (grant No. 2584/21). 
CAFG was supported by NSF through grants AST-2108230, AST-2307327, and CAREER award AST-1652522; by NASA through grants 17-ATP17-0067 and 21-ATP21-0036; by STScI through grants HST-GO-16730.016-A and JWSTAR-03252.001-A; and by CXO through grant TM2-23005X.
This work was supported in part by a Simons Investigator award from the Simons Foundaton (EQ) and by NSF grant AST-2107872. JSB was supported by the National Science Foundation (NSF) grant AST-1910965 and NASA grant 80NSSC22K0827. 
The computations in this work were run at facilities supported by the Scientific Computing Core at the Flatiron Institute, a division of the Simons Foundation.

\section*{Data availability}

The simulation data underlying this article will be shared on reasonable request to the corresponding author (JS). A public version of the GIZMO simulation code is available \href{http://www.tapir.caltech.edu/phopkins/Site/GIZMO.html}{http://www.tapir.caltech.edu/phopkins/Site/GIZMO.html}.

\bibliographystyle{mnras}
\bibliography{main}

\begin{thebibliography}{}
\makeatletter
\relax
\def\mn@urlcharsother{\let\do\@makeother \do\$\do\&\do\#\do\^\do\_\do\%\do\~}
\def\mn@doi{\begingroup\mn@urlcharsother \@ifnextchar [ {\mn@doi@}
  {\mn@doi@[]}}
\def\mn@doi@[#1]#2{\def\@tempa{#1}\ifx\@tempa\@empty \href
  {http://dx.doi.org/#2} {doi:#2}\else \href {http://dx.doi.org/#2} {#1}\fi
  \endgroup}
\def\mn@eprint#1#2{\mn@eprint@#1:#2::\@nil}
\def\mn@eprint@arXiv#1{\href {http://arxiv.org/abs/#1} {{\tt arXiv:#1}}}
\def\mn@eprint@dblp#1{\href {http://dblp.uni-trier.de/rec/bibtex/#1.xml}
  {dblp:#1}}
\def\mn@eprint@#1:#2:#3:#4\@nil{\def\@tempa {#1}\def\@tempb {#2}\def\@tempc
  {#3}\ifx \@tempc \@empty \let \@tempc \@tempb \let \@tempb \@tempa \fi \ifx
  \@tempb \@empty \def\@tempb {arXiv}\fi \@ifundefined
  {mn@eprint@\@tempb}{\@tempb:\@tempc}{\expandafter \expandafter \csname
  mn@eprint@\@tempb\endcsname \expandafter{\@tempc}}}

\bibitem[\protect\citeauthoryear{{Afruni}, {Pezzulli}  \&
  {Fraternali}}{{Afruni} et~al.}{2022}]{Afruni22}
{Afruni} A.,  {Pezzulli} G.,   {Fraternali} F.,  2022, \mn@doi [\mnras]
  {10.1093/mnras/stab3237}, \href
  {https://ui.adsabs.harvard.edu/abs/2022MNRAS.509.4849A} {509, 4849}

\bibitem[\protect\citeauthoryear{{Afruni}, {Pezzulli}, {Fraternali}  \&
  {Gr{\o}nnow}}{{Afruni} et~al.}{2023}]{Afruni23}
{Afruni} A.,  {Pezzulli} G.,  {Fraternali} F.,   {Gr{\o}nnow} A.,  2023,
  \mn@doi [\mnras] {10.1093/mnras/stad1963}, \href
  {https://ui.adsabs.harvard.edu/abs/2023MNRAS.524.2351A} {524, 2351}

\bibitem[\protect\citeauthoryear{{Agertz} et~al.,}{{Agertz}
  et~al.}{2007}]{Agertz07}
{Agertz} O.,  et~al., 2007, \mn@doi [\mnras]
  {10.1111/j.1365-2966.2007.12183.x}, \href
  {https://ui.adsabs.harvard.edu/abs/2007MNRAS.380..963A} {380, 963}

\bibitem[\protect\citeauthoryear{{Anderson} \& {Bregman}}{{Anderson} \&
  {Bregman}}{2010}]{Anderson10}
{Anderson} M.~E.,  {Bregman} J.~N.,  2010, \mn@doi [\apj]
  {10.1088/0004-637X/714/1/320}, \href
  {https://ui.adsabs.harvard.edu/abs/2010ApJ...714..320A} {714, 320}

\bibitem[\protect\citeauthoryear{{Anderson}, {Gaspari}, {White}, {Wang}  \&
  {Dai}}{{Anderson} et~al.}{2015}]{Anderson15}
{Anderson} M.~E.,  {Gaspari} M.,  {White} S. D.~M.,  {Wang} W.,   {Dai} X.,
  2015, \mn@doi [\mnras] {10.1093/mnras/stv437}, \href
  {https://ui.adsabs.harvard.edu/abs/2015MNRAS.449.3806A} {449, 3806}

\bibitem[\protect\citeauthoryear{{Anderson}, {Churazov}  \&
  {Bregman}}{{Anderson} et~al.}{2016}]{Anderson16}
{Anderson} M.~E.,  {Churazov} E.,   {Bregman} J.~N.,  2016, \mn@doi [\mnras]
  {10.1093/mnras/stv2314}, \href
  {https://ui.adsabs.harvard.edu/abs/2016MNRAS.455..227A} {455, 227}

\bibitem[\protect\citeauthoryear{{Armillotta}, {Fraternali}  \&
  {Marinacci}}{{Armillotta} et~al.}{2016}]{Armillotta16}
{Armillotta} L.,  {Fraternali} F.,   {Marinacci} F.,  2016, \mn@doi [\mnras]
  {10.1093/mnras/stw1930}, \href
  {https://ui.adsabs.harvard.edu/abs/2016MNRAS.462.4157A} {462, 4157}

\bibitem[\protect\citeauthoryear{{Armillotta}, {Fraternali}, {Werk},
  {Prochaska}  \& {Marinacci}}{{Armillotta} et~al.}{2017}]{Armillotta17}
{Armillotta} L.,  {Fraternali} F.,  {Werk} J.~K.,  {Prochaska} J.~X.,
  {Marinacci} F.,  2017, \mn@doi [\mnras] {10.1093/mnras/stx1239}, \href
  {https://ui.adsabs.harvard.edu/abs/2017MNRAS.470..114A} {470, 114}

\bibitem[\protect\citeauthoryear{{Balbus} \& {Hawley}}{{Balbus} \&
  {Hawley}}{1998}]{Balbus98}
{Balbus} S.~A.,  {Hawley} J.~F.,  1998, \mn@doi [Reviews of Modern Physics]
  {10.1103/RevModPhys.70.1}, \href
  {https://ui.adsabs.harvard.edu/abs/1998RvMP...70....1B} {70, 1}

\bibitem[\protect\citeauthoryear{{Balbus} \& {Soker}}{{Balbus} \&
  {Soker}}{1989}]{Balbus89}
{Balbus} S.~A.,  {Soker} N.,  1989, \mn@doi [\apj] {10.1086/167521}, \href
  {https://ui.adsabs.harvard.edu/abs/1989ApJ...341..611B} {341, 611}

\bibitem[\protect\citeauthoryear{{Balbus}, {Hawley}  \& {Stone}}{{Balbus}
  et~al.}{1996}]{Balbus96}
{Balbus} S.~A.,  {Hawley} J.~F.,   {Stone} J.~M.,  1996, \mn@doi [\apj]
  {10.1086/177585}, \href
  {https://ui.adsabs.harvard.edu/abs/1996ApJ...467...76B} {467, 76}

\bibitem[\protect\citeauthoryear{{Barnab{\`e}}, {Ciotti}, {Fraternali}  \&
  {Sancisi}}{{Barnab{\`e}} et~al.}{2006}]{Barnabe06}
{Barnab{\`e}} M.,  {Ciotti} L.,  {Fraternali} F.,   {Sancisi} R.,  2006,
  \mn@doi [\aap] {10.1051/0004-6361:20053386}, \href
  {https://ui.adsabs.harvard.edu/abs/2006A&A...446...61B} {446, 61}

\bibitem[\protect\citeauthoryear{{Bertschinger}}{{Bertschinger}}{1989}]{Bertschinger89}
{Bertschinger} E.,  1989, \mn@doi [\apj] {10.1086/167428}, \href
  {https://ui.adsabs.harvard.edu/abs/1989ApJ...340..666B} {340, 666}

\bibitem[\protect\citeauthoryear{{Birnboim} \& {Dekel}}{{Birnboim} \&
  {Dekel}}{2003}]{Birnboim03}
{Birnboim} Y.,  {Dekel} A.,  2003, \mn@doi [\mnras]
  {10.1046/j.1365-8711.2003.06955.x}, \href
  {https://ui.adsabs.harvard.edu/abs/2003MNRAS.345..349B} {345, 349}

\bibitem[\protect\citeauthoryear{{Bland-Hawthorn} \&
  {Gerhard}}{{Bland-Hawthorn} \& {Gerhard}}{2016}]{BlandHawthorn16}
{Bland-Hawthorn} J.,  {Gerhard} O.,  2016, \mn@doi [\araa]
  {10.1146/annurev-astro-081915-023441}, \href
  {https://ui.adsabs.harvard.edu/abs/2016ARA&A..54..529B} {54, 529}

\bibitem[\protect\citeauthoryear{{Bogd{\'a}n}, {Bourdin}, {Forman}, {Kraft},
  {Vogelsberger}, {Hernquist}  \& {Springel}}{{Bogd{\'a}n}
  et~al.}{2017}]{Bogdan17}
{Bogd{\'a}n} {\'A}.,  {Bourdin} H.,  {Forman} W.~R.,  {Kraft} R.~P.,
  {Vogelsberger} M.,  {Hernquist} L.,   {Springel} V.,  2017, \mn@doi [\apj]
  {10.3847/1538-4357/aa9523}, \href
  {https://ui.adsabs.harvard.edu/abs/2017ApJ...850...98B} {850, 98}

\bibitem[\protect\citeauthoryear{{Bondi}}{{Bondi}}{1952}]{Bondi52}
{Bondi} H.,  1952, \mn@doi [\mnras] {10.1093/mnras/112.2.195}, \href
  {https://ui.adsabs.harvard.edu/abs/1952MNRAS.112..195B} {112, 195}

\bibitem[\protect\citeauthoryear{{Bregman}}{{Bregman}}{1980}]{Bregman80}
{Bregman} J.~N.,  1980, \mn@doi [\apj] {10.1086/157776}, \href
  {https://ui.adsabs.harvard.edu/abs/1980ApJ...236..577B} {236, 577}

\bibitem[\protect\citeauthoryear{{Bregman}, {Hodges-Kluck}, {Qu}, {Pratt}, {Li}
   \& {Yun}}{{Bregman} et~al.}{2022}]{Bregman22}
{Bregman} J.~N.,  {Hodges-Kluck} E.,  {Qu} Z.,  {Pratt} C.,  {Li} J.-T.,
  {Yun} Y.,  2022, \mn@doi [\apj] {10.3847/1538-4357/ac51de}, \href
  {https://ui.adsabs.harvard.edu/abs/2022ApJ...928...14B} {928, 14}

\bibitem[\protect\citeauthoryear{{Bullock}, {Dekel}, {Kolatt}, {Kravtsov},
  {Klypin}, {Porciani}  \& {Primack}}{{Bullock} et~al.}{2001}]{Bullock01}
{Bullock} J.~S.,  {Dekel} A.,  {Kolatt} T.~S.,  {Kravtsov} A.~V.,  {Klypin}
  A.~A.,  {Porciani} C.,   {Primack} J.~R.,  2001, \mn@doi [\apj]
  {10.1086/321477}, \href
  {https://ui.adsabs.harvard.edu/abs/2001ApJ...555..240B} {555, 240}

\bibitem[\protect\citeauthoryear{{Cassen} \& {Pettibone}}{{Cassen} \&
  {Pettibone}}{1976}]{Cassen76}
{Cassen} P.,  {Pettibone} D.,  1976, \mn@doi [\apj] {10.1086/154632}, \href
  {https://ui.adsabs.harvard.edu/abs/1976ApJ...208..500C} {208, 500}

\bibitem[\protect\citeauthoryear{{Chadayammuri}, {Bogd{\'a}n}, {Oppenheimer},
  {Kraft}, {Forman}  \& {Jones}}{{Chadayammuri} et~al.}{2022}]{Chadayammuri22}
{Chadayammuri} U.,  {Bogd{\'a}n} {\'A}.,  {Oppenheimer} B.~D.,  {Kraft} R.~P.,
  {Forman} W.~R.,   {Jones} C.,  2022, \mn@doi [\apjl]
  {10.3847/2041-8213/ac8936}, \href
  {https://ui.adsabs.harvard.edu/abs/2022ApJ...936L..15C} {936, L15}

\bibitem[\protect\citeauthoryear{{Comer{\'o}n} et~al.,}{{Comer{\'o}n}
  et~al.}{2012}]{Comeron12}
{Comer{\'o}n} S.,  et~al., 2012, \mn@doi [\apj] {10.1088/0004-637X/759/2/98},
  \href {https://ui.adsabs.harvard.edu/abs/2012ApJ...759...98C} {759, 98}

\bibitem[\protect\citeauthoryear{{Comparat} et~al.,}{{Comparat}
  et~al.}{2022}]{Comparat22}
{Comparat} J.,  et~al., 2022, \mn@doi [\aap] {10.1051/0004-6361/202243101},
  \href {https://ui.adsabs.harvard.edu/abs/2022A&A...666A.156C} {666, A156}

\bibitem[\protect\citeauthoryear{{Cowie}, {Fabian}  \& {Nulsen}}{{Cowie}
  et~al.}{1980}]{Cowie80}
{Cowie} L.~L.,  {Fabian} A.~C.,   {Nulsen} P.~E.~J.,  1980, \mn@doi [\mnras]
  {10.1093/mnras/191.2.399}, \href
  {https://ui.adsabs.harvard.edu/abs/1980MNRAS.191..399C} {191, 399}

\bibitem[\protect\citeauthoryear{{Das}, {Mathur}, {Gupta}, {Nicastro},
  {Krongold}  \& {Null}}{{Das} et~al.}{2019}]{Das19}
{Das} S.,  {Mathur} S.,  {Gupta} A.,  {Nicastro} F.,  {Krongold} Y.,   {Null}
  C.,  2019, \mn@doi [\apj] {10.3847/1538-4357/ab48df}, \href
  {https://ui.adsabs.harvard.edu/abs/2019ApJ...885..108D} {885, 108}

\bibitem[\protect\citeauthoryear{{DeFelippis}, {Genel}, {Bryan}, {Nelson},
  {Pillepich}  \& {Hernquist}}{{DeFelippis} et~al.}{2020}]{DeFelippis20}
{DeFelippis} D.,  {Genel} S.,  {Bryan} G.~L.,  {Nelson} D.,  {Pillepich} A.,
  {Hernquist} L.,  2020, \mn@doi [\apj] {10.3847/1538-4357/ab8a4a}, \href
  {https://ui.adsabs.harvard.edu/abs/2020ApJ...895...17D} {895, 17}

\bibitem[\protect\citeauthoryear{{Donahue} \& {Voit}}{{Donahue} \&
  {Voit}}{2022}]{DonahueVoit22}
{Donahue} M.,  {Voit} G.~M.,  2022, \mn@doi [\physrep]
  {10.1016/j.physrep.2022.04.005}, \href
  {https://ui.adsabs.harvard.edu/abs/2022PhR...973....1D} {973, 1}

\bibitem[\protect\citeauthoryear{{Fabian}, {Nulsen}  \& {Canizares}}{{Fabian}
  et~al.}{1984}]{Fabian84}
{Fabian} A.~C.,  {Nulsen} P.~E.~J.,   {Canizares} C.~R.,  1984, \mn@doi [\nat]
  {10.1038/310733a0}, \href
  {https://ui.adsabs.harvard.edu/abs/1984Natur.310..733F} {310, 733}

\bibitem[\protect\citeauthoryear{{Fabian}, {Reynolds}, {Taylor}  \&
  {Dunn}}{{Fabian} et~al.}{2005}]{Fabian05}
{Fabian} A.~C.,  {Reynolds} C.~S.,  {Taylor} G.~B.,   {Dunn} R.~J.~H.,  2005,
  \mn@doi [\mnras] {10.1111/j.1365-2966.2005.09484.x}, \href
  {https://ui.adsabs.harvard.edu/abs/2005MNRAS.363..891F} {363, 891}

\bibitem[\protect\citeauthoryear{{Faerman}, {Sternberg}  \& {McKee}}{{Faerman}
  et~al.}{2017}]{Faerman17}
{Faerman} Y.,  {Sternberg} A.,   {McKee} C.~F.,  2017, \mn@doi [\apj]
  {10.3847/1538-4357/835/1/52}, \href
  {https://ui.adsabs.harvard.edu/abs/2017ApJ...835...52F} {835, 52}

\bibitem[\protect\citeauthoryear{{Faerman}, {Sternberg}  \& {McKee}}{{Faerman}
  et~al.}{2020}]{Faerman20}
{Faerman} Y.,  {Sternberg} A.,   {McKee} C.~F.,  2020, \mn@doi [\apj]
  {10.3847/1538-4357/ab7ffc}, \href
  {https://ui.adsabs.harvard.edu/abs/2020ApJ...893...82F} {893, 82}

\bibitem[\protect\citeauthoryear{{Fall} \& {Rees}}{{Fall} \&
  {Rees}}{1985}]{Fall85}
{Fall} S.~M.,  {Rees} M.~J.,  1985, \mn@doi [\apj] {10.1086/163585}, \href
  {https://ui.adsabs.harvard.edu/abs/1985ApJ...298...18F} {298, 18}

\bibitem[\protect\citeauthoryear{{Faucher-Gigu{\`e}re}}{{Faucher-Gigu{\`e}re}}{2018}]{FaucherGiguere18}
{Faucher-Gigu{\`e}re} C.-A.,  2018, \mn@doi [\mnras] {10.1093/mnras/stx2595},
  \href {https://ui.adsabs.harvard.edu/abs/2018MNRAS.473.3717F} {473, 3717}

\bibitem[\protect\citeauthoryear{{Fielding}, {McKee}, {Socrates}, {Cunningham}
  \& {Klein}}{{Fielding} et~al.}{2015}]{Fielding15}
{Fielding} D.~B.,  {McKee} C.~F.,  {Socrates} A.,  {Cunningham} A.~J.,
  {Klein} R.~I.,  2015, \mn@doi [\mnras]
  {10.1093/mnras/stv83610.48550/arXiv.1409.5148}, \href
  {https://ui.adsabs.harvard.edu/abs/2015MNRAS.450.3306F} {450, 3306}

\bibitem[\protect\citeauthoryear{{Fielding}, {Quataert}, {McCourt}  \&
  {Thompson}}{{Fielding} et~al.}{2017}]{Fielding17}
{Fielding} D.,  {Quataert} E.,  {McCourt} M.,   {Thompson} T.~A.,  2017,
  \mn@doi [\mnras] {10.1093/mnras/stw3326}, \href
  {https://ui.adsabs.harvard.edu/abs/2017MNRAS.466.3810F} {466, 3810}

\bibitem[\protect\citeauthoryear{{Fielding} et~al.,}{{Fielding}
  et~al.}{2020}]{Fielding20b}
{Fielding} D.~B.,  et~al., 2020, \mn@doi [\apj] {10.3847/1538-4357/abbc6d},
  \href {https://ui.adsabs.harvard.edu/abs/2020ApJ...903...32F} {903, 32}

\bibitem[\protect\citeauthoryear{{Fox}, {Richter}, {Ashley}, {Heckman},
  {Lehner}, {Werk}, {Bordoloi}  \& {Peeples}}{{Fox} et~al.}{2019}]{Fox19}
{Fox} A.~J.,  {Richter} P.,  {Ashley} T.,  {Heckman} T.~M.,  {Lehner} N.,
  {Werk} J.~K.,  {Bordoloi} R.,   {Peeples} M.~S.,  2019, \mn@doi [\apj]
  {10.3847/1538-4357/ab40ad}, \href
  {https://ui.adsabs.harvard.edu/abs/2019ApJ...884...53F} {884, 53}

\bibitem[\protect\citeauthoryear{{Fraternali}}{{Fraternali}}{2017}]{Fraternali17}
{Fraternali} F.,  2017, in {Fox} A.,  {Dav{\'e}} R.,  eds,  Astrophysics and
  Space Science Library Vol. 430, Gas Accretion onto Galaxies. p.~323
  (\mn@eprint {arXiv} {1612.00477}), \mn@doi{10.1007/978-3-319-52512-9_14}

\bibitem[\protect\citeauthoryear{{Fraternali} \& {Binney}}{{Fraternali} \&
  {Binney}}{2006}]{Fraternali06}
{Fraternali} F.,  {Binney} J.~J.,  2006, \mn@doi [\mnras]
  {10.1111/j.1365-2966.2005.09816.x}, \href
  {https://ui.adsabs.harvard.edu/abs/2006MNRAS.366..449F} {366, 449}

\bibitem[\protect\citeauthoryear{{Fraternali} \& {Binney}}{{Fraternali} \&
  {Binney}}{2008}]{Fraternali08}
{Fraternali} F.,  {Binney} J.~J.,  2008, \mn@doi [\mnras]
  {10.1111/j.1365-2966.2008.13071.x}, \href
  {https://ui.adsabs.harvard.edu/abs/2008MNRAS.386..935F} {386, 935}

\bibitem[\protect\citeauthoryear{{Gaensler}, {Madsen}, {Chatterjee}  \&
  {Mao}}{{Gaensler} et~al.}{2008}]{Gaensler08}
{Gaensler} B.~M.,  {Madsen} G.~J.,  {Chatterjee} S.,   {Mao} S.~A.,  2008,
  \mn@doi [\pasa] {10.1071/AS08004}, \href
  {https://ui.adsabs.harvard.edu/abs/2008PASA...25..184G} {25, 184}

\bibitem[\protect\citeauthoryear{{Gurvich} et~al.,}{{Gurvich}
  et~al.}{2023}]{Gurvich23}
{Gurvich} A.~B.,  et~al., 2023, \mn@doi [\mnras] {10.1093/mnras/stac3712},
  \href {https://ui.adsabs.harvard.edu/abs/2023MNRAS.519.2598G} {519, 2598}

\bibitem[\protect\citeauthoryear{{Hafen} et~al.,}{{Hafen}
  et~al.}{2022}]{Hafen22}
{Hafen} Z.,  et~al., 2022, arXiv e-prints, \href
  {https://ui.adsabs.harvard.edu/abs/2022arXiv220107235H} {p. arXiv:2201.07235}

\bibitem[\protect\citeauthoryear{{Heckman} \& {Thompson}}{{Heckman} \&
  {Thompson}}{2017}]{Heckman17}
{Heckman} T.~M.,  {Thompson} T.~A.,  2017, in {Alsabti} A.~W.,  {Murdin} P.,
  eds, , Handbook of Supernovae.
p.~2431, \mn@doi{10.1007/978-3-319-21846-5_23}

\bibitem[\protect\citeauthoryear{{Heesen} et~al.,}{{Heesen}
  et~al.}{2023}]{Heesen23}
{Heesen} V.,  et~al., 2023, \mn@doi [\aap] {10.1051/0004-6361/202346008}, \href
  {https://ui.adsabs.harvard.edu/abs/2023A&A...670L..23H} {670, L23}

\bibitem[\protect\citeauthoryear{{Heitsch} \& {Putman}}{{Heitsch} \&
  {Putman}}{2009}]{Heitsch09}
{Heitsch} F.,  {Putman} M.~E.,  2009, \mn@doi [\apj]
  {10.1088/0004-637X/698/2/148510.48550/arXiv.0904.1995}, \href
  {https://ui.adsabs.harvard.edu/abs/2009ApJ...698.1485H} {698, 1485}

\bibitem[\protect\citeauthoryear{{Hodges-Kluck}, {Miller}  \&
  {Bregman}}{{Hodges-Kluck} et~al.}{2016}]{HodgesKluck16}
{Hodges-Kluck} E.~J.,  {Miller} M.~J.,   {Bregman} J.~N.,  2016, \mn@doi [\apj]
  {10.3847/0004-637X/822/1/21}, \href
  {https://ui.adsabs.harvard.edu/abs/2016ApJ...822...21H} {822, 21}

\bibitem[\protect\citeauthoryear{{Hopkins}}{{Hopkins}}{2015}]{hopkins15}
{Hopkins} P.~F.,  2015, \mn@doi [\mnras] {10.1093/mnras/stv195}, \href
  {https://ui.adsabs.harvard.edu/abs/2015MNRAS.450...53H} {450, 53}

\bibitem[\protect\citeauthoryear{{Hopkins} et~al.,}{{Hopkins}
  et~al.}{2018}]{Hopkins18}
{Hopkins} P.~F.,  et~al., 2018, \mn@doi [\mnras] {10.1093/mnras/sty1690}, \href
  {https://ui.adsabs.harvard.edu/abs/2018MNRAS.480..800H} {480, 800}

\bibitem[\protect\citeauthoryear{{Huscher}, {Oppenheimer}, {Lonardi}, {Crain},
  {Richings}  \& {Schaye}}{{Huscher} et~al.}{2021}]{Huscher21}
{Huscher} E.,  {Oppenheimer} B.~D.,  {Lonardi} A.,  {Crain} R.~A.,  {Richings}
  A.~J.,   {Schaye} J.,  2021, \mn@doi [\mnras] {10.1093/mnras/staa3203}, \href
  {https://ui.adsabs.harvard.edu/abs/2021MNRAS.500.1476H} {500, 1476}

\bibitem[\protect\citeauthoryear{{Kacprzak}, {Muzahid}, {Churchill}, {Nielsen}
  \& {Charlton}}{{Kacprzak} et~al.}{2015}]{Kacprzak15}
{Kacprzak} G.~G.,  {Muzahid} S.,  {Churchill} C.~W.,  {Nielsen} N.~M.,
  {Charlton} J.~C.,  2015, \mn@doi [\apj] {10.1088/0004-637X/815/1/22}, \href
  {https://ui.adsabs.harvard.edu/abs/2015ApJ...815...22K} {815, 22}

\bibitem[\protect\citeauthoryear{{Kamphuis} et~al.,}{{Kamphuis}
  et~al.}{2022}]{Kamphuis22}
{Kamphuis} P.,  et~al., 2022, \mn@doi [\aap] {10.1051/0004-6361/202140704},
  \href {https://ui.adsabs.harvard.edu/abs/2022A&A...668A.182K} {668, A182}

\bibitem[\protect\citeauthoryear{{Kaufmann}, {Mayer}, {Wadsley}, {Stadel}  \&
  {Moore}}{{Kaufmann} et~al.}{2006}]{Kaufmann06}
{Kaufmann} T.,  {Mayer} L.,  {Wadsley} J.,  {Stadel} J.,   {Moore} B.,  2006,
  \mn@doi [\mnras] {10.1111/j.1365-2966.2006.10599.x}, \href
  {https://ui.adsabs.harvard.edu/abs/2006MNRAS.370.1612K} {370, 1612}

\bibitem[\protect\citeauthoryear{{Kaufmann}, {Bullock}, {Maller}, {Fang}  \&
  {Wadsley}}{{Kaufmann} et~al.}{2009}]{Kaufmann09}
{Kaufmann} T.,  {Bullock} J.~S.,  {Maller} A.~H.,  {Fang} T.,   {Wadsley} J.,
  2009, \mn@doi [\mnras] {10.1111/j.1365-2966.2009.14744.x}, \href
  {https://ui.adsabs.harvard.edu/abs/2009MNRAS.396..191K} {396, 191}

\bibitem[\protect\citeauthoryear{{Kraft} et~al.,}{{Kraft} et~al.}{2022}]{LEM}
{Kraft} R.,  et~al., 2022, arXiv e-prints, \href
  {https://ui.adsabs.harvard.edu/abs/2022arXiv221109827K} {p. arXiv:2211.09827}

\bibitem[\protect\citeauthoryear{{Kravtsov}}{{Kravtsov}}{2013}]{Kravtsov13}
{Kravtsov} A.~V.,  2013, \mn@doi [\apjl] {10.1088/2041-8205/764/2/L31}, \href
  {https://ui.adsabs.harvard.edu/abs/2013ApJ...764L..31K} {764, L31}

\bibitem[\protect\citeauthoryear{{Kregel}, {van der Kruit}  \& {de
  Grijs}}{{Kregel} et~al.}{2002}]{Kregel02}
{Kregel} M.,  {van der Kruit} P.~C.,   {de Grijs} R.,  2002, \mn@doi [\mnras]
  {10.1046/j.1365-8711.2002.05556.x}, \href
  {https://ui.adsabs.harvard.edu/abs/2002MNRAS.334..646K} {334, 646}

\bibitem[\protect\citeauthoryear{{Krumholz}, {Burkhart}, {Forbes}  \&
  {Crocker}}{{Krumholz} et~al.}{2018}]{Krumholz18}
{Krumholz} M.~R.,  {Burkhart} B.,  {Forbes} J.~C.,   {Crocker} R.~M.,  2018,
  \mn@doi [\mnras] {10.1093/mnras/sty852}, \href
  {https://ui.adsabs.harvard.edu/abs/2018MNRAS.477.2716K} {477, 2716}

\bibitem[\protect\citeauthoryear{{Lan} \& {Prochaska}}{{Lan} \&
  {Prochaska}}{2020}]{Lan20}
{Lan} T.-W.,  {Prochaska} J.~X.,  2020, \mn@doi [\mnras]
  {10.1093/mnras/staa1750}, \href
  {https://ui.adsabs.harvard.edu/abs/2020MNRAS.496.3142L} {496, 3142}

\bibitem[\protect\citeauthoryear{{Li} \& {Bregman}}{{Li} \&
  {Bregman}}{2017}]{LiBregman17}
{Li} Y.,  {Bregman} J.,  2017, \mn@doi [\apj] {10.3847/1538-4357/aa92c6}, \href
  {https://ui.adsabs.harvard.edu/abs/2017ApJ...849..105L} {849, 105}

\bibitem[\protect\citeauthoryear{{Li} \& {Wang}}{{Li} \& {Wang}}{2013}]{Li13b}
{Li} J.-T.,  {Wang} Q.~D.,  2013, \mn@doi [\mnras] {10.1093/mnras/stt1501},
  \href {https://ui.adsabs.harvard.edu/abs/2013MNRAS.435.3071L} {435, 3071}

\bibitem[\protect\citeauthoryear{{Li}, {Crain}  \& {Wang}}{{Li}
  et~al.}{2014}]{Li14}
{Li} J.-T.,  {Crain} R.~A.,   {Wang} Q.~D.,  2014, \mn@doi [\mnras]
  {10.1093/mnras/stu329}, \href
  {https://ui.adsabs.harvard.edu/abs/2014MNRAS.440..859L} {440, 859}

\bibitem[\protect\citeauthoryear{{Lilly}, {Carollo}, {Pipino}, {Renzini}  \&
  {Peng}}{{Lilly} et~al.}{2013}]{Lilly13}
{Lilly} S.~J.,  {Carollo} C.~M.,  {Pipino} A.,  {Renzini} A.,   {Peng} Y.,
  2013, \mn@doi [\apj] {10.1088/0004-637X/772/2/119}, \href
  {https://ui.adsabs.harvard.edu/abs/2013ApJ...772..119L} {772, 119}

\bibitem[\protect\citeauthoryear{{Maller} \& {Bullock}}{{Maller} \&
  {Bullock}}{2004}]{Maller04}
{Maller} A.~H.,  {Bullock} J.~S.,  2004, \mn@doi [\mnras]
  {10.1111/j.1365-2966.2004.08349.x}, \href
  {https://ui.adsabs.harvard.edu/abs/2004MNRAS.355..694M} {355, 694}

\bibitem[\protect\citeauthoryear{{Marasco}, {Fraternali}  \&
  {Binney}}{{Marasco} et~al.}{2012}]{Marasco12}
{Marasco} A.,  {Fraternali} F.,   {Binney} J.~J.,  2012, \mn@doi [\mnras]
  {10.1111/j.1365-2966.2011.19771.x}, \href
  {https://ui.adsabs.harvard.edu/abs/2012MNRAS.419.1107M} {419, 1107}

\bibitem[\protect\citeauthoryear{{Marasco} et~al.,}{{Marasco}
  et~al.}{2019}]{Marasco19}
{Marasco} A.,  et~al., 2019, \mn@doi [\aap] {10.1051/0004-6361/201936338},
  \href {https://ui.adsabs.harvard.edu/abs/2019A&A...631A..50M} {631, A50}

\bibitem[\protect\citeauthoryear{{Mart{\'\i}n-Navarro}
  et~al.,}{{Mart{\'\i}n-Navarro} et~al.}{2012}]{MartinNavarro12}
{Mart{\'\i}n-Navarro} I.,  et~al., 2012, \mn@doi [\mnras]
  {10.1111/j.1365-2966.2012.21929.x}, \href
  {https://ui.adsabs.harvard.edu/abs/2012MNRAS.427.1102M} {427, 1102}

\bibitem[\protect\citeauthoryear{{Mart{\'\i}n-Navarro}, {Trujillo}, {Knapen},
  {Bakos}  \& {Fliri}}{{Mart{\'\i}n-Navarro} et~al.}{2014}]{MartinNavarro14}
{Mart{\'\i}n-Navarro} I.,  {Trujillo} I.,  {Knapen} J.~H.,  {Bakos} J.,
  {Fliri} J.,  2014, \mn@doi [\mnras] {10.1093/mnras/stu767}, \href
  {https://ui.adsabs.harvard.edu/abs/2014MNRAS.441.2809M} {441, 2809}

\bibitem[\protect\citeauthoryear{{Martizzi}, {Quataert}, {Faucher-Gigu{\`e}re}
  \& {Fielding}}{{Martizzi} et~al.}{2019}]{Martizzi19}
{Martizzi} D.,  {Quataert} E.,  {Faucher-Gigu{\`e}re} C.-A.,   {Fielding} D.,
  2019, \mn@doi [\mnras] {10.1093/mnras/sty3273}, \href
  {https://ui.adsabs.harvard.edu/abs/2019MNRAS.483.2465M} {483, 2465}

\bibitem[\protect\citeauthoryear{{Masada} \& {Sano}}{{Masada} \&
  {Sano}}{2008}]{Masada08}
{Masada} Y.,  {Sano} T.,  2008, \mn@doi [\apj] {10.1086/592601}, \href
  {https://ui.adsabs.harvard.edu/abs/2008ApJ...689.1234M} {689, 1234}

\bibitem[\protect\citeauthoryear{{Mathews} \& {Bregman}}{{Mathews} \&
  {Bregman}}{1978}]{Mathews78}
{Mathews} W.~G.,  {Bregman} J.~N.,  1978, \mn@doi [\apj] {10.1086/156379},
  \href {https://ui.adsabs.harvard.edu/abs/1978ApJ...224..308M} {224, 308}

\bibitem[\protect\citeauthoryear{{McCourt}, {Sharma}, {Quataert}  \&
  {Parrish}}{{McCourt} et~al.}{2012}]{McCourt12}
{McCourt} M.,  {Sharma} P.,  {Quataert} E.,   {Parrish} I.~J.,  2012, \mn@doi
  [\mnras] {10.1111/j.1365-2966.2011.19972.x}, \href
  {https://ui.adsabs.harvard.edu/abs/2012MNRAS.419.3319M} {419, 3319}

\bibitem[\protect\citeauthoryear{{McDonald}, {Gaspari}, {McNamara}  \&
  {Tremblay}}{{McDonald} et~al.}{2018}]{McDonald18}
{McDonald} M.,  {Gaspari} M.,  {McNamara} B.~R.,   {Tremblay} G.~R.,  2018,
  \mn@doi [\apj] {10.3847/1538-4357/aabace}, \href
  {https://ui.adsabs.harvard.edu/abs/2018ApJ...858...45M} {858, 45}

\bibitem[\protect\citeauthoryear{{McQuinn} \& {Werk}}{{McQuinn} \&
  {Werk}}{2018}]{McQuinn18}
{McQuinn} M.,  {Werk} J.~K.,  2018, \mn@doi [\apj] {10.3847/1538-4357/aa9d3f},
  \href {https://ui.adsabs.harvard.edu/abs/2018ApJ...852...33M} {852, 33}

\bibitem[\protect\citeauthoryear{{Miller} \& {Bregman}}{{Miller} \&
  {Bregman}}{2015}]{Miller15}
{Miller} M.~J.,  {Bregman} J.~N.,  2015, \mn@doi [\apj]
  {10.1088/0004-637X/800/1/14}, \href
  {https://ui.adsabs.harvard.edu/abs/2015ApJ...800...14M} {800, 14}

\bibitem[\protect\citeauthoryear{{Muratov}, {Kere{\v{s}}},
  {Faucher-Gigu{\`e}re}, {Hopkins}, {Quataert}  \& {Murray}}{{Muratov}
  et~al.}{2015}]{Muratov15}
{Muratov} A.~L.,  {Kere{\v{s}}} D.,  {Faucher-Gigu{\`e}re} C.-A.,  {Hopkins}
  P.~F.,  {Quataert} E.,   {Murray} N.,  2015, \mn@doi [\mnras]
  {10.1093/mnras/stv2126}, \href
  {https://ui.adsabs.harvard.edu/abs/2015MNRAS.454.2691M} {454, 2691}

\bibitem[\protect\citeauthoryear{{Murray} \& {Chang}}{{Murray} \&
  {Chang}}{2015}]{Murray15}
{Murray} N.,  {Chang} P.,  2015, \mn@doi [\apj] {10.1088/0004-637X/804/1/44},
  \href {https://ui.adsabs.harvard.edu/abs/2015ApJ...804...44M} {804, 44}

\bibitem[\protect\citeauthoryear{{Murray}, {Chang}, {Murray}  \&
  {Pittman}}{{Murray} et~al.}{2017}]{Murray17}
{Murray} D.~W.,  {Chang} P.,  {Murray} N.~W.,   {Pittman} J.,  2017, \mn@doi
  [\mnras] {10.1093/mnras/stw2796}, \href
  {https://ui.adsabs.harvard.edu/abs/2017MNRAS.465.1316M} {465, 1316}

\bibitem[\protect\citeauthoryear{{Narayan} \& {Fabian}}{{Narayan} \&
  {Fabian}}{2011}]{NarayanFabian11}
{Narayan} R.,  {Fabian} A.~C.,  2011, \mn@doi [\mnras]
  {10.1111/j.1365-2966.2011.18987.x}, \href
  {https://ui.adsabs.harvard.edu/abs/2011MNRAS.415.3721N} {415, 3721}

\bibitem[\protect\citeauthoryear{{Narayan} \& {Medvedev}}{{Narayan} \&
  {Medvedev}}{2001}]{Narayan01}
{Narayan} R.,  {Medvedev} M.~V.,  2001, \mn@doi [\apjl] {10.1086/338325}, \href
  {https://ui.adsabs.harvard.edu/abs/2001ApJ...562L.129N} {562, L129}

\bibitem[\protect\citeauthoryear{{Nelson}, {Vogelsberger}, {Genel}, {Sijacki},
  {Kere{\v{s}}}, {Springel}  \& {Hernquist}}{{Nelson} et~al.}{2013}]{Nelson13}
{Nelson} D.,  {Vogelsberger} M.,  {Genel} S.,  {Sijacki} D.,  {Kere{\v{s}}} D.,
   {Springel} V.,   {Hernquist} L.,  2013, \mn@doi [\mnras]
  {10.1093/mnras/sts595}, \href
  {https://ui.adsabs.harvard.edu/abs/2013MNRAS.429.3353N} {429, 3353}

\bibitem[\protect\citeauthoryear{{Nelson} et~al.,}{{Nelson}
  et~al.}{2019}]{Nelson19}
{Nelson} D.,  et~al., 2019, \mn@doi [\mnras] {10.1093/mnras/stz2306}, \href
  {https://ui.adsabs.harvard.edu/abs/2019MNRAS.490.3234N} {490, 3234}

\bibitem[\protect\citeauthoryear{{Nica}, {Oppenheimer}, {Crain}, {Bogd{\'a}n},
  {Davies}, {Forman}, {Kraft}  \& {ZuHone}}{{Nica} et~al.}{2021}]{Nica21}
{Nica} A.,  {Oppenheimer} B.~D.,  {Crain} R.~A.,  {Bogd{\'a}n} {\'A}.,
  {Davies} J.~J.,  {Forman} W.~R.,  {Kraft} R.~P.,   {ZuHone} J.~A.,  2021,
  arXiv e-prints, \href {https://ui.adsabs.harvard.edu/abs/2021arXiv211214919N}
  {p. arXiv:2112.14919}

\bibitem[\protect\citeauthoryear{{Oppenheimer}}{{Oppenheimer}}{2018}]{Oppenheimer18}
{Oppenheimer} B.~D.,  2018, \mn@doi [\mnras] {10.1093/mnras/sty1918}, \href
  {https://ui.adsabs.harvard.edu/abs/2018MNRAS.480.2963O} {480, 2963}

\bibitem[\protect\citeauthoryear{{Pandya} et~al.,}{{Pandya}
  et~al.}{2021}]{Pandya21}
{Pandya} V.,  et~al., 2021, \mn@doi [\mnras] {10.1093/mnras/stab2714}, \href
  {https://ui.adsabs.harvard.edu/abs/2021MNRAS.508.2979P} {508, 2979}

\bibitem[\protect\citeauthoryear{{Pezzulli}, {Fraternali}  \&
  {Binney}}{{Pezzulli} et~al.}{2017}]{Pezzulli17}
{Pezzulli} G.,  {Fraternali} F.,   {Binney} J.,  2017, \mn@doi [\mnras]
  {10.1093/mnras/stx029}, \href
  {https://ui.adsabs.harvard.edu/abs/2017MNRAS.467..311P} {467, 311}

\bibitem[\protect\citeauthoryear{{Popescu, C. C.}, {Tuffs, R. J.}, {Kylafis, N.
  D.}  \& {Madore, B. F.}}{{Popescu, C. C.} et~al.}{2004}]{Popescu04}
{Popescu, C. C.} {Tuffs, R. J.} {Kylafis, N. D.}  {Madore, B. F.} 2004, \mn@doi
  [A\&A] {10.1051/0004-6361:20031581}, 414, 45

\bibitem[\protect\citeauthoryear{{Prochaska} \& {Zheng}}{{Prochaska} \&
  {Zheng}}{2019}]{XYZ19}
{Prochaska} J.~X.,  {Zheng} Y.,  2019, \mn@doi [\mnras] {10.1093/mnras/stz261},
  \href {https://ui.adsabs.harvard.edu/abs/2019MNRAS.485..648P} {485, 648}

\bibitem[\protect\citeauthoryear{{Prochaska} et~al.,}{{Prochaska}
  et~al.}{2019}]{Prochaska19}
{Prochaska} J.~X.,  et~al., 2019, \mn@doi [\apjs]
  {10.3847/1538-4365/ab2b9a10.48550/arXiv.1908.07675}, \href
  {https://ui.adsabs.harvard.edu/abs/2019ApJS..243...24P} {243, 24}

\bibitem[\protect\citeauthoryear{{Putman}, {Peek}  \& {Joung}}{{Putman}
  et~al.}{2012}]{Putman12}
{Putman} M.~E.,  {Peek} J.~E.~G.,   {Joung} M.~R.,  2012, \mn@doi [\araa]
  {10.1146/annurev-astro-081811-125612}, \href
  {https://ui.adsabs.harvard.edu/abs/2012ARA&A..50..491P} {50, 491}

\bibitem[\protect\citeauthoryear{{Quataert} \& {Narayan}}{{Quataert} \&
  {Narayan}}{2000}]{Quataert00}
{Quataert} E.,  {Narayan} R.,  2000, \mn@doi [\apj] {10.1086/308171}, \href
  {https://ui.adsabs.harvard.edu/abs/2000ApJ...528..236Q} {528, 236}

\bibitem[\protect\citeauthoryear{Ramachandran \& Varoquaux}{Ramachandran \&
  Varoquaux}{2011}]{ramachandran2011mayavi}
Ramachandran P.,  Varoquaux G.,  2011, Computing in Science \& Engineering, 13,
  40

\bibitem[\protect\citeauthoryear{{Ravi} et~al.,}{{Ravi} et~al.}{2023}]{Ravi23}
{Ravi} V.,  et~al., 2023, \mn@doi [arXiv e-prints] {10.48550/arXiv.2301.01000},
  \href {https://ui.adsabs.harvard.edu/abs/2023arXiv230101000R} {p.
  arXiv:2301.01000}

\bibitem[\protect\citeauthoryear{{Robertson} \& {Goldreich}}{{Robertson} \&
  {Goldreich}}{2012}]{Robertson12}
{Robertson} B.,  {Goldreich} P.,  2012, \mn@doi [\apjl]
  {10.1088/2041-8205/750/2/L31}, \href
  {https://ui.adsabs.harvard.edu/abs/2012ApJ...750L..31R} {750, L31}

\bibitem[\protect\citeauthoryear{{Rodr{\'\i}guez-Puebla}, {Behroozi},
  {Primack}, {Klypin}, {Lee}  \& {Hellinger}}{{Rodr{\'\i}guez-Puebla}
  et~al.}{2016}]{rodriguezpuebla16}
{Rodr{\'\i}guez-Puebla} A.,  {Behroozi} P.,  {Primack} J.,  {Klypin} A.,  {Lee}
  C.,   {Hellinger} D.,  2016, \mn@doi [\mnras] {10.1093/mnras/stw1705}, \href
  {https://ui.adsabs.harvard.edu/abs/2016MNRAS.462..893R} {462, 893}

\bibitem[\protect\citeauthoryear{{Ro{\v{s}}kar}, {Debattista}, {Brooks},
  {Quinn}, {Brook}, {Governato}, {Dalcanton}  \& {Wadsley}}{{Ro{\v{s}}kar}
  et~al.}{2010}]{Roskar10}
{Ro{\v{s}}kar} R.,  {Debattista} V.~P.,  {Brooks} A.~M.,  {Quinn} T.~R.,
  {Brook} C.~B.,  {Governato} F.,  {Dalcanton} J.~J.,   {Wadsley} J.,  2010,
  \mn@doi [\mnras] {10.1111/j.1365-2966.2010.17178.x}, \href
  {https://ui.adsabs.harvard.edu/abs/2010MNRAS.408..783R} {408, 783}

\bibitem[\protect\citeauthoryear{{Sancisi}, {Fraternali}, {Oosterloo}  \& {van
  der Hulst}}{{Sancisi} et~al.}{2008}]{Sancisi08}
{Sancisi} R.,  {Fraternali} F.,  {Oosterloo} T.,   {van der Hulst} T.,  2008,
  \mn@doi [\aapr] {10.1007/s00159-008-0010-0}, \href
  {https://ui.adsabs.harvard.edu/abs/2008A&ARv..15..189S} {15, 189}

\bibitem[\protect\citeauthoryear{{Schneider}, {Ostriker}, {Robertson}  \&
  {Thompson}}{{Schneider} et~al.}{2020}]{Schneider20}
{Schneider} E.~E.,  {Ostriker} E.~C.,  {Robertson} B.~E.,   {Thompson} T.~A.,
  2020, \mn@doi [\apj] {10.3847/1538-4357/ab8ae8}, \href
  {https://ui.adsabs.harvard.edu/abs/2020ApJ...895...43S} {895, 43}

\bibitem[\protect\citeauthoryear{{Sch{\"o}nrich} \& {Binney}}{{Sch{\"o}nrich}
  \& {Binney}}{2009}]{Schonrich09}
{Sch{\"o}nrich} R.,  {Binney} J.,  2009, \mn@doi [\mnras]
  {10.1111/j.1365-2966.2009.14750.x}, \href
  {https://ui.adsabs.harvard.edu/abs/2009MNRAS.396..203S} {396, 203}

\bibitem[\protect\citeauthoryear{{Shapiro}}{{Shapiro}}{1973}]{Shapiro73}
{Shapiro} S.~L.,  1973, \mn@doi [\apj] {10.1086/152396}, \href
  {https://ui.adsabs.harvard.edu/abs/1973ApJ...185...69S} {185, 69}

\bibitem[\protect\citeauthoryear{{Shapiro} \& {Field}}{{Shapiro} \&
  {Field}}{1976}]{Shapiro76}
{Shapiro} P.~R.,  {Field} G.~B.,  1976, \mn@doi [\apj] {10.1086/154332}, \href
  {https://ui.adsabs.harvard.edu/abs/1976ApJ...205..762S} {205, 762}

\bibitem[\protect\citeauthoryear{{Sharma}, {Steinmetz}  \&
  {Bland-Hawthorn}}{{Sharma} et~al.}{2012a}]{Sharma12b}
{Sharma} S.,  {Steinmetz} M.,   {Bland-Hawthorn} J.,  2012a, \mn@doi [\apj]
  {10.1088/0004-637X/750/2/107}, \href
  {https://ui.adsabs.harvard.edu/abs/2012ApJ...750..107S} {750, 107}

\bibitem[\protect\citeauthoryear{{Sharma}, {McCourt}, {Quataert}  \&
  {Parrish}}{{Sharma} et~al.}{2012b}]{Sharma12}
{Sharma} P.,  {McCourt} M.,  {Quataert} E.,   {Parrish} I.~J.,  2012b, \mn@doi
  [\mnras] {10.1111/j.1365-2966.2011.20246.x}, \href
  {https://ui.adsabs.harvard.edu/abs/2012MNRAS.420.3174S} {420, 3174}

\bibitem[\protect\citeauthoryear{{Sormani} \& {Sobacchi}}{{Sormani} \&
  {Sobacchi}}{2019}]{Sormani19}
{Sormani} M.~C.,  {Sobacchi} E.,  2019, \mn@doi [\mnras]
  {10.1093/mnras/stz793}, \href
  {https://ui.adsabs.harvard.edu/abs/2019MNRAS.486..215S} {486, 215}

\bibitem[\protect\citeauthoryear{{Sormani}, {Sobacchi}, {Pezzulli}, {Binney}
  \& {Klessen}}{{Sormani} et~al.}{2018}]{Sormani18}
{Sormani} M.~C.,  {Sobacchi} E.,  {Pezzulli} G.,  {Binney} J.,   {Klessen}
  R.~S.,  2018, \mn@doi [\mnras] {10.1093/mnras/sty2500}, \href
  {https://ui.adsabs.harvard.edu/abs/2018MNRAS.481.3370S} {481, 3370}

\bibitem[\protect\citeauthoryear{Springel, di Matteo  \& Hernquist}{Springel
  et~al.}{2005}]{Springel05_2}
Springel V.,  di Matteo T.,   Hernquist L.,  2005, \mn@doi [Monthly Notices of
  the Royal Astronomical Society] {10.1111/j.1365-2966.2005.09238.x}, 361, 776

\bibitem[\protect\citeauthoryear{{Stern}, {Fielding}, {Faucher-Gigu{\`e}re}  \&
  {Quataert}}{{Stern} et~al.}{2019}]{Stern19}
{Stern} J.,  {Fielding} D.,  {Faucher-Gigu{\`e}re} C.-A.,   {Quataert} E.,
  2019, \mn@doi [\mnras] {10.1093/mnras/stz1859}, \href
  {https://ui.adsabs.harvard.edu/abs/2019MNRAS.488.2549S} {488, 2549}

\bibitem[\protect\citeauthoryear{{Stern}, {Fielding}, {Faucher-Gigu{\`e}re}  \&
  {Quataert}}{{Stern} et~al.}{2020}]{Stern20}
{Stern} J.,  {Fielding} D.,  {Faucher-Gigu{\`e}re} C.-A.,   {Quataert} E.,
  2020, \mn@doi [\mnras] {10.1093/mnras/staa198}, \href
  {https://ui.adsabs.harvard.edu/abs/2020MNRAS.492.6042S} {492, 6042}

\bibitem[\protect\citeauthoryear{{Stern} et~al.,}{{Stern}
  et~al.}{2021a}]{Stern21a}
{Stern} J.,  et~al., 2021a, \mn@doi [\apj] {10.3847/1538-4357/abd776}, \href
  {https://ui.adsabs.harvard.edu/abs/2021ApJ...911...88S} {911, 88}

\bibitem[\protect\citeauthoryear{{Stern} et~al.,}{{Stern}
  et~al.}{2021b}]{Stern21b}
{Stern} J.,  et~al., 2021b, \mn@doi [\mnras] {10.1093/mnras/stab2240}, \href
  {https://ui.adsabs.harvard.edu/abs/2021MNRAS.507.2869S} {507, 2869}

\bibitem[\protect\citeauthoryear{{Stevens}, {Lagos}, {Contreras}, {Croton},
  {Padilla}, {Schaller}, {Schaye}  \& {Theuns}}{{Stevens}
  et~al.}{2017}]{Stevens17}
{Stevens} A. R.~H.,  {Lagos} C. d.~P.,  {Contreras} S.,  {Croton} D.~J.,
  {Padilla} N.~D.,  {Schaller} M.,  {Schaye} J.,   {Theuns} T.,  2017, \mn@doi
  [\mnras] {10.1093/mnras/stx243}, \href
  {https://ui.adsabs.harvard.edu/abs/2017MNRAS.467.2066S} {467, 2066}

\bibitem[\protect\citeauthoryear{{Stewart}, {Brooks}, {Bullock}, {Maller},
  {Diemand}, {Wadsley}  \& {Moustakas}}{{Stewart} et~al.}{2013}]{Stewart13}
{Stewart} K.~R.,  {Brooks} A.~M.,  {Bullock} J.~S.,  {Maller} A.~H.,  {Diemand}
  J.,  {Wadsley} J.,   {Moustakas} L.~A.,  2013, \mn@doi [\apj]
  {10.1088/0004-637X/769/1/74}, \href
  {https://ui.adsabs.harvard.edu/abs/2013ApJ...769...74S} {769, 74}

\bibitem[\protect\citeauthoryear{{Stewart} et~al.,}{{Stewart}
  et~al.}{2017}]{Stewart17}
{Stewart} K.~R.,  et~al., 2017, \mn@doi [\apj] {10.3847/1538-4357/aa6dff},
  \href {https://ui.adsabs.harvard.edu/abs/2017ApJ...843...47S} {843, 47}

\bibitem[\protect\citeauthoryear{{Su} et~al.,}{{Su} et~al.}{2019}]{Su19}
{Su} K.-Y.,  et~al., 2019, \mn@doi [\mnras] {10.1093/mnras/stz1494}, \href
  {https://ui.adsabs.harvard.edu/abs/2019MNRAS.487.4393S} {487, 4393}

\bibitem[\protect\citeauthoryear{{Su} et~al.,}{{Su} et~al.}{2020}]{Su20}
{Su} K.-Y.,  et~al., 2020, \mn@doi [\mnras] {10.1093/mnras/stz3011}, \href
  {https://ui.adsabs.harvard.edu/abs/2020MNRAS.491.1190S} {491, 1190}

\bibitem[\protect\citeauthoryear{{Sweet}}{{Sweet}}{1950}]{Sweet50}
{Sweet} P.~A.,  1950, \mn@doi [\mnras] {10.1093/mnras/110.6.548}, \href
  {https://ui.adsabs.harvard.edu/abs/1950MNRAS.110..548S} {110, 548}

\bibitem[\protect\citeauthoryear{{Tan}, {Oh}  \& {Gronke}}{{Tan}
  et~al.}{2023}]{Tan23}
{Tan} B.,  {Oh} S.~P.,   {Gronke} M.,  2023, \mn@doi [\mnras]
  {10.1093/mnras/stad236}, \href
  {https://ui.adsabs.harvard.edu/abs/2023MNRAS.520.2571T} {520, 2571}

\bibitem[\protect\citeauthoryear{{Tassoul}}{{Tassoul}}{2007}]{Tassoul07}
{Tassoul} J.-L.,  2007, {Stellar Rotation}

\bibitem[\protect\citeauthoryear{{Thompson}, {Quataert}, {Zhang}  \&
  {Weinberg}}{{Thompson} et~al.}{2016}]{Thompson16}
{Thompson} T.~A.,  {Quataert} E.,  {Zhang} D.,   {Weinberg} D.~H.,  2016,
  \mn@doi [\mnras] {10.1093/mnras/stv2428}, \href
  {https://ui.adsabs.harvard.edu/abs/2016MNRAS.455.1830T} {455, 1830}

\bibitem[\protect\citeauthoryear{{Trapp} et~al.,}{{Trapp}
  et~al.}{2022}]{Trapp22}
{Trapp} C.~W.,  et~al., 2022, \mn@doi [\mnras] {10.1093/mnras/stab3251}, \href
  {https://ui.adsabs.harvard.edu/abs/2022MNRAS.509.4149T} {509, 4149}

\bibitem[\protect\citeauthoryear{{Truong}, {Pillepich}, {Nelson}, {Werner}  \&
  {Hernquist}}{{Truong} et~al.}{2021}]{Truong21}
{Truong} N.,  {Pillepich} A.,  {Nelson} D.,  {Werner} N.,   {Hernquist} L.,
  2021, \mn@doi [\mnras] {10.1093/mnras/stab2638}, \href
  {https://ui.adsabs.harvard.edu/abs/2021MNRAS.508.1563T} {508, 1563}

\bibitem[\protect\citeauthoryear{Truong et~al.,}{Truong
  et~al.}{2023}]{Truong23}
Truong N.,  et~al., 2023, \mn@doi [Monthly Notices of the Royal Astronomical
  Society] {10.1093/mnras/stad2216}, 525, 1976

\bibitem[\protect\citeauthoryear{{Voit}, {Donahue}, {O'Shea}, {Bryan}, {Sun}
  \& {Werner}}{{Voit} et~al.}{2015}]{Voit15}
{Voit} G.~M.,  {Donahue} M.,  {O'Shea} B.~W.,  {Bryan} G.~L.,  {Sun} M.,
  {Werner} N.,  2015, \mn@doi [\apjl] {10.1088/2041-8205/803/2/L21}, \href
  {https://ui.adsabs.harvard.edu/abs/2015ApJ...803L..21V} {803, L21}

\bibitem[\protect\citeauthoryear{{Voit}, {Meece}, {Li}, {O'Shea}, {Bryan}  \&
  {Donahue}}{{Voit} et~al.}{2017}]{Voit17}
{Voit} G.~M.,  {Meece} G.,  {Li} Y.,  {O'Shea} B.~W.,  {Bryan} G.~L.,
  {Donahue} M.,  2017, \mn@doi [\apj] {10.3847/1538-4357/aa7d04}, \href
  {https://ui.adsabs.harvard.edu/abs/2017ApJ...845...80V} {845, 80}

\bibitem[\protect\citeauthoryear{{Wang} \& {Lilly}}{{Wang} \&
  {Lilly}}{2022}]{Wang22}
{Wang} E.,  {Lilly} S.~J.,  2022, \mn@doi [\apj] {10.3847/1538-4357/ac49ed},
  \href {https://ui.adsabs.harvard.edu/abs/2022ApJ...927..217W} {927, 217}

\bibitem[\protect\citeauthoryear{{White} \& {Rees}}{{White} \&
  {Rees}}{1978}]{White78}
{White} S.~D.~M.,  {Rees} M.~J.,  1978, \mn@doi [\mnras]
  {10.1093/mnras/183.3.341}, \href
  {https://ui.adsabs.harvard.edu/abs/1978MNRAS.183..341W} {183, 341}

\bibitem[\protect\citeauthoryear{{Wiersma}, {Schaye}  \& {Smith}}{{Wiersma}
  et~al.}{2009}]{Wiersma09}
{Wiersma} R. P.~C.,  {Schaye} J.,   {Smith} B.~D.,  2009, \mn@doi [\mnras]
  {10.1111/j.1365-2966.2008.14191.x}, \href
  {https://ui.adsabs.harvard.edu/abs/2009MNRAS.393...99W} {393, 99}

\bibitem[\protect\citeauthoryear{{Williams}, {Khan}  \& {McQuinn}}{{Williams}
  et~al.}{2023}]{Williams23}
{Williams} I.~M.,  {Khan} A.,   {McQuinn} M.,  2023, \mn@doi [\mnras]
  {10.1093/mnras/stad293}, \href
  {https://ui.adsabs.harvard.edu/abs/2023MNRAS.520.3626W} {520, 3626}

\bibitem[\protect\citeauthoryear{{Yang} \& {Reynolds}}{{Yang} \&
  {Reynolds}}{2016}]{Yang16}
{Yang} H. Y.~K.,  {Reynolds} C.~S.,  2016, \mn@doi [\apj]
  {10.3847/0004-637X/829/2/90}, \href
  {https://ui.adsabs.harvard.edu/abs/2016ApJ...829...90Y} {829, 90}

\bibitem[\protect\citeauthoryear{{Yang}, {Dav{\'e}}, {Cui}, {Cai}, {Peacock}
  \& {Sorini}}{{Yang} et~al.}{2023}]{Yang23}
{Yang} T.,  {Dav{\'e}} R.,  {Cui} W.,  {Cai} Y.-C.,  {Peacock} J.~A.,
  {Sorini} D.,  2023, \mn@doi [\mnras] {10.1093/mnras/stad3223}, \href
  {https://ui.adsabs.harvard.edu/abs/2023MNRAS.tmp.3089Y} {}

\bibitem[\protect\citeauthoryear{{Yu} et~al.,}{{Yu} et~al.}{2021}]{Yu21}
{Yu} S.,  et~al., 2021, arXiv e-prints, \href
  {https://ui.adsabs.harvard.edu/abs/2021arXiv210303888Y} {p. arXiv:2103.03888}

\bibitem[\protect\citeauthoryear{{Yu} et~al.,}{{Yu} et~al.}{2022}]{Yu22}
{Yu} S.,  et~al., 2022, arXiv e-prints, \href
  {https://ui.adsabs.harvard.edu/abs/2022arXiv221003845Y} {p. arXiv:2210.03845}

\bibitem[\protect\citeauthoryear{{ZuHone} \& {Hallman}}{{ZuHone} \&
  {Hallman}}{2016}]{ZuHone16}
{ZuHone} J.~A.,  {Hallman} E.~J.,  2016, {pyXSIM: Synthetic X-ray observations
  generator}, Astrophysics Source Code Library, record ascl:1608.002
  (\mn@eprint {ascl} {1608.002})

\bibitem[\protect\citeauthoryear{{ZuHone} et~al.,}{{ZuHone}
  et~al.}{2023}]{ZuHone23}
{ZuHone} J.~A.,  et~al., 2023, \mn@doi [arXiv e-prints]
  {10.48550/arXiv.2307.01269}, \href
  {https://ui.adsabs.harvard.edu/abs/2023arXiv230701269Z} {p. arXiv:2307.01269}

\bibitem[\protect\citeauthoryear{{van der Kruit}}{{van der
  Kruit}}{2007}]{VanderKruit07}
{van der Kruit} P.~C.,  2007, \mn@doi [\aap] {10.1051/0004-6361:20066941},
  \href {https://ui.adsabs.harvard.edu/abs/2007A&A...466..883V} {466, 883}

\makeatother
\end{thebibliography}

\appendix

\section{$\theta$-dependent rotation profiles}\label{a:F}

In this section we give analytic solutions for hot CGM inflows in the slow-rotation limit as derived in section~\ref{s: analytic sol}, for different assumptions on how the gas rotates at large radii. This dependence of $\Omega_1$ on $\theta$ is set by the outer boundary condition of the solution and is parameterized using the function $F(\theta)$ (see eqn.~\ref{e:forms}):
\begin{equation}
\Omega_1 = \frac{\vc}{\Rcirc}\left(\frac{r}{\Rcirc}\right)^{-2}F(\theta)~,
\end{equation}
where $F(\pi/2)=1$. The solution in the main text (eqn.~\ref{e:solution}) is for $F(\theta)=1$. For $F=\sin\theta$ we find
\begin{eqnarray}
f_{P}&=& \frac{3}{8} \sin^{4}{\left(\theta \right)} - \frac{3}{10}  \nonumber \\
f_{\rho}&=&\frac{15}{8} \sin^{4}{\left(\theta \right)} - \frac{7}{10}  \nonumber \\
f_{T}&=&-  \frac{3}{2} \sin^{4}{\left(\theta \right)} +  \frac{2}{5}  ~.
\end{eqnarray}
For $F(\theta)=\sin^2\theta$ we get
\begin{eqnarray}
f_{P}&=&\frac{1}{4} \sin^{6}{\left(\theta \right)} - \frac{1}{5} \nonumber \\
f_{\rho}&=&\frac{19}{12} \sin^{6}{\left(\theta \right)}  - \frac{7}{15} \nonumber \\
f_{T}&=&- \frac{4}{3} \sin^{6}{\left(\theta \right)}+ \frac{4}{15} ~,
\end{eqnarray}
while for $F(\theta)=\sin^3\theta$ we get
\begin{eqnarray}
f_{P}&=&\frac{3}{16} \sin^{8}{\left(\theta \right)} - \frac{16}{105} \nonumber \\
f_{\rho}&=&\frac{23}{16} \sin^{8}{\left(\theta \right)} - \frac{16}{45} \nonumber \\
f_{T}&=& -\frac{5}{4} \sin^{8}{\left(\theta \right)}+\frac{64}{315} ~.
\end{eqnarray}

\section{Dimensionless flow equations}\label{a:Mdotcrit}

\newcommand{\rtag}{r'}
\newcommand{\Ptag}{P'}
\renewcommand{\rhotag}{\rho'}
\newcommand{\cstag}{\cs'}
\newcommand{\vrtag}{\vr'}
\newcommand{\vthetatag}{v_\theta'}
\newcommand{\vphitag}{\vphi'}
\renewcommand{\mp}{m_{\rm p}}
The flow equations~(\ref{e:HSE r}) -- (\ref{e:K with AM}) can be dedimensionalized, if we approximate the cooling function as a power-law $\Lambda=\Lambda_{10^6}(T/10^6)^{-l}$. We define the dimensionless variables as
\begin{eqnarray}
\rtag &\equiv& \frac{r}{\Rcirc} \nonumber\\
\{\vrtag,\vthetatag,\vphitag,\cstag\}&\equiv&\frac{\{\vr,v_\theta,\vphi,\cs\}}{\vc}\nonumber\\
\rhotag \equiv \frac{\rho}{\rho^\star},& &\ \Ptag\equiv \frac{P}{\frac{3}{5}\rho^\star\vc^2}
\end{eqnarray}
with the density normalization equal to 
\begin{equation}
\rho^\star = \frac{9\mp^2\vc^{3+2l}}{10X^2\Lambda_{10^6}\Rcirc}
\end{equation}
where $\mp/X=\rho/\nH$ ($\mp$ and $X$ are the proton mass and hydrogen mass fraction, respectively). This gives for the flow equations
\begin{eqnarray}
\frac{\partial \Ptag}{\partial \rtag} &=& -\frac{\rhotag}{\rtag} + \frac{\vphitag^2}{\rtag} \nonumber\\
\frac{\partial\Ptag}{\partial \theta} &=& -\rhotag\vphitag^2\frac{\cos\theta}{\sin\theta} \nonumber\\
\vrtag\frac{\partial(\vphitag\rtag\sin\theta)}{\partial \rtag} +\vthetatag\frac{\partial(\vphitag\rtag\sin\theta)}{\rtag\partial \theta} &=& 0\nonumber\\
\frac{1}{\rtag^2}\frac{\partial(\rhotag\vrtag\rtag^2)}{\partial \rtag} +\frac{1}{\rtag\sin\theta}\frac{\partial(\rhotag\vthetatag\rtag^2)}{\partial \theta} &=& 0\nonumber\\
\vrtag\frac{\partial(\ln K)}{\partial \rtag} + \vthetatag\frac{\partial(\ln K)}{\rtag\partial \theta}  &=&\rhotag\cstag^{-2(1+l)}
\end{eqnarray}
These dimensionless equations imply that changing $\Rcirc$ and/or $\vc$ does not change the solution beyond a scaling, if the gas density is also scaled by $\rho^\star$.  

The value of $\rho^\star$ is related to the critical accretion rate $\Mdotcrit$ discussed in \cite{Stern20}:
\begin{equation}\label{e:Mdotcrit}
4\pi \Rcirc^2\rho^\star \vc = \frac{18\pi\mp^2\vc^{4+2l}\Rcirc}{5X^2\Lambda_{10^6}}=\Mdotcrit(\Rcirc,\vc) 
\end{equation}
(see eqn.~8 there).\footnote{In \cite{Stern20} we effectively assumed $l=0$ and neglected a factor of $9/10$ in the calculation of $\Mdotcrit$.} When $\Mdot=\Mdotcrit$ the ratio $\tcool/\tff$ equals unity at $r=\Rcirc$ in the spherical solution (eqn.~\ref{e:rad solution}), and hence the sonic radius of the flow coincides with the circularization radius. The relation (\ref{e:Mdotcrit}) between $\rho^\star$ and $\Mdotcrit$ thus implies that changing $\Rcirc$ and/or $\vc$ does not change the solution beyond a scaling, if $\Mdot$ is also scaled by $\Mdotcrit$. Hot rotating inflows thus form a family of solutions characterized by a single parameter, $\Mdot/\Mdotcrit$, up to a scaling of the physical dimensions.

\section{Gas properties near the time of cooling and accretion}

\begin{figure*}
    \centering
    \includegraphics{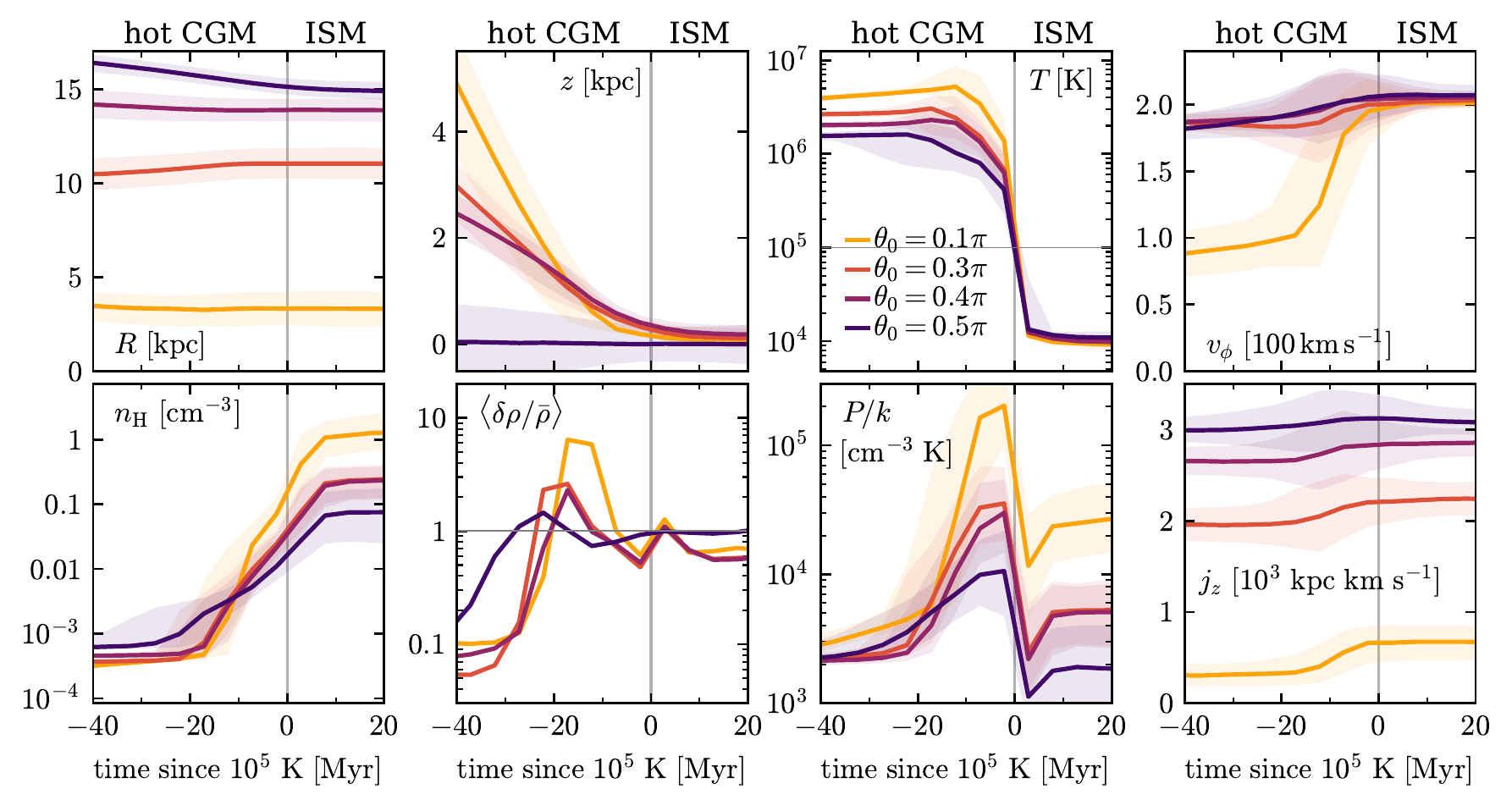}
    \caption{Gas properties along flowlines as in Fig.~\ref{f:track properties t5}, zoomed in on times near the time of cooling. 
Color denotes flowlines originating at different polar angles $\thetalarge$. Solid lines and bands correspond to medians and $16^{\rm th}-84^{\rm th}$ percentiles. 
Top panels show $\Rcyl, z, T$ and $v_\phi$, while bottom panels show $\nH$, $\langle\delta\rho/\bar{\rho}\rangle$, $P/k$ and $j_z$. The $\Rcyl$ and $z$ panels demonstrate that cooling occurs just above the disk, at the `disc-halo interface'. Except for the midplane flowline ($\thetalarge=0.5\pi$) the flow direction is almost vertical at the shown times, with $z$ decreasing by $2-4\kpc$ and $\Rcyl$ changing by $\lesssim0.5\kpc$. At $\approx20\Myr$ prior to cooling gas densities start to increase while density fluctuations become significant. This rapidly shortens the cooling time which allows the eventual drop in temperature to occur over a short timescale of  $\sim10\Myr$, as can be seen in the $T$ panel. 
}
    \label{f:track properties t5 zoom}
\end{figure*}
Figure~\ref{f:track properties t5 zoom} plots gas properties along flowlines as in Fig.~\ref{f:track properties t5}, zoomed in on times near $\tfive$, which corresponds to the time of accretion from the hot CGM onto the cool ISM. 

\section{Magnetic fields in hot and rotating CGM inflows}\label{a:B fields}

In this section we derive an estimate of the magnetic field $B$ in a rotating hot CGM inflow, by assuming ideal MHD conditions and neglecting any potential dynamical effects of the field on the flow. We employ the analytic solution in eqn.~(\ref{e:solution}), disregarding terms of order $(\Rcirc/r)^2$ or higher. Combined with the assumption of axisymmetry and steady-state we then have $v_\theta=\partial/\partial\phi=\partial/\partial t=0$, and the ideal MHD equations for the magnetic field
\begin{equation}\label{e:ideal MHD}
\nabla\cdot\vec{B} = 0, ~~\nabla\times(v\times\vec{B}) = \frac{\partial\vec{B}}{\partial t} 
\end{equation}
reduce to
\begin{eqnarray}\label{e:MHD rotating}
\frac{1}{r^2}\frac{\partial}{\partial r} (r^2 B_r) &=& -\frac{1}{r\sin\theta}\frac{\partial}{\partial \theta} (B_\theta\sin\theta)\nonumber \\
\frac{\partial}{\partial r} (v_r B_\theta r) &=& 0 \nonumber  \\
\frac{\partial}{\partial \theta}(v_r B_\theta \sin\theta) &=& 0 \nonumber  \\ 
\frac{\partial}{\partial r}(v_\phi B_r r -  v_r B_\phi r) &=& -\frac{\partial  (v_\phi B_\theta)}{\partial \theta}  ~.
\end{eqnarray}
Assuming for simplicity that the field is isotropic at the outer radius of the inflow $r_0$ with $B_r(r_0)=B_\theta(r_0)=B_\phi(r_0)\equiv B_0$, the first two equations imply
\begin{eqnarray}
\label{e:B_r}
B_r &=& B_0\left(\frac{r}{r_0}\right)^{-2} \\
\label{e:B_theta}
B_\theta &=& B_0\left(\frac{v_r r}{v_{r}(r_0) r_0}\right)^{-1} ~.
\end{eqnarray}
Equations~(\ref{e:B_r}) and (\ref{e:B_theta})   are the same as in a non-rotating, spherical inflow solution \citep{Shapiro73}.  
To solve for $B_\phi$, we note that the right hand side in the last equation in (\ref{e:MHD rotating}) equals zero. This follows from the third condition in eqn.~(\ref{e:MHD rotating}) together with $v_\phi\propto \sin\theta$ and $\partial v_r/\partial \theta=0$ (eqn.~\ref{e:solution}, up to corrections of order $(r/\Rcirc)^{-2}$). It thus follows that 
\begin{equation}
\frac{\partial(v_\phi B_r r - v_r B_\phi r)}{\partial \theta} = 0 ~,
\end{equation}
and integrating we get
\begin{equation}
v_\phi B_r r - v_r B_\phi r = B_0r_0 (v_{\phi}(r_0)-v_{r}(r_0)) ~.
 \end{equation}
Using eqn.~(\ref{e:B_r}) for $B_r$ then gives
\begin{equation}\label{e:Bphi 0}
 B_\phi = B_0\left(\frac {v_r r}{v_{r}(r_0)r_0}\right)^{-1}\left[1+\frac{v_\phi r_0}{v_{r}(r_0) r}-\frac{v_{\phi}(r_0)}{v_r(r_0)}\right] ~.
 \end{equation}
For $v_\phi=0$ equation~(\ref{e:Bphi 0}) reduces to $B_\phi\propto (v_r r)^{-1}$, as in the \cite{Shapiro73} non-rotating solution. For finite $v_\phi$ the second term in the brackets scales as $\approx r^{-3/2}$ (since at large radii $v_{\phi}\sim r^{-1}$ and $v_{r}\sim r^{-0.5}$), and hence for $r$ somewhat smaller than $r_0$ this term will dominate the other two terms. We thus get
\begin{equation}
B_\phi \approx B_0\frac{v_\phi}{v_r}\left(\frac{r}{r_0}\right)^{-2} = \sqrt{2}\frac{\tcool}{\tff}\frac{r_0^2\Rcirc}{r^3}\sin\theta ~,
\end{equation}
where in the last approximation we use $\vphi=\vc\Rcirc\sin\theta/r$, $v_{r} = r/\tcool$, and $\tff=\sqrt{2}r/\vc$. 
Specifically, at $r\approx\Rcirc$ just prior to accretion we get 
\begin{equation}\label{e:Bphi Rcirc}
 \frac{B_\phi(\Rcirc)}{B_0} \approx \sqrt{2}\frac{\tcool}{\tff}(\Rcirc)\left(\frac{\Rcirc}{r_0}\right)^{-2}\sin\theta ~.
 \end{equation} 
This result suggests that the enhancement of $B_\phi$ at $\Rcirc$ is a product of $\tcool/\tff(\Rcirc)$ and $(r_0/\Rcirc)^{2}$, where the former tracks the number of radians rotated by the inflow (eqn.~\ref{e:Nrot}) and the latter tracks the contraction of an inflowing shell. 

\end{document}